\documentclass[]{pasj01}
\usepackage{url}
\usepackage{lineno}
\pdfoutput=0
\def\uncatcodespecials{\def\do##1{\catcode`##1=12}\dospecials}%
{\catcode`\`=\active\gdef`{\relax\lq}}
\def\setupcode 
    {\tt %
     \spaceskip=0pt \xspaceskip=0pt 
     \catcode`\`=\active
     \uncatcodespecials \obeyspaces
     \catcode`\{=1\catcode`\}=2
    }%
\def\SETUPCODE 
    {\tt %
     \spaceskip=0pt \xspaceskip=0pt 
     \catcode`\`=\active
     \uncatcodespecials \obeyspaces
    }%

\begin{document}
\SetRunningHead{GALAXY CRUISE}{GC}

\title{GALAXY CRUISE: Deep Insights into Interacting Galaxies in the Local Universe\footnote{aaa}} 

\author{
  Masayuki Tanaka\altaffilmark{1,2}, 
  Michitaro Koike\altaffilmark{1},
  Sei'ichiro Naito\altaffilmark{1},
  Junko Shibata\altaffilmark{1},
  Kumiko Usuda-Sato\altaffilmark{3},
  Hitoshi Yamaoka\altaffilmark{1},
  Makoto Ando\altaffilmark{4},
  Kei Ito\altaffilmark{4},
  Umi Kobayashi\altaffilmark{1,2},
  Yutaro Kofuji\altaffilmark{4,5},
  Atsuki Kuwata\altaffilmark{4},
  Suzuka Nakano\altaffilmark{1,2},
  Rhythm Shimakawa\altaffilmark{1},
  Ken-ichi Tadaki\altaffilmark{1},
  Suguru Takebayashi\altaffilmark{4},
  Chie Tsuchiya\altaffilmark{1},
  Tomofumi Umemoto\altaffilmark{1},
  Connor Bottrell\altaffilmark{6}
}
\altaffiltext{1}{National Astronomical Observatory of Japan, 2-21-1 Osawa, Mitaka, Tokyo 181-8588, Japan}
\altaffiltext{2}{Department of Astronomical Science, The Graduate University for Advanced Studies, SOKENDAI, 2-21-1 Osawa, Mitaka, Tokyo, 181-8588, Japan}
\altaffiltext{3}{Subaru Telescope, National Astronomical Observatory of Japan, 650 North A`ohoku Pl, Hilo, HI 96720, USA}
\altaffiltext{4}{Department of Astronomy, Graduate School of Science, The University of Tokyo, 7-3-1, Hongo, Bunkyo-ku, Tokyo, 113-0033, Japan}
\altaffiltext{5}{Mizusawa VLBI Observatory, National Astronomical Observatory of Japan, 2-12 Hoshigaoka, Mizusawa, Oshu, Iwate 023-0861, Japan}
\altaffiltext{6}{Kavli Institute for the Physics and Mathematics of the Universe (WPI), UTIAS, University of Tokyo, Kashiwa, Chiba 277-8583, Japan}


\email{masayuki.tanaka@nao.ac.jp}


\maketitle
\definecolor{gray}{rgb}{0.6, 0.6, 0.6}
\newcommand{\commentblue}[1]{\textcolor{blue} {\textbf{#1}}}
\newcommand{\commentred}[1]{\textcolor{red} {\textbf{#1}}}
\newcommand{\commentgray}[1]{\textcolor{gray} {\textbf{#1}}}


\begin{abstract}
  We present the first results from GALAXY CRUISE, a community (or citizen) science project based on data from
  the Hyper Suprime-Cam Subaru Strategic Program (HSC-SSP).  The current paradigm of galaxy
  evolution suggests that galaxies grow hierarchically via mergers, but our observational
  understanding of the role of mergers is still limited.  The data from HSC-SSP are ideally
  suited to improve our understanding with improved identifications of 
  interacting galaxies thanks to the superb depth and image quality of HSC-SSP.  We have launched
  a community science project, GALAXY CRUISE, in 2019 and collected over 2 million independent classifications of
  20,686 galaxies at $z<0.2$.  We first characterize the accuracy of the participants' classifications
  and demonstrate that it surpasses previous studies based on shallower imaging data.
  We then investigate various aspects of interacting galaxies in detail.  We show that
  there is a clear sign of enhanced activities of super massive black holes and star formation in interacting
  galaxies compared to those in isolated galaxies.
  The enhancement seems particularly strong for galaxies undergoing violent merger.
  We also show that the mass growth rate inferred
  from our results is roughly consistent with the observed evolution of the stellar mass function.
  The 2nd season of GALAXY CRUISE is currently under way and we conclude with future prospects.
  We make the morphological classification catalog used in this paper publicly available at the GALAXY CRUISE website,
  which will be particularly useful for machine-learning applications.
\end{abstract}
\linenumbers

\nolinenumbers
\section{Introduction}
\label{sec:introduction}

\renewcommand{\thefootnote}{\fnsymbol{footnote}}
\footnote[0]{* This paper is made possible by about 10,000 participants and they are individually acknowledged at \url{https://galaxycruise.mtk.nao.ac.jp/en/commendations.html}.}

\renewcommand{\thefootnote}{\arabic{footnote}}
\setcounter{footnote}{0}

Astronomy is one of the research fields where not only professional researchers
but the general public as well have made significant contributions to advance the field.
An old example can be found back in the 18th century when Edmond Halley asked
‘the Curious' to observe the total solar eclipse over England \citep{pasachoff99}.
Astronomy has a long tradition of active contributions of the general public and
amateur astronomers, and new discoveries of interesting objects such as comets
and supernovae are often reported even today, some of which are followed up by
professional astronomers with a larger telescope for detailed studies.
This teamwork between non-professional and professional researchers has been
extremely fruitful to improve our understanding of the Universe.

A new area of such collaboration emerged with the advent of large sky surveys, which
observe billions of celestial objects in a large area of the sky.  While objects
from these surveys are detected and measured in an automated fashion, not all
properties of the objects can be fully captured and cataloged.  Galaxy morphology
is one such property; there is no easy measurement that fully describes the entire
variation of galaxy morphology.
Morphology has traditionally been classified by human eyes as they are good at
recognizing characteristic patterns in an image such as spiral arms. A number of
automated methods have been proposed such as concentration index
\citep{morgan1958,doi93,abraham94}, Sersic index \citep{sersic68}, Gini coefficient
\citep{abraham03}, etc, but human classifications still remain highly
complementary to these automated classifications.

Galaxy Zoo \citep{lintott08} is the first large-scale citizen science (hereafter community science)
project based on
the imaging data from the Sloan Digital Sky Survey (SDSS; \cite{york00}).  Participants classify
morphologies of objects in the SDSS image and a large number of independent classifications
have been collected for nearly a million objects.  There are many scientific papers based on
the Galaxy Zoo's classifications published to date.  Galaxy Zoo is undoubtedly the most
successful community science project in astronomy and the successor and spin-off
projects such as Galaxy Zoo 2 \citep{willett13} have also been successful.

The Galaxy Zoo’s classifications \citep{lintott08} are fundamentally limited by the imaging quality of SDSS.
While SDSS delivered the state of the art imaging data at that time, recent imaging
surveys such as Dark Energy Survey (DES; \cite{des05}), PanSTARRS1 (PS1; \cite{chambers16}),
and Hyper Suprime-Cam Subaru Strategic Program (HSC-SSP; \cite{aihara18b}) yield deeper images, which help improve
the classification accuracy.  Among these surveys, HSC-SSP is
particularly interesting due to its superb image quality and depth; the median seeing in
the $i$-band is 0.6 arcsec and the images reach 26-27th magnitude, i.e., 2-4 magnitudes
deeper than SDSS, DES and PS1.  The depth is particularly important to detect faint spirals
arms as well as diffuse tidal features as we demonstrate in this paper.

Based on the imaging data from HSC-SSP, we have launched GALAXY CRUISE\footnote[1]{
We launched the Japanese site on November 1st, 2019 and English site on February 19, 2020.
The 1st season ran through April 25, 2022, and the 2nd season is currently in progress.
The GALAXY CRUISE website is \url{https://galaxycruise.mtk.nao.ac.jp/}.
}, the first
Japanese community science project in astronomy.  In order to fully exploit the superb
imaging data, we put an emphasis on interacting galaxies.  In the bottom-up galaxy formation
scenario, we expect that galaxy-galaxy mergers play a fundamental role in the evolution
of galaxies over the cosmic time.  However, our understanding of mergers is
is still limited; frequency of mergers, triggering of starburst
and active galactic nuclei (AGN) still remain unclear.  This is largely due to the difficulty of
identifying interacting/merging galaxies.  A traditional way to identify mergers is to look
for tidal features around galaxies, but these features are typically very faint
and deep imaging data is essential to identify them (e.g., \cite{lotz08,tal09,duc15,bottrell19,wilkinson22}).  To avoid this
difficulty, close pairs are also often used to estimate the merger rate \citep{patton00,patton02,lin04}.
But, close pairs have their own difficulties such as contamination of line-of-sight pairs that
are not physically bound and a lack of sensitivity to post-merger phases.
We focus on the former approach to fully exploit the imaging depth of HSC-SSP in this paper and present the first results from GALAXY CRUISE.

The paper is organized as follows.  We introduce GALAXY CRUISE and discuss classification
statistics in Section 2.  We then characterize the classification accuracy in Section 3.
Based on the participants' classifications, we make comparisons to previous community science projects
and also examine properties of interacting galaxies in Section 4.
Finally, the paper is summarized in Section 5.  Unless otherwise stated, we use the AB magnitude
system \citep{oke83} and the adopt the flat cosmology with $\rm H_0=70\ km\ s^{-1}\ Mpc^{-1}$,
$\rm \Omega_M=0.3$, and $\Omega_\Lambda=0.7$.

\section{GALAXY CRUISE}
\label{sec:galaxy_cruise}

\subsection{Motivation}
\label{sec:motivation}

We focus on interacting galaxies to address  some of the outstanding questions about galaxy mergers
as discussed in the previous Section.
As interacting galaxies are a rare population, we need to inspect many galaxies
to gain sufficient statistics to study their properties.  However, the visual
inspection of tens of thousands of galaxies is difficult to perform by
professional astronomers as astronomers are a rare population.  This is
where the interested general public can significantly contribute as Galaxy Zoo
demonstrated.  We follow the same approach and ask participants to
classify galaxies and find interacting galaxies.

\begin{figure}
  \begin{center}
    \includegraphics[width=9cm]{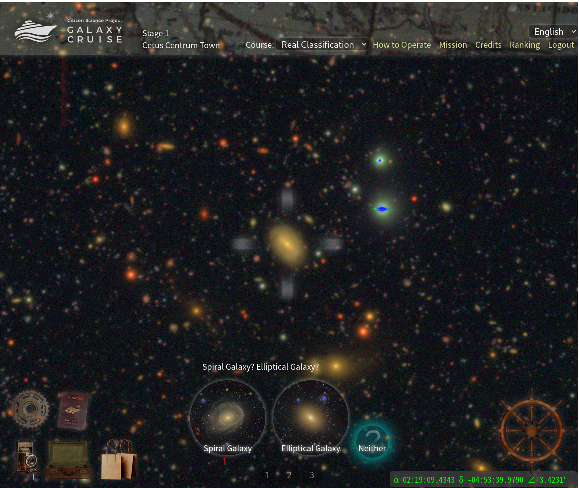}
  \end{center}
  \caption{
    GALAXY CRUISE's classification screen.  Participants classify the target at the center,
    and then 'sail' to the next target.  Passport stamps, souvenirs and other features are
    available to motivate the participants.
  }
  \label{fig:season1}
\end{figure}

A community science project is not just a science project, but it also has a major
aspect of public engagement.  We put an emphasis on this aspect in GALAXY
CRUISE and implemented a lot of gamification factors to motivate and entertain
participants as they classify objects.  As part of this,  we made the theme of
the classification site a cruise ship sailing across the cosmic ocean, hence the name
GALAXY CRUISE. Participants collect passport stamps and souvenirs as they make progress.
There are also seasonal campaigns to make, e.g., a snowman as they classify.
We encourage participants not just classify targets but also freely explore the vast Universe.
For this, we do not to show postage stamps of galaxies and instead show the entire sky observed
by HSC-SSP as a contiguous image.
The contiguous image is essential also for identifications of tidal features; we do not
know a priori where tidal features are and some of them may be located far away from the central galaxy.
A fixed-scale postage-stamp may miss such features, but the contiguous image does not.
A screenshot of our classification site is shown in Fig.~\ref{fig:season1}.

As mentioned above, we use publicly available images from HSC-SSP.
HSC-SSP is a wide and deep survey of the sky using HSC installed at the Subaru Telescope.
It has three survey layers.  The Wide survey covers $\sim1,100$ square degrees of the sky
located mostly around the equator in five broad-band filters ($grizy$) down to $r\sim26.5$
for point sources at $5\sigma$.
The Deep survey has four fields separated roughly equally in R.A., so that one of them
is observable at any time.  The Deep survey typically goes a magnitude deeper than Wide.
Finally, the UltraDeep survey is 2 fields, COSMOS and SXDS,
reaching down to $\sim28$th mag.  Further details of the survey can be found in
\citet{aihara18a}.  The HSC-SSP data are routinely made public and we use mulit-band images from
HSC-SSP Public Data Relase 2 (PDR2; \cite{aihara19}), which was the latest release when our project started.
Most of our targets are from the Wide survey, but we also include objects from the Deep and UltraDeep surveys
allowing for duplication, so that we can characterize the dependence on the depth in our classifications
(see the next Section for details).
Table 2 of \citet{aihara19} summarizes the quality of the imaging data.
We combine the $g$, $r$, and $i$-bands to generate color images of the sky using
\citet{lupton04} scheme.  Participants can adjust levels so that bright/faint parts
of targets can be easily seen.  This feature is particularly useful for identifying
interaction features, which are often diffuse and extended.
Also, single-filter black/white images are available for detailed inspections.

While our primary objective is to identify interacting galaxies, we are also interested
in the morphology of normal (non-interacting) galaxies because the depth and quality of
HSC may bring a difference in our basic understanding of galaxy morphology in the local
Universe.  We choose to make two-step classifications; first is to classify morphology
of galaxies such as elliptical and spiral, and the second is to look for interaction
features.  There are multiple morphology classification schemes.  Some are rather
complicated and difficult, while others are more straightforward.  
How detailed classification participants can make with what accuracy is not a trivial question to answer.

In order to address the question, we made public experiments at National Museum of
Emerging Science and Innovation (Miraikan) in Tokyo.
We display a large-format poster of a contiguous region of
the sky taken by HSC-SSP.  We made an introductory lecture about galaxy morphology to
participants and asked them to classify target galaxies marked on the poster.
We also directly talked to them to see how they recognize various features of galaxies.

Lessons learned from the experiment include: (1) the general public can classify
bright galaxies reasonably well, but (2) faint galaxies (for which basic classifications
can still be made by professional astronomers) are very difficult, and (3) subtle differences such as presence of
weak bars, elliptical vs S0, etc are difficult to recognize.  Following these lessons,
we choose to focus on bright galaxies and adopt a simple classification scheme,
which we detail in the next subsection.
Another lesson learned from the experiment is that edge-on
spirals tend to be classified as ellipticals.  While we made an attempt to make
the difference clear in the tutorial, this tendency is still observed in
our classifications in GALAXY CRUISE.
The introductory lecture used in the experiment turned into training courses for
first-time participants at the GALAXY CRUISE website.
Overall, the public experiment was an essential piece of our project.

\subsection{Classification Scheme}
\label{sec:classificaition_scheme}

\begin{figure}
  \begin{center}
    \includegraphics[width=8cm]{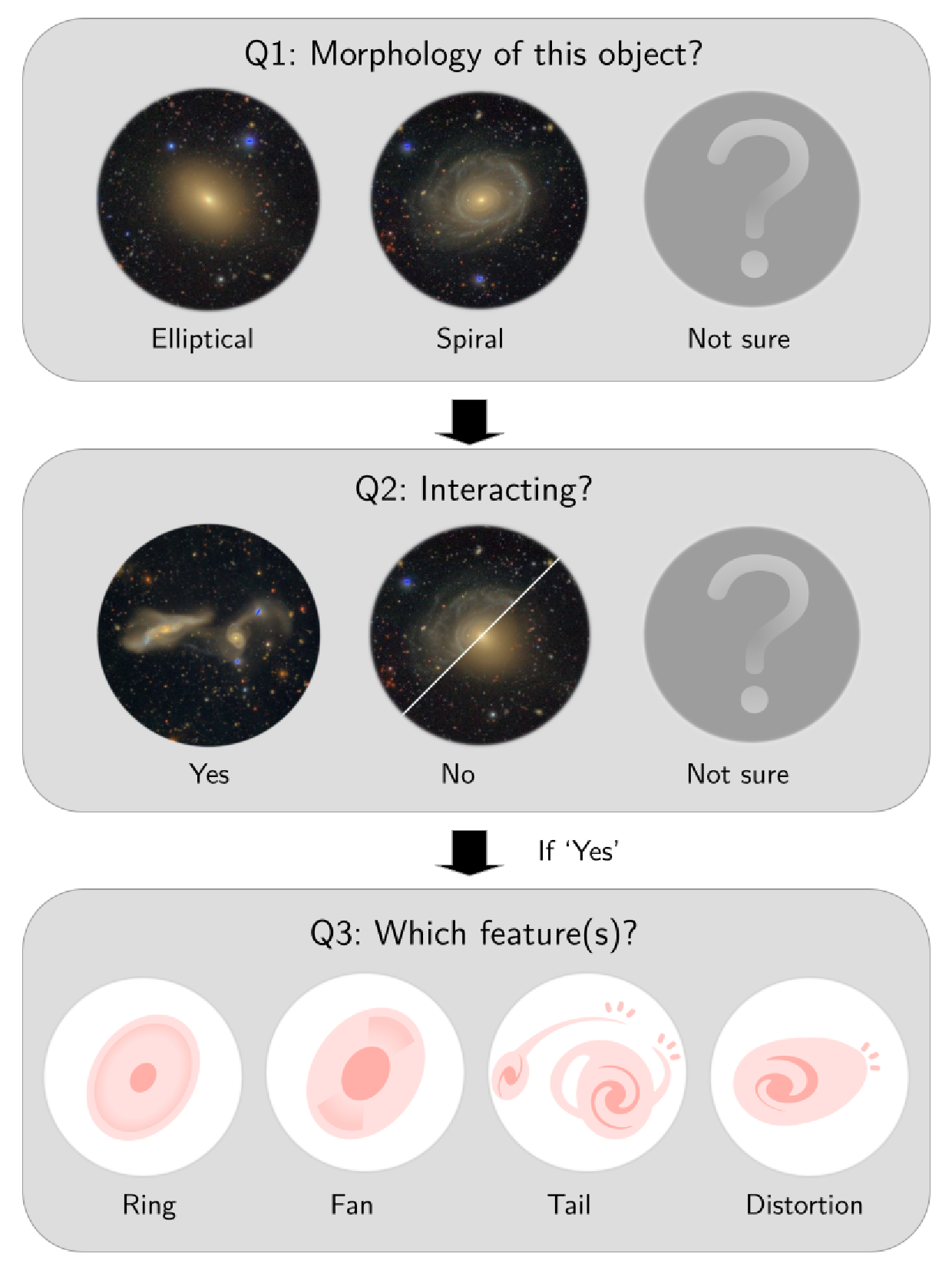}
  \end{center}
  \caption{
    GALAXY CRUISE's classification scheme.  The first question is whether a galaxy is
    spiral or elliptical.  The second question asks if the galaxy is interacting or not.
    If yes, the participant can choose observed feature(s).
  }
  \label{fig:scheme}
\end{figure}

Fig.~\ref{fig:scheme} shows our classification flow.  The first question is whether
a galaxy is an elliptical galaxy or spiral galaxy.  As described earlier, subtle differences
between ellipticals and S0’s and bar structure are difficult to distinguish for the general public.
We thus choose to adopt the simplified classification scheme.  Some interacting galaxies show
significantly distorted shapes and their original morphology can be difficult to infer.
In such a case, the participant can choose ‘not sure’.  When a target galaxy is too close to a nearby
bright star and a fair classification is difficult, the participant can choose ‘not sure’ as well.

The 2nd question is whether there is a sign of interaction or not.  Interacting galaxies exhibit
a variety of features, but we define 4 typical features; (1) tidal tail/stream, (2) distorted shape,
(3) fan/shell structure, and (4) ring structure.  The first feature is perhaps the most prominent
feature seen around interacting galaxies; leading/trailing streams are formed by stars
escaping from an infalling galaxy mostly through the Lagrange points $L_1$ and $L_2$ due to
tidal disruption \citep{eyre11}.  A distorted shape is also a common feature and is also
caused by tidal forces.  A fan/shell feature is likely a special case of collision with
a small impact parameter (i.e., low angular momentum), which results in two-sided caustics with
large opening angles \citep{pop18}.  Finally, the ring structure can be formed by multiple processes
including secular process, but galaxy-galaxy interactions are a strong theory for the origin
of ring galaxies (e.g., \cite{elagali18}) and we interpret the ring structure as a possible sign of interaction.

One can choose multiple interaction features for a given object.  This is because multiple features
are often seen (especially distortion and tail) and also because some of the features are
not easily distinguishable (e.g., arc-like tail vs. fan).  As we will discuss below,
tail and distortion features are the most common features and ring and fan features are
less often observed.

Before a participant joins GALAXY CRUISE and start to make classifications, we ask the participant
to go over an online introductory course, which is based on the tutorial from
the public experiment as mentioned above, to understand our classification scheme.
The online course is a three-step tutorial.
We first explain the primary differences between elliptical and spiral
galaxies such as the presence of spiral arms and disk structure.
We then show that galaxies are normally symmetric in shape and deviations from the symmetry is
often an indication of interaction.  Finally, we introduce the four typical interaction features
discussed above.
We ask several questions to the participant at each step to make
sure the participant understands the scheme.
These three steps are identical to the questions shown in Fig.~\ref{fig:scheme}.
In other words, we explain each question in the tutorial.
After the tutorial, the participant gets a 'boarding pass' to join GALAXY CRUISE.
For further details, the reader is referred to the GALAXY CRUISE website.

\subsection{Target Galaxies}
\label{sec:target_galaxies}

GALAXY CRUISE is based on deep imaging data from HSC-SSP PDR2 \citep{aihara19}, in which the global sky subtraction was applied
to keep extended wings of bright objects.  This is an ideal feature for our purpose as interaction
features are often diffuse and extended.

The target galaxies for classifications are drawn from spectroscopic data from
the Sloan Digital Sky Survey (SDSS; \cite{york00}) and the GAMA survey \citep{driver09}.
The reasons why we draw targets from the spectrscopic data include (1) they are bright objects for
which the participants can make good classifications (see Section \ref{sec:motivation}), and
(2) the distance information allows us to accurately infer physical properties of galaxies such
as star formation rates (SFRs) from emission line fluxes, and (3) the spectra allow us to identify active galactic nuclei (AGN)
from emission line intensity ratios.
The SDSS main galaxy sample \citep{strauss02} is primarily used here, but any objects with spectroscopic
redshifts from Data Release 15 \citep{aguado19} located at $z<0.2$ are included.
We imposed this redshift cut to make sure we focus on nearby galaxies for which
we expect to have sufficient spatial resolution to classify.
We supplement the SDSS redshifts with those from GAMA DR2 \citep{lieske15} to achieve
a higher sampling rate.
The same redshift cut of $z<0.2$ is imposed here as well.
We then apply
a magnitude cut to the cModel photometry in the $z$-band ($z_{cmodel}<17.0$)
to the entire sample to ensure that we focus on very bright galaxies.

The magnitude cut was applied using the HSC photometry, which later turned out not to be the best choice;
due to the high spatial resolution of HSC, bright galaxies, especially face-on spiral galaxies,
are occasionally deblended into multiple pieces \citep{bosch18,aihara19}.  This resulted in reduced brightness of
the central galaxy, and a fraction of bright galaxies were excluded from the target sample
by the magnitude cut.
The same problem occurred in SDSS as well but to a lesser extent.  A comparison between HSC and SDSS
photometry indicates that this effect is not significant for most sources, but about 10\% of
the sources have fainter HSC photometry by more than 1 mag.  In order
to mitigate the problem, GALAXY CRUISE 2nd season has been running since April 2022.
The 2nd season is an extension of the 1st season, but we target fainter galaxies without
any magnitude cut including those that should have been targeted in the 1st season.
They are still bright galaxies with spectroscopic redshifts following the lesson learned from
the public experiment described in Section \ref{sec:motivation}.
We note that the all photometry used in what follows is from SDSS.

The targets are drawn from SDSS (18,960 galaxies) and GAMA (1,726 galaxies).
Among them, 20,116 galaxies are in the Wide layer, and 1,410 galaxies are in the D/UD layer.
As three of the four HSC-SSP D/UD fields are embedded in the Wide fields, 840 galaxies are
duplicated and classified twice; once at the D/UD depth and once at the Wide depth.
We evaluate how classifications change depending on the image depth in Section 3.1.

\subsection{Combined Classifications}
\label{sec:combined_classifications}

\begin{figure}
  \begin{center}
    \includegraphics[width=8cm]{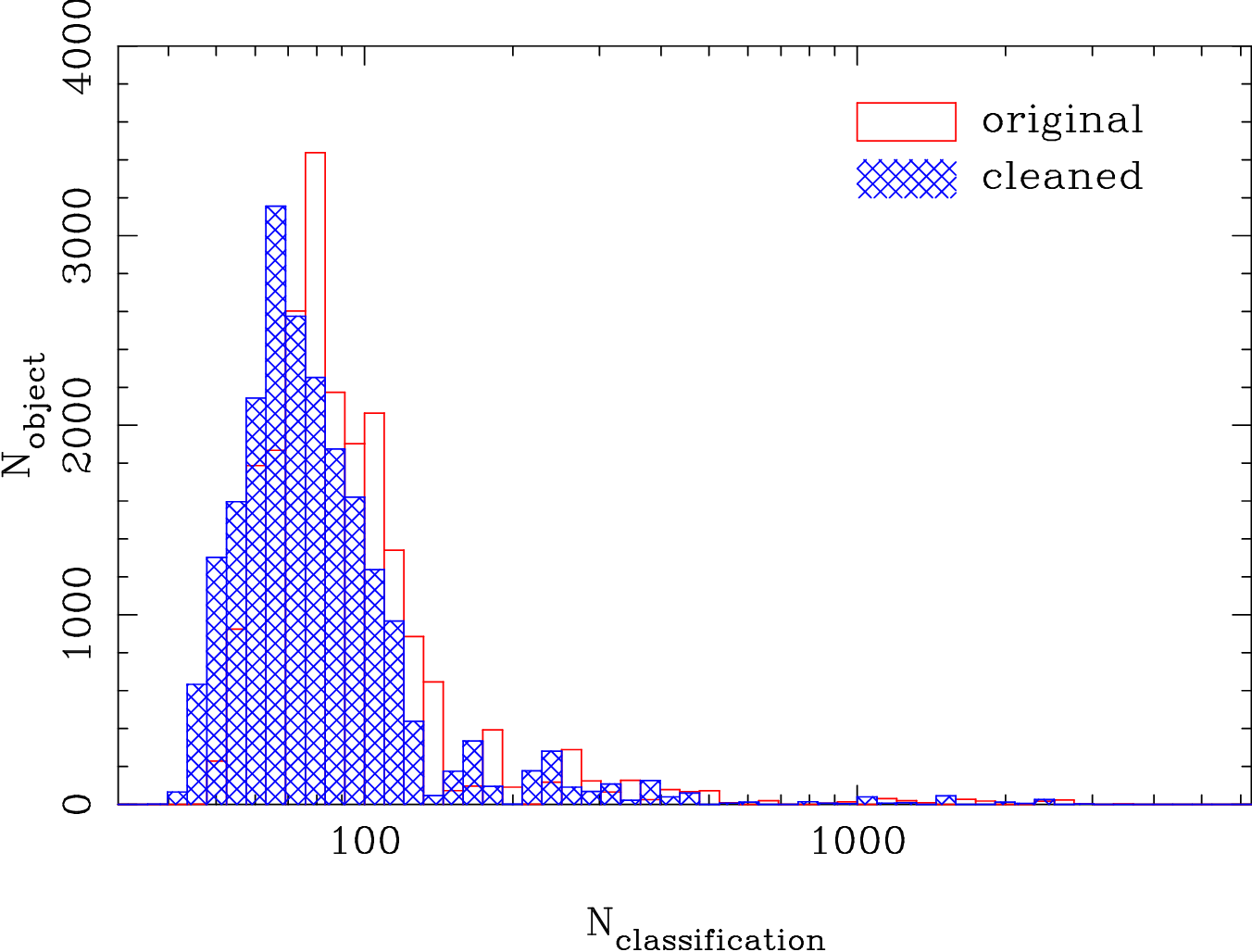}
  \end{center}
  \caption{
    Distribution of the number of classifications for each object.
    The open histogram is the original classifications, while the shaded one is after
    the cleaning of bad classifications (see text for details).  The median of the distribution is
    83 for the original and 74 for the cleaned catalogs, respectively.
  }
  \label{fig:class_num}
\end{figure}

In this paper, we base our analysis on the classifications collected during the 1st season
between November 2019 and April 2022.  Due to the way classifications were made, a small
number of participants classified the same objects multiple times.  We use only the last classification.
Fig.~\ref{fig:class_num} shows the number of classifications
made for each object.   The median of the distribution is 83, which is
sufficient for statistical classifications of objects.
We define a probability that an object is a spiral galaxy as

\begin{equation}
  P(spiral)=\frac{N_{spiral}}{N_{total}},
\end{equation}

\noindent
where $N_{total}$ is the total number of classifications made for that object, and $N_{spiral}$
is the number of spiral classifications among them.  If $P(spiral)=1$, everyone agrees
that an object is a spiral galaxy.  If $P(spiral)=0$ instead, then everyone agrees that
it is an elliptical galaxy.  Of course, there are other types of galaxies in the Universe
such as S0 galaxies, but we adopt this simplified classification scheme as discussed in 
Section~\ref{sec:classificaition_scheme}.
Strictly speaking, $P(spiral)$ is not a probability but just
a fraction of classifications.  However, we will later discuss fraction of spiral galaxies
among our sample.  In order to reduce confusions about which fraction we are discussing,
we refer to the fraction of spiral classifications
by the participants as spiral probability or $P(spiral)$.  In the same manner,
we define $P(int.)$, which indicates the fraction of people who vote for interaction in
the 2nd question.

\begin{figure}
  \begin{center}
    \includegraphics[width=8cm]{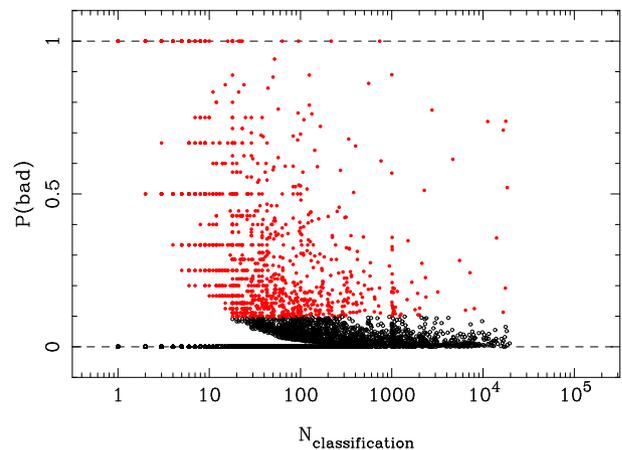}
  \end{center}
  \caption{
    $P(bad)$ plotted against the total number of classification for each participant.  Participants
    with $P(bad)>0.1$ (the red points) are excluded from the main analyses.
    The other participants shown in black are included.
  }
  \label{fig:random_class}
\end{figure}

We first merge all of the individual classifications and compute $P(spiral)$.
While most participants carefully classify objects, there are a small number of people whose
classifications are less accurate.  To make a reliable morphological catalog,
we make an attempt to exclude those less-accurate participants.  
To this aim, we focus on galaxies with obvious morphologies, i.e., galaxies for which
the vast majority of people agree on their morphological types.  To be specific,
we select objects with $P(spiral)<0.05$ or $P(spiral)>0.95$.  These are clearly either 
elliptical or spiral.  Such objects comprise about 10\% of the entire sample.
We then compile a list of participants whose classifications fall in the minority
(i.e., likely incorrect).  We define a probability that a participant makes an inaccurate
classification as

\begin{equation}
  P(bad)=\frac{N(bad)}{N(obvious)},
\end{equation}

\noindent
where $N(obvious)$ is the number of objects with obvious morphology that a participant
classified, and $N(bad)$ is a subset of $N(obvious)$ and is the number of classifications that
the participant falls in the minority.  Fig.~\ref{fig:random_class} shows $P(bad)$ against the total number of
classification that a participant made.  As can be seen, most participants make good
classifications with low $P(bad)$, but there are people whose classifications are less accurate.
We choose to exclude classifications by participants with $P(bad)>0.1$ in this work.
This excludes 276,530 classifications out of total 2,431,455 classifications (11 \%).
We have confirmed that our results do not change if we include them.
We note that Galaxy Zoo adopted a similar approach; they down-weighted participants who
consistently disagreed with the majority.
Note as well that there is a group of participants at $N_{\rm classification}\sim1,000$ in Fig.~\ref{fig:random_class}.
It is likely due to
classification campaigns we have promoted, during which 1,000 classifications give
participants a special image.

The hatched histogram in Fig.~\ref{fig:class_num} shows the number of classifications
for each object after the less accurate participants are excluded.  The median number
of classification is reduced from 83 to 74, but it is still sufficient for statistical analyses in this paper.
We use $P(spiral)$ and $P(int.)$ after excluding these inaccurate classifiers in what follows.

For the first and second questions, we include the 'not sure' option as shown in Fig.~\ref{fig:scheme}.
This is intended to flag objects that are difficult to classify because
the target is very strongly disturbed.
We find that the majority (92\%) of objects have $P(not\_sure)<0.1$ for the first question.
Careful inspections of individual cases suggest that most participants did not choose 'not sure'
even when the target is strongly disturbed. For instance, objects with $P(not\_sure)\sim0.3$ are often
undergoing violent interactions, but $\sim70\%$ of the participants classified them into elliptical/spiral.
Also, it seems a small fraction of people used the 'not sure' option for simply difficult cases
(e.g., galaxies that appear small and hard to classify into ellitical or spiral).
This was not our original intention for this option.
The trend is similar for the second question; many of the objects with
$P(not\_sure)>0.1$ are simply difficult cases such as targets in dense clusters,
where the inracluster light is clearly visible.

As some fraction of participants used the 'not sure' option in the way
that we originally did not intend, we choose to exclude 'not sure' votes when
we compute $P(spiral)$ and $P(int.)$ (i.e., $N_{total}$ in Eq.~1 is equal to $N_{spiral}+N_{elliptical}$)
because it is the least harmful way to handle them.
We have confirmed that our conclusions remain
the same if we include them (i.e., $N_{total}=N_{spiral} + N_{elliptical} + N_{not\_sure}$).
We should emphasize, however, that the 'not sure' votes are not useless; they actually turn out to be quite useful to
identify violent mergers as we demonstrate below.

\section{Classification Accuracy}
\label{sec:classification_results}

Visual classifications of galaxy morphology is dependent on the imaging depth, e.g.,
faint arms of a spiral galaxy may not be clearly identified in a shallow image.
Diffuse tidal streams may be missed for the same reason.  There is also redshift
dependence as a faint feature is harder to identify at higher redshift.
Furthermore, there may be yet another effect arising from
systematic biases in the participants' classifications.  We make a few tests
in this Section to characterize the depth and redshift dependence as well as
classification biases.

\subsection{Depth Dependence}
\label{sec:depth_dependence}

We first focus on the dependence on the imaging depth.  As we noted above,
a small number of the targets are classified in both of the Wide and D/UD layers.
This is done exactly for the purpose of the depth dependence characterization;
as the D/UD layer is deeper than Wide layer (the typical exposure time is about 5 times longer; \cite{aihara19}),
we can make a direct comparison between them for the same set of objects.

\begin{figure*}
  \begin{center}
    \includegraphics[width=7.8cm]{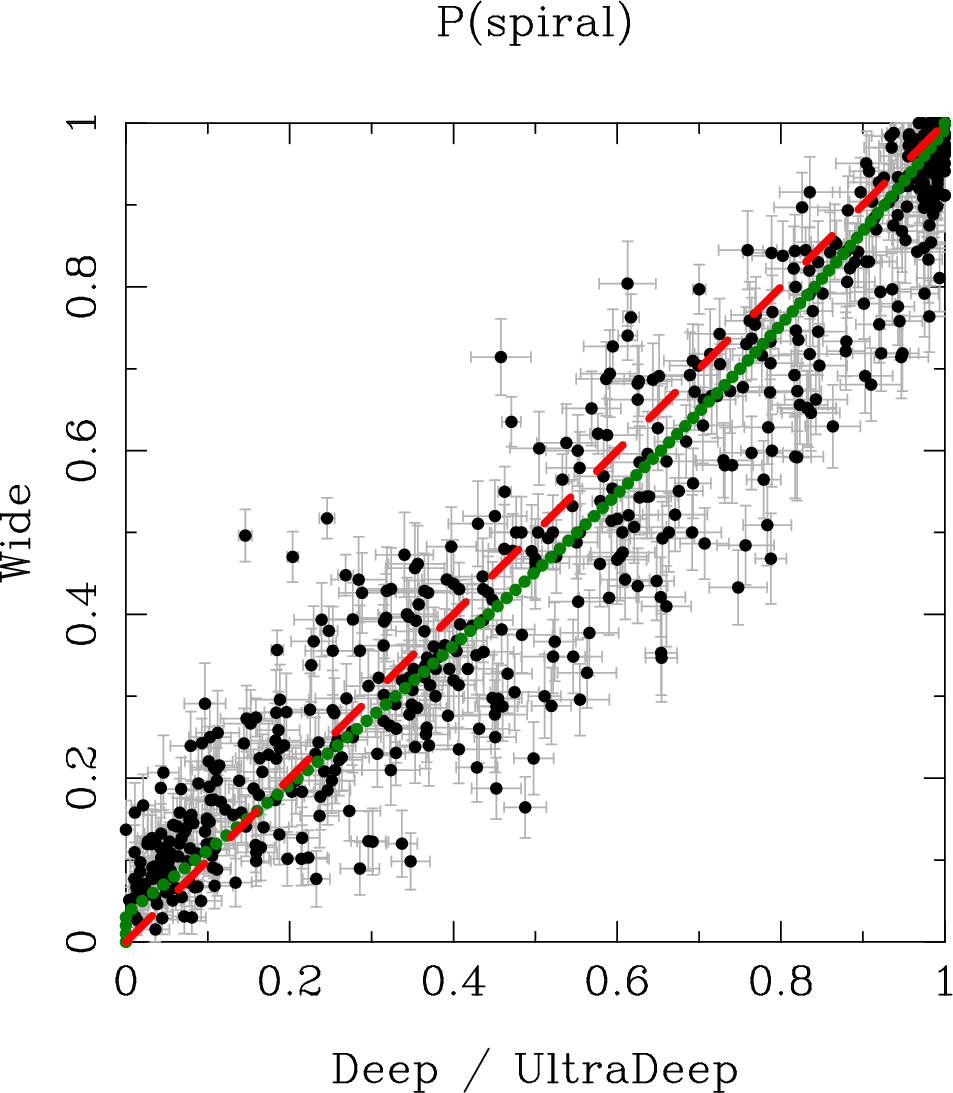}\hspace{1cm}
    \includegraphics[width=7.8cm]{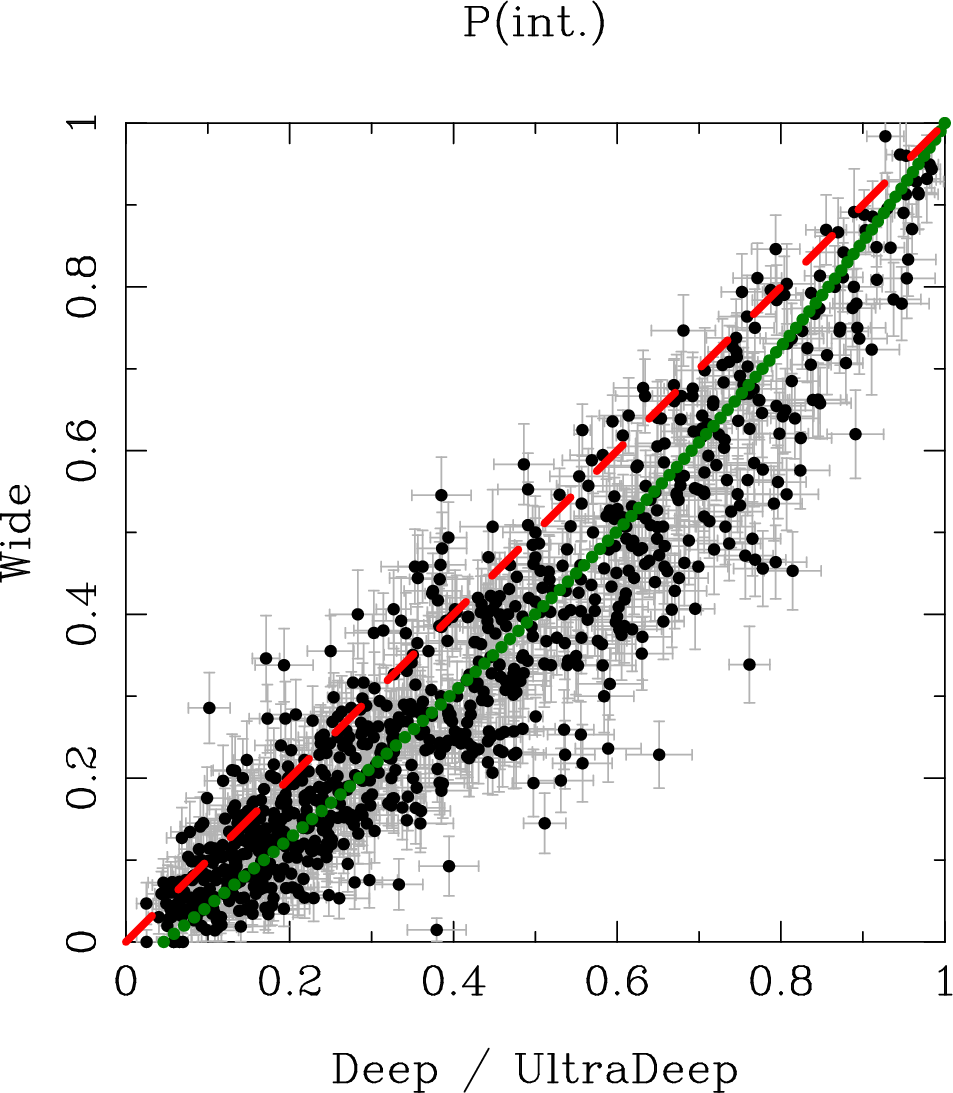}
  \end{center}
  \caption{
    {\bf Left:} Comparison of $P(spiral)$ between the Wide and Deep/UltraDeep layers for the same set of objects.
    A point is an object and the associated error bars indicate the statistical uncertainty.
    The dashed line is the one-to-one relationship, while the green dotted curve is the median of the distribution
    running over $P(spiral)$ for Wide.
    {\bf Right:} As in the left figure but for $P(int.)$.
  }
  \label{fig:wide_vs_dud}
\end{figure*}

Fig.~\ref{fig:wide_vs_dud} makes this comparison and illustrates the role of depth at
fixed resolution by comparing the Wide layer with the D/UD layer.
$P(spiral)$ in the left panel shows that, while there is a good agreement at $P(spiral)\sim0$ and 1,
the different depths give a slightly different $P(spiral)$ at intermediate $P(spiral)$.
We interpret this trend as spiral galaxies being more robustly identified in the deeper images.
As we demonstrate below, GALAXY CRUISE finds a higher fraction of spiral galaxies
compared to previous projects because of the much improved depth and quality.  That is,
spiral features can be more securely identified in deeper and sharper images.  This explains
the trend seen in Fig.~\ref{fig:wide_vs_dud}; $P(spiral)$ tends to be larger in D/UD than in Wide
for ambiguous cases.
The depth difference is less important when a target galaxy is obviously elliptical or spiral,
and $P(spiral)$ agrees well at 0 and 1.

The right panel, which shows $P(int.)$ for D/UD and Wide,  can be explained in the same way.
As interaction features are often faint and diffuse, deeper images reveal a larger
fraction of interacting galaxies.  The green points in both panels show the running median
of the distribution and we use this curve to statistically correct $P(spiral)$
and $P(int.)$ from Wide to the D/UD depth.  The correction would be larger if we had deeper
images than D/UD (i.e., we would identify more interacting features if
we could go deeper), but the green points in the figure are the correction we can make with the data in hand.
We adopt the D/UD classifications whenever available
(i.e., we exclude the Wide classifications for the duplicated objects), and apply the correction
as shown by the green dots to objects classified only at the Wide depth
to increase their $P(spiral)$ and $P(int.)$ to the D/UD depth.
This leaves us 20,686 unique objects.

It is interesting to note that the $P(spiral)$ distribution shows a clear concentration
at $P(spiral)\sim0$ and 1, and galaxies are relatively sparse at intermediate $P(spiral)$.
In contrast, the distribution of $P(int.)$ is more contiguous and there are many
galaxies with intermediate $P(int.)$, illustrating the difficulty of identifying
interacting galaxies compared to the elliptical vs. spiral classifications.

\subsection{Redshift Dependence}
\label{sec:redshift_dependence}

\begin{figure*}
  \begin{center}
    \includegraphics[width=7.8cm]{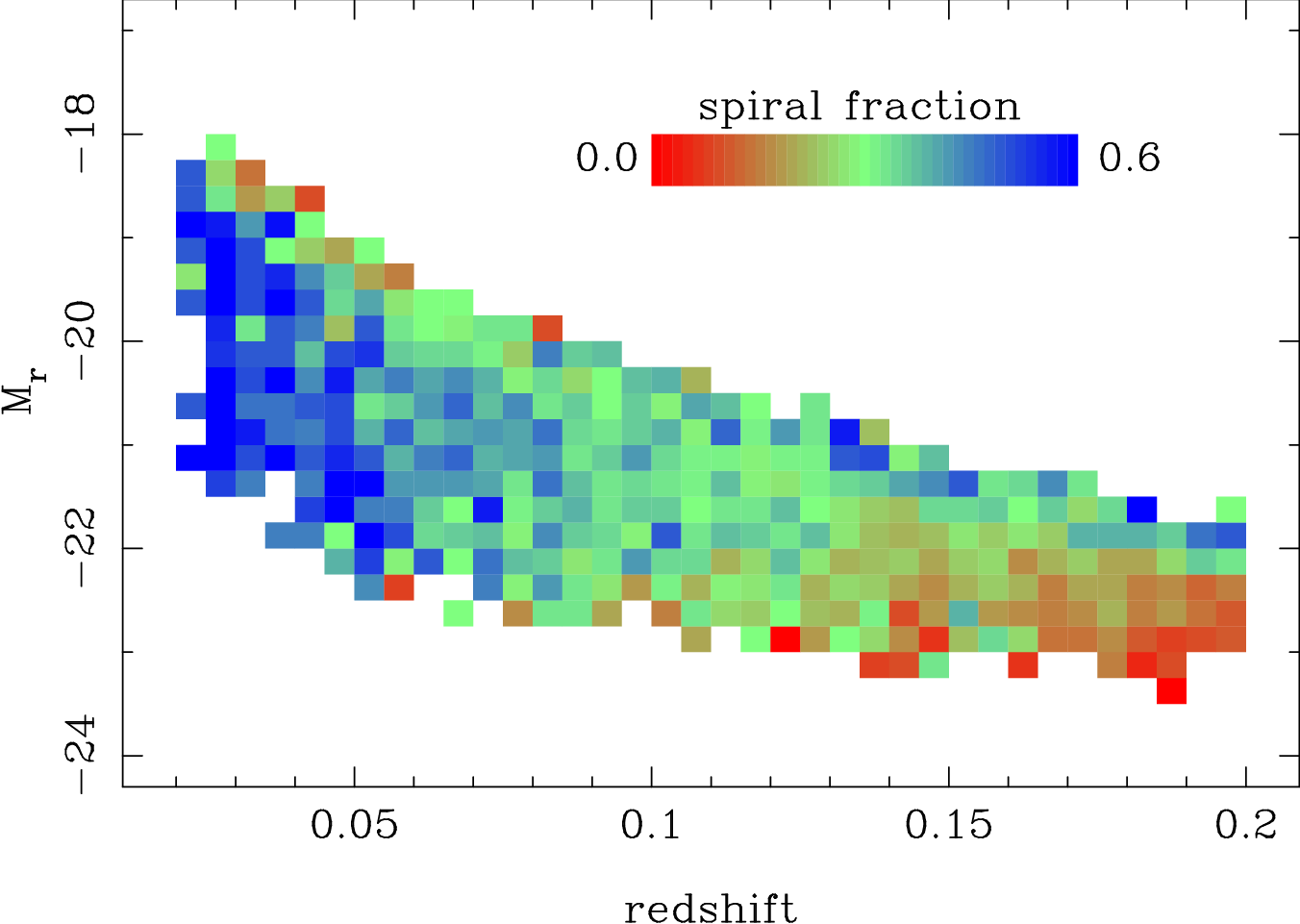}\hspace{1cm}
    \includegraphics[width=7.8cm]{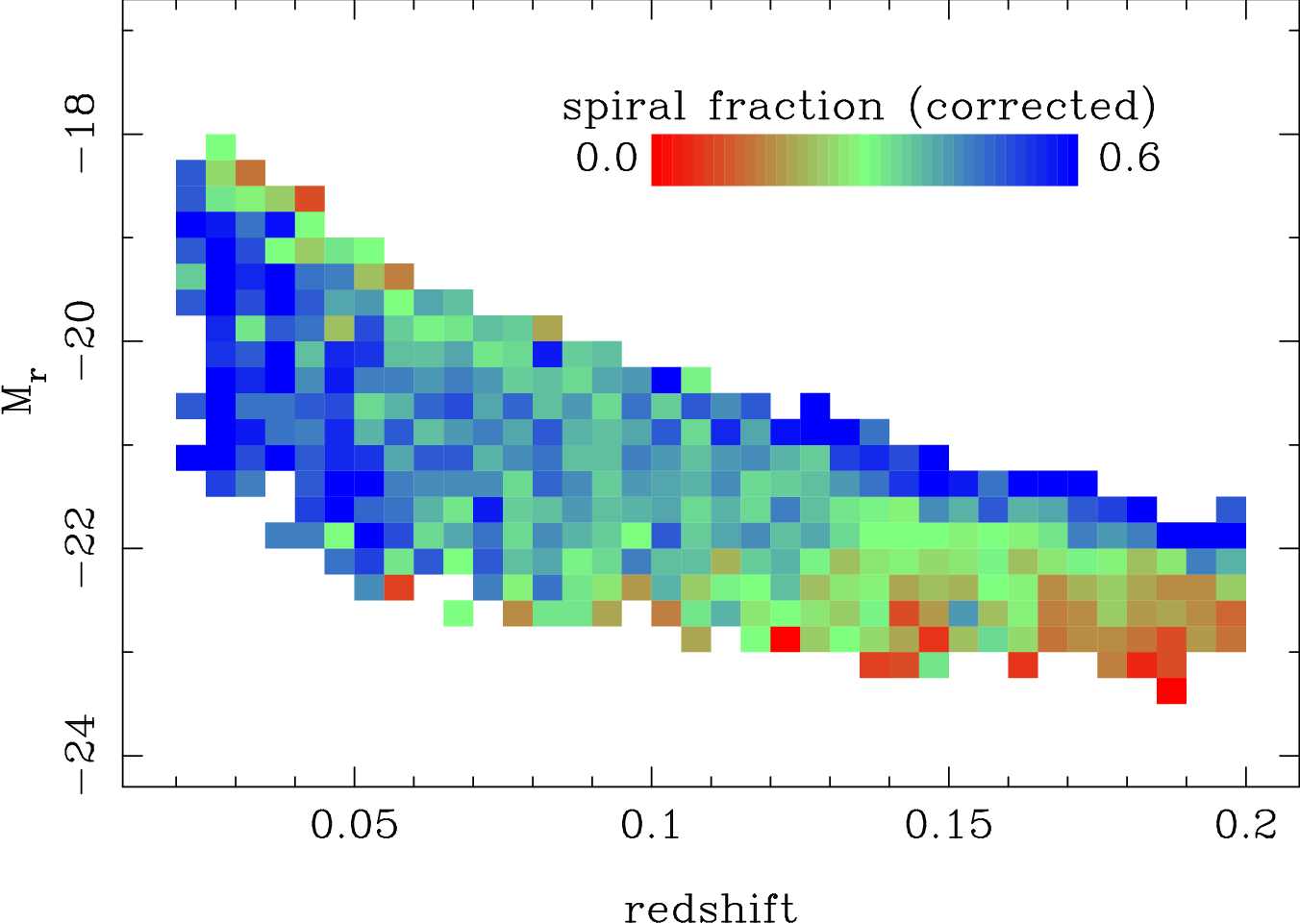}\\\vspace{0.5cm}
    \includegraphics[width=7.8cm]{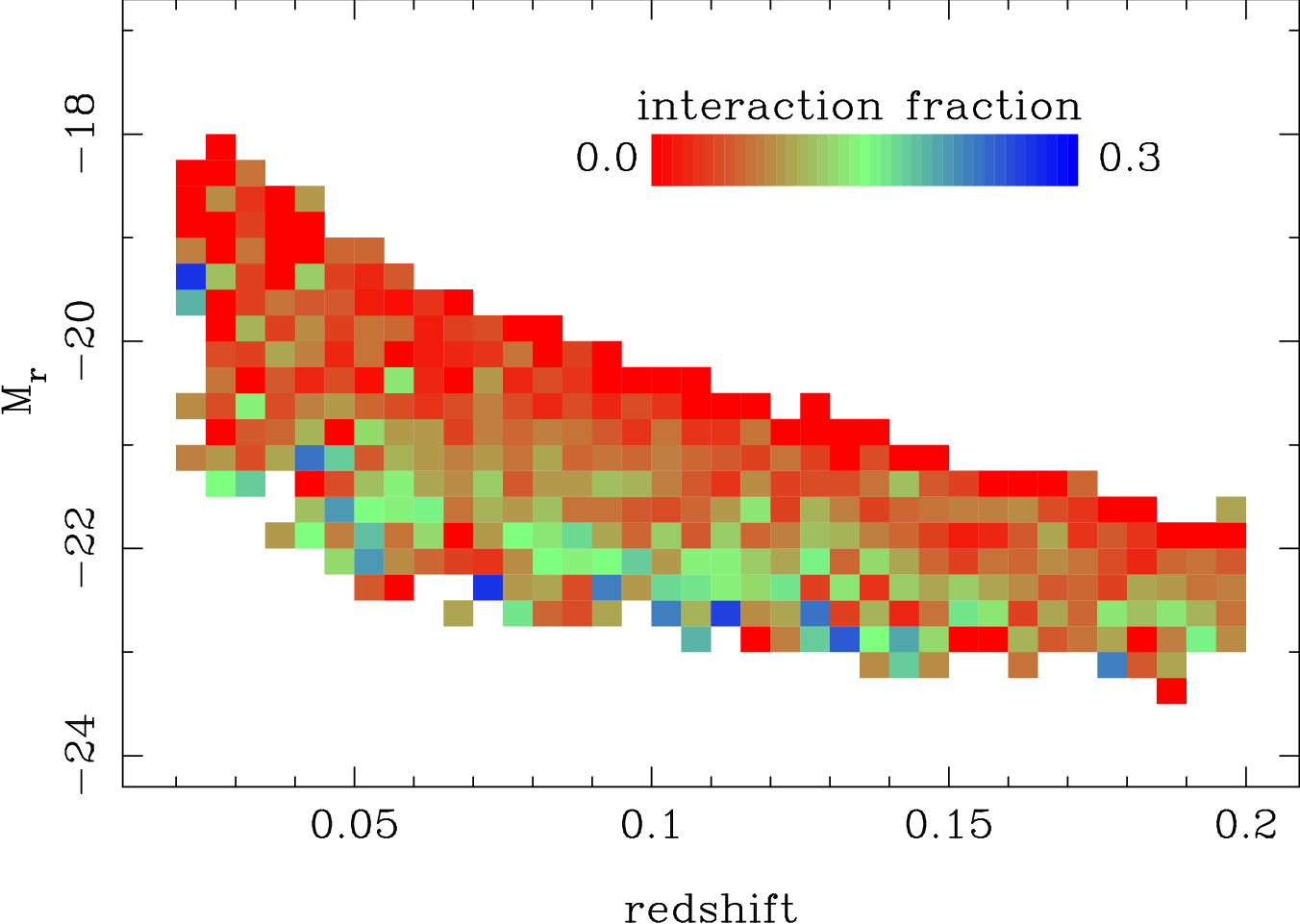}\hspace{1cm}
    \includegraphics[width=7.8cm]{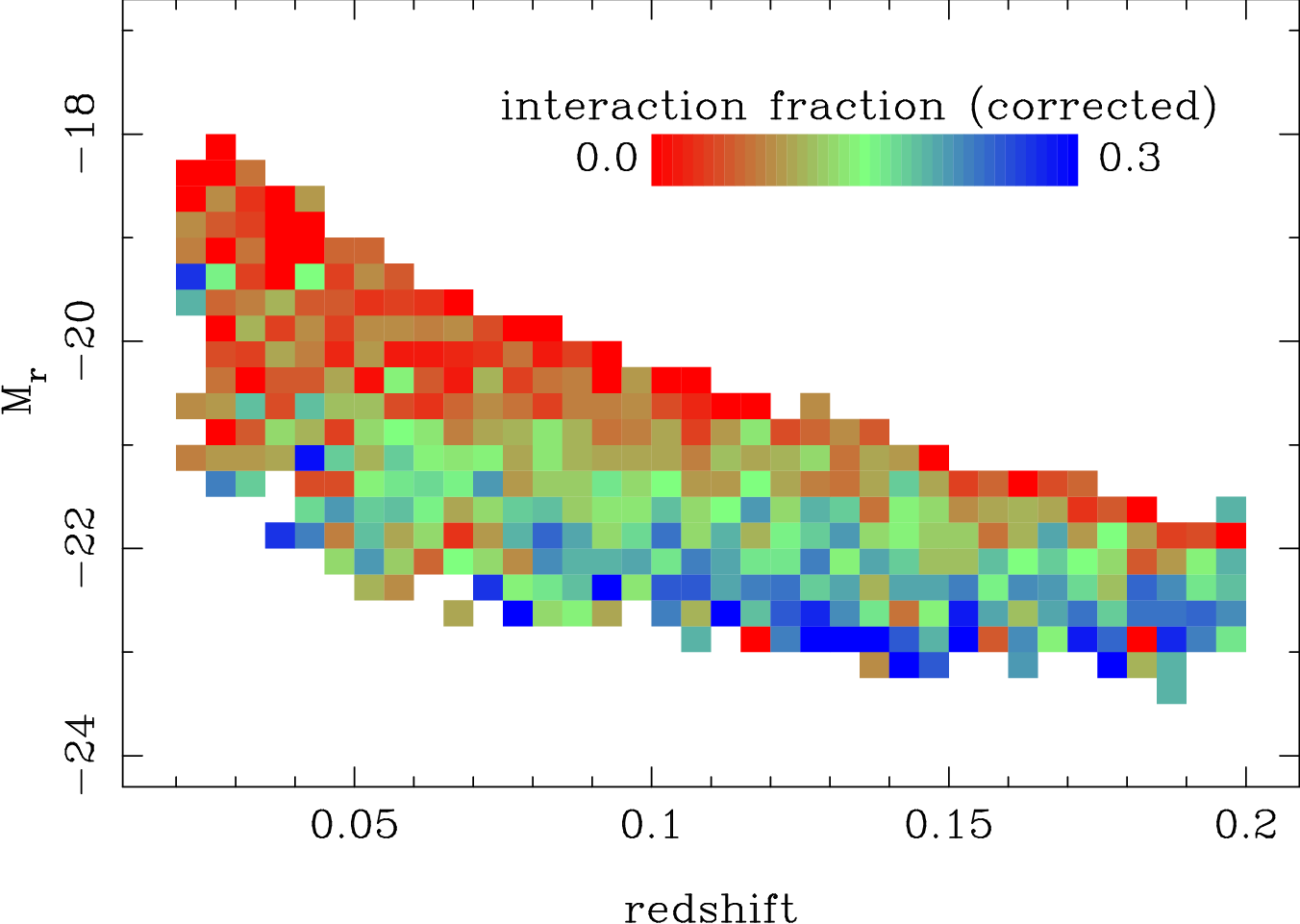}\\
  \end{center}
  \caption{
    Fraction of spiral galaxies (top) and interacting galaxies (bottom) plotted as a function of
    redshift and absolute magnitude.  The left  is the original classification, and the right
    is after applying the redshift corrections (Eqs.~\ref{eq:typecorr} and \ref{eq:intcorr} ).
    The color bar in each panel shows the mapping between color and fraction.
    Only pixels with more than 10 objects are plotted.  In each panel, the upper right part is missing objects because
    both SDSS and GAMA are fluxed-limited survey (i.e., intrinsically fainter galaxies at higher redshifts are
    not targeted).  The missing bottom-left part is due to the small volume probed at low redshifts.
  }
  \label{fig:redshift_dep}
\end{figure*}

Next, we look into the redshift dependence as galaxies at higher redshifts suffer more significantly from
cosmological dimming.  We evaluate the dependence as a function of absolute magnitude; the reduction
in the angular resolution on the physical scale at high redshift likely affects faint (and small) objects
more severely than bright objects and thus the effect is strongly dependent on magnitude as well.
Fig.~\ref{fig:redshift_dep} shows the fractions of spiral and
interacting galaxies as a function of redshift and absolute magnitude in the $r$-band.
We define spiral galaxies as those with $P(spiral)>0.79$ and interacting galaxies as those with
$P(int.)>0.79$ (see the next subsection for the choice of the threshold).

We first discuss the spiral fraction shown in the top panels.  The spiral fraction is a function
of magnitude in the sense that the spiral fraction decreases at brighter magnitudes.
But, that is not our focus here as it is an intrinsic trend.  We are interested in
the redshift dependence.  Over the narrow redshift range of $0<z<0.2$, we can reasonably assume that
the intrinsic spiral fraction does not significantly change (i.e., no significant evolution since $z=0.2$).
Thus, any trend with redshift can be attributed to a classification bias.  \citet{willett13} indeed
observed strong redshift dependence in Galaxy Zoo 2 (GZ2) in the sense that the spiral fraction decreases at higher redshifts
and they made a statistical correction for it.
Our spiral classification shows a similar but weaker trend with redshift.
GALAXY CRUISE appears less biased than GZ2.  This is likely because the HSC images are sufficiently
deep to robustly classify bright galaxies across this redshift range \citep{tadaki20}.
We apply only a weak correction to the spiral fraction:

\begin{equation}\label{eq:typecorr}
  P(spiral)_{corr}=\frac{P(spiral)}{1-0.5z},
\end{equation}

\noindent
where $z$ is redshift.  The maximum correction factor applied is only 11\%, but
it does reduce the redshift trend as shown in the top-right panel of Fig.~\ref{fig:redshift_dep}.
Note that, when $P(spiral)_{corr}>1$, we set $P(spiral)_{corr}=1$.
For reference, the correction applied in GZ2 spans over a wide range but is typically $\sim50\%$.
There is still some redshift trend left in the top-right panel in the sense that the spiral
fraction is very large at $z\lesssim0.05$.  It might be due to the small volume probed.
We find that there are few massive clusters at $z<0.05$, and galaxies at this redshift range are
predominantly field galaxies ($\sim1\%$ galaxies in are in dense environments).  On the other hand,
about 10\% of galaxies at $0.05<z<0.1$ are in dense environments.
We may be suffering from the morphology-density relation, which we will discuss in Section~\ref{sec:dependence_on_environment},
in the top-right panel.

The bottom panels in Fig.~\ref{fig:redshift_dep} shows the interaction fraction.  
There is a weak redshift trend even for very bright galaxies in the sense that
the interaction fraction decreases towards higher redshifts.  This is expected because tidal features
are often diffuse and extended and we miss those features more easily at higher redshifts.
We correct for the redshift dependence in the following way:

\begin{equation}\label{eq:intcorr}
  P(int.)_{corr}=\frac{P(int.)}{1-z}.
\end{equation}

\noindent
The correction is stronger than that for the spiral fraction and the maximum correction applied is 25\%.
As shown in the bottom-right panel, this correction significantly reduces the redshift trend for interacting galaxies.
We note that we need both the redshift and depth corrections.
The former reduces the redshift dependence of our classifications and the latter changes
the overall normalization of $P(spiral)$ and $P(int.)$.

We have confirmed that our main results in this paper are not affected significantly by whether or not we apply the corrections
for the spiral and interaction fractions.
We denote $P(spiral)_{corr}$ and $P(int.)_{corr}$ as simply $P(spiral)$ and $P(int.)$ respectively in what follows
for the sake of simplicity.

\subsection{Color-Magnitude Diagram}
\label{sec:color_magnitude_diagram}

There may also be a bias in the participants' classifications themselves as visual classifications
are subjective.  In order to reduce the subjectivity, it is a common practice to
classify objects by multiple people independently (e.g., \cite{fukugita07}).
Although this is already done in GALAXY CRUISE as we merge classifications by many participants
for each galaxy (see Fig.~\ref{fig:class_num}), it is still
useful to check if there is a systematic bias especially because the classifications are made
by non-professional astronomers.  However, this is not a trivial question because
there is no truth table of morphology for our sample and edge-cases are difficult even for
professional astronomers.

In order to roughly evaluate the classification accuracy, we use colors
as the truth table.  That is, we assume that elliptical galaxies are on the red sequence and
spiral galaxies are in the blue cloud.  There are of course green-valley galaxies, which are
often bulge-dominated spiral galaxies, and also spiral galaxies are not always blue; 
edge-on spirals are red, and there is a population of spiral galaxies with no
sign of ongoing star formation \citep{vandenbergh76,couch98,masters10,shimakawa22}.
However, there is overall a good correlation between morphology and color.  For instance,
\citet{schawinski09} showed that the vast majority ($96\%$) of the visually selected elliptical galaxies are red.
Colors are a different attribute than morphology, but we use them as a proxy for morphology for now.

\begin{figure}
  \begin{center}
    \includegraphics[width=8cm]{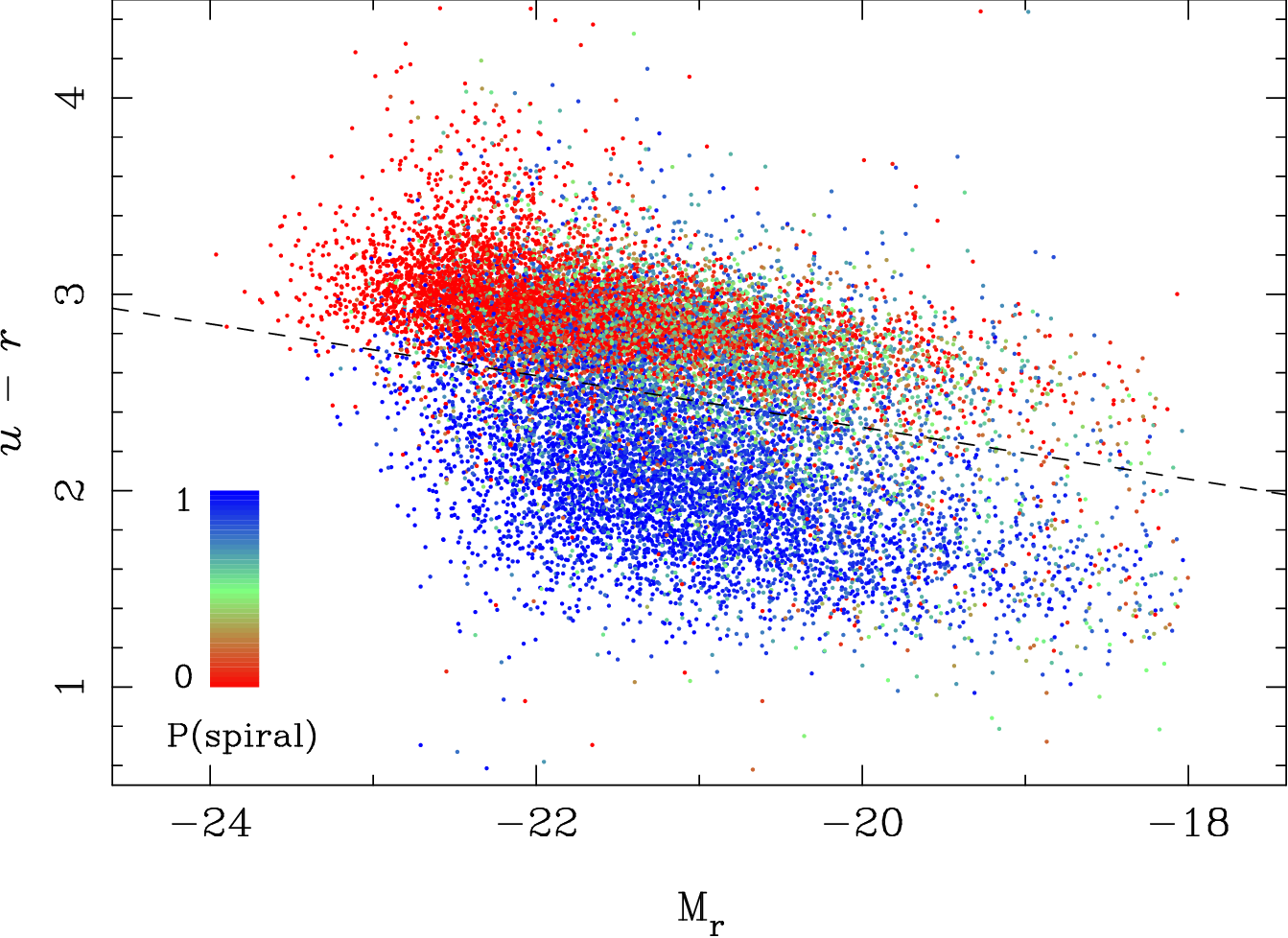}\hspace{1cm}
  \end{center}
  \caption{
    Rest-frame $u-r$ color plotted against $r$-band absolute magnitude.
    Galaxies are color-coded according to their $P(spiral)$ as shown by the color bar.
    We split the red/blue galaxies at the dashed line.
  }
  \label{fig:color_magnitude}
\end{figure}

Fig.~\ref{fig:color_magnitude} shows the rest-frame $u-r$ color k-corrected to $z=0.1$ using
the k-correction code by \citet{blanton07} plotted against absolute magnitude
in the $r$-band with galaxies color-coded according to their $P(spiral)$.  As can be seen,
the red sequence and blue cloud are fairly well populated by elliptical and spiral galaxies,
respectively.  This demonstrates that the classifications are good at least to the first order.
There are, however, galaxies with large $P(spiral)$ on the red sequence and
those with low $P(spiral)$ in the blue cloud.
Galaxies with intermediate $P(spiral)$ are located both on the red sequence and the blue cloud.
They may not necessarily be wrong classifications
as discussed above, but we for now assume that the colors are the truth table and make
an attempt to quantify the classification accuracy below.

\begin{figure*}
  \begin{center}
    \includegraphics[width=7.8cm]{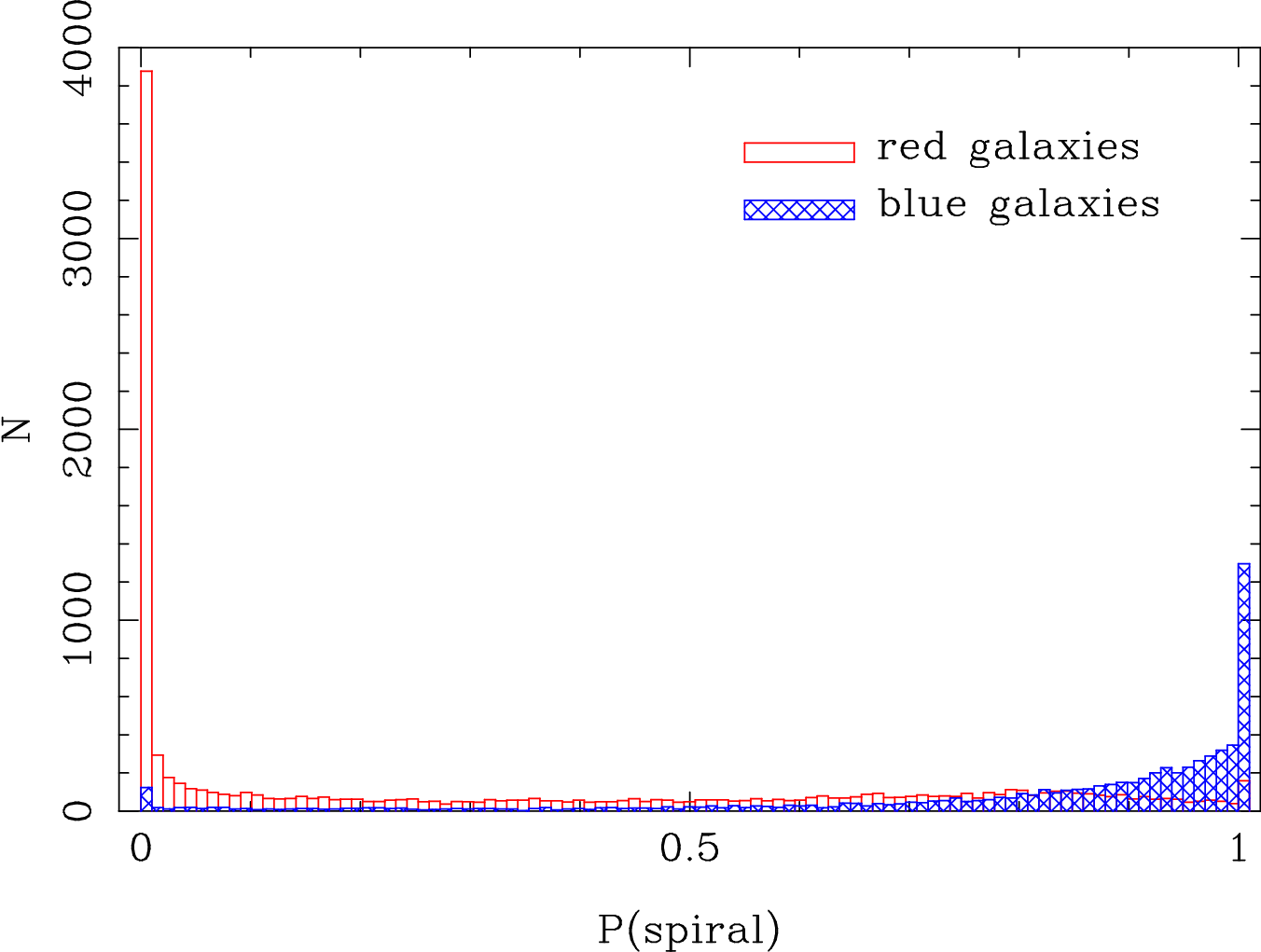}\hspace{1cm}
    \includegraphics[width=7.8cm]{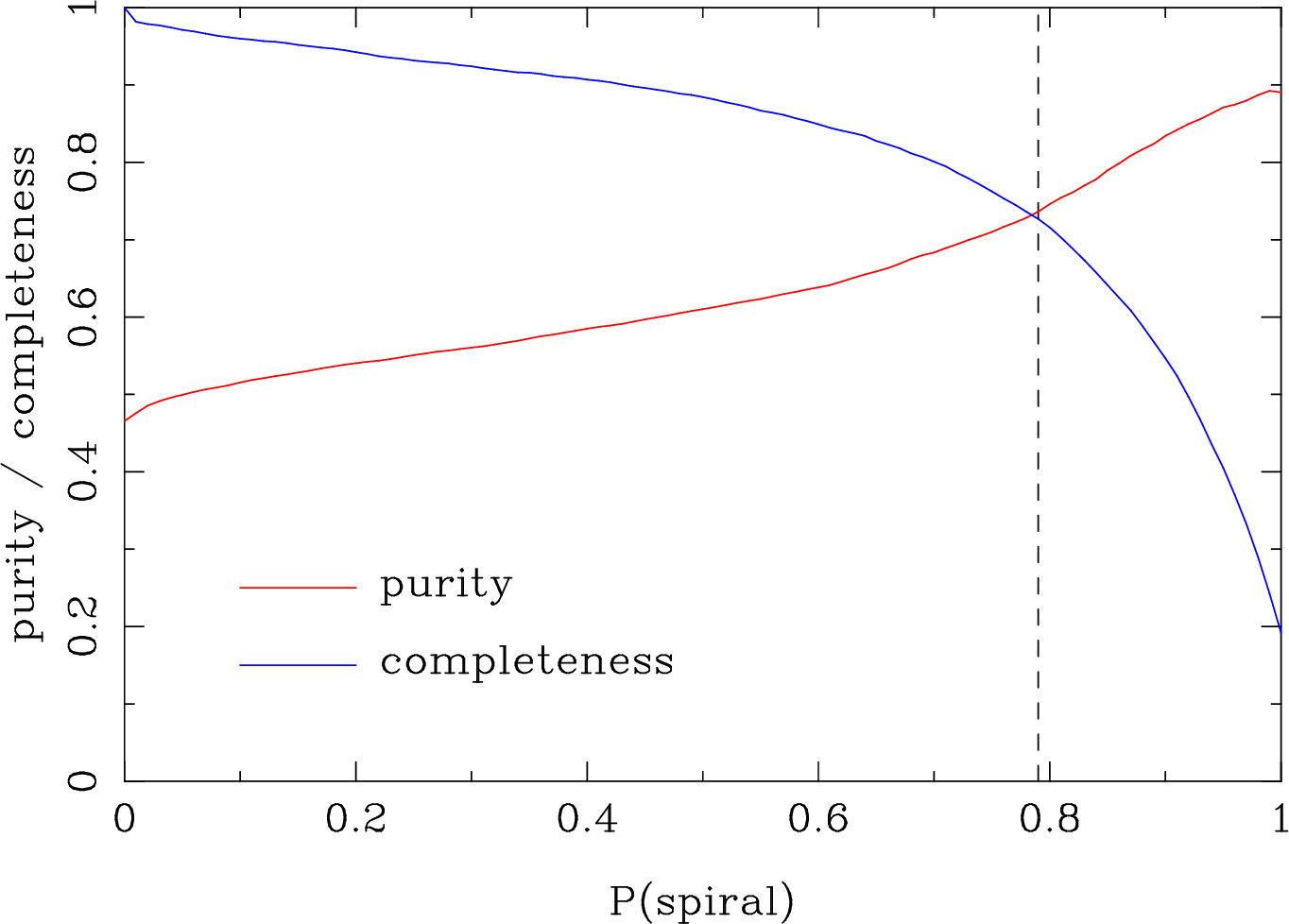}
  \end{center}
  \caption{
    {\bf Left:} Distribution of $P(spiral)$ for red (open histogram) and blue (shaded histogram) galaxies.
    {\bf Right:} Purity and contamination rate plotted as a function of threshold $P(spiral)$.  For instance,
    the purity and contamination at $P(spiral)=0.2$ are for spiral galaxies defined as $P(spiral)>0.2$.
    The vertical dashed line is the threshold adopted in this paper.
  }
  \label{fig:color_magnitude2}
\end{figure*}

The left panel of Fig.~\ref{fig:color_magnitude2} shows the $P(spiral)$ distributions of
the red and blue galaxies.  $P(spiral)$ of red galaxies is sharply peaked at $P(spiral)=0$
with a long and flat tail towards larger $P(spiral)$.  The distribution of blue galaxies is
similar albeit with a broader wing around $P(spiral)=1$.  The tails indicate that there may be
misclassifications.  To better characterize it, we show in the right panel the purity and
completeness of spiral galaxies as a function of threshold $P(spiral)$.
Here we define purity as

\begin{equation}
  {\rm purity} = \frac{N_{blue,\ selected}}{N_{selected}},
\end{equation}

\noindent
where $N_{selected}$ is the number of galaxies with $P(spiral)$ above a certain threshold.
The threshold here is a free parameter and is taken in the horizontal axis.
$N_{blue,\ selected}$ is the number of blue galaxies among the selected galaxies.
Thus, the purity is the fraction of blue galaxies among galaxies selected with a threshold $P(spiral)$.
Likewise, we define completeness as

\begin{equation}
  {\rm completeness} = \frac{N_{blue,\ selected}}{N_{blue}},
\end{equation}

\noindent
where $N_{blue}$ is the total number of blue galaxies and $N_{blue,\ selected}$ is
the number of blue galaxies with $P(spiral)$ above the threshold.

The purity shown in the right panel increases with increasing the threshold $P(spiral)$.
This is expected because the contamination of elliptical galaxies decreases with more
stringent $P(spiral)$ cuts.   On the other hand, the completeness decreases with
increasing threshold $P(spiral)$.  We want to be both pure and complete, but
we need to make a compromise between the two in the presence of contaminating galaxies.
Here we choose to take $P(spiral)=0.79$ indicated as the vertical dashed line as
the threshold.  This is the point where both purity and completeness is about 75\%.
We have confirmed that our primary conclusions in this paper remain unchanged if
we perturb the threshold within a reasonable range, e.g., $P(spiral)>0.5$.

A similar analysis will be useful for $P(int.)$, but it is more difficult because
there is no truth table or even a good proxy for interaction features.  Rest-frame colors
or other well-measured quantities do not work as a proxy.  Furthermore, interaction
features are intrinsically more difficult to identify than elliptical vs spiral.
Future missions with deeper and sharper imaging data over a wide area will allow us
to better characterize the classification accuracy for interacting galaxies.
For now, we choose to conservatively adopt the same threshold, $P(int.)>0.79$, to
define interacting galaxies.  Once again, we have confirmed that main conclusions of
the paper are not sensitive to the particular choice of the threshold within
a reasonable range.
We plan to carry out a campaign to classify galaxies from cosmological hydrodynamical
simulations; we visualize these simulated galaxies accounting for various observational effects
and isert them to the real HSC images, so that the participants can classify them
(Bottrell et al. in prep.).
Because we know the merger histories of the simulated galaxies, we expect that
we will be able to evaluate how well the partipants can identify interacting galaxies
as functions of, e.g., mass ratio and merger phase,  and discover an optimal $P(int.)$
threshold.

An approach alternative to probability thresholding adopted here is to simply use
the probabilities as weights; a galaxy with $P(spiral)=0.3$ is counted as 0.3 of a spiral galaxy.
Which approach to adopt depends on science applications, and we briefly discuss this point
in Appendix A.

\subsection{Gallery of Selected Objects}
\label{sec:gallery_of_selected_objects}

\begin{figure*}
  \begin{center}
    \includegraphics[height=12cm, angle=90]{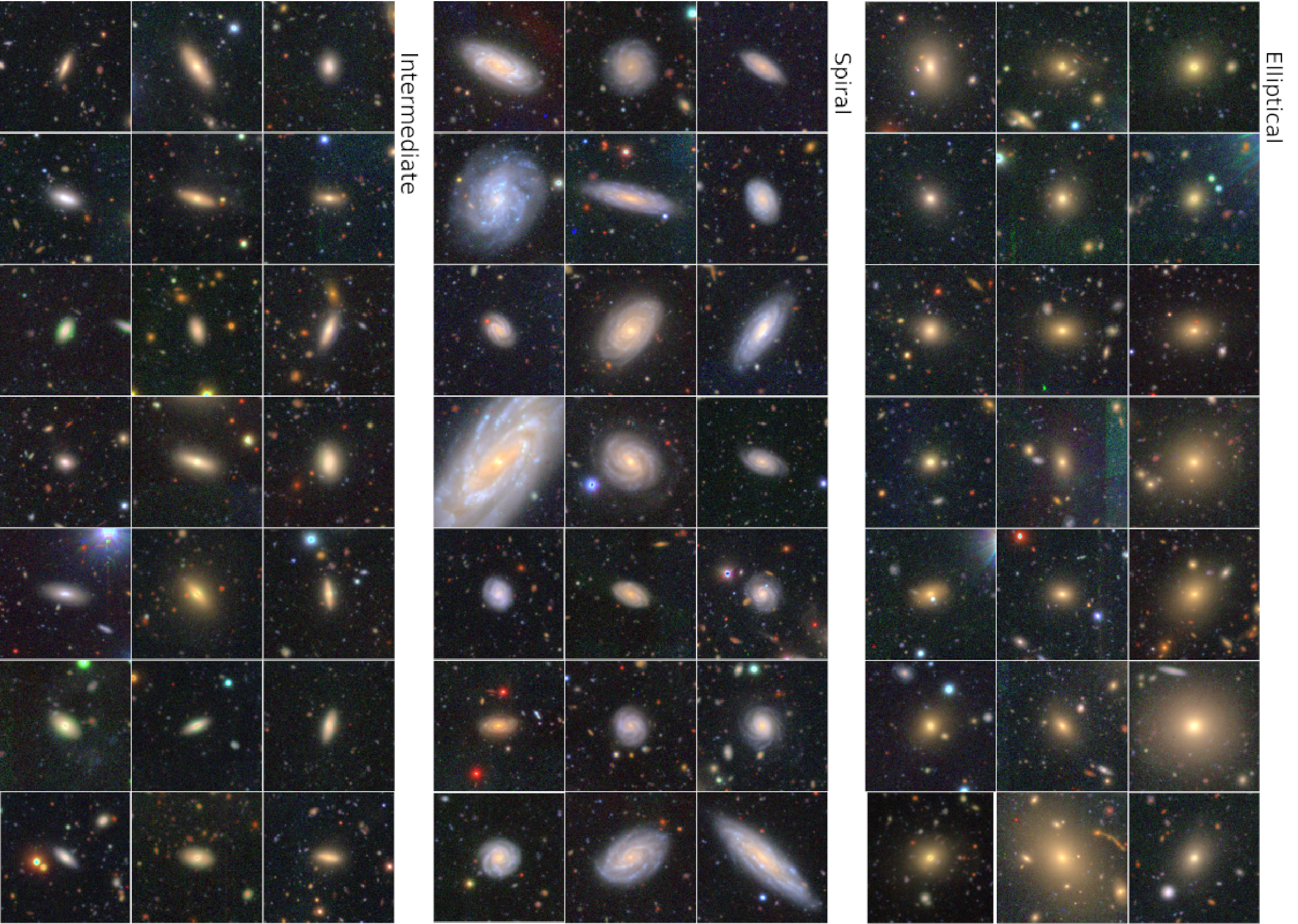}
  \end{center}
  \caption{
    {\bf Top:} Gallery of galaxies with $P(spiral)<0.05$ (i.e., elliptical galaxies).
    Each image is 1 arcmin on a side.
    {\bf Middle:} Same as the top panel, but for spiral galaxies with $P(spiral)>0.95$.
    {\bf Bottom:}  Galaxies with $P(spiral)\sim0.5$. These objects tend to be edge-on S0 galaxies.
  }
  \label{fig:gallery1}
\end{figure*}

To visualize the participants' classifications, we present a gallery of
elliptical and spiral galaxies in Fig.~\ref{fig:gallery1}.
They are high-confidence elliptical and spiral galaxies and the classifications are indeed
accurate.  It will also be instructive to look at galaxies with intermediate $P(spiral)$
shown in the bottom panel.  Interestingly, these objects tend to be
S0-like galaxies; we do not explicitly include S0's as a galaxy type in our classification
scheme as they are difficult even for professional astronomers, but the participants'
classifications are actually useful to identify them and they are rightly classified as
an intermediate type between elliptical and spiral.  The color of these objects seem to
be a mixture of red and blue and we find that they are indeed located on both the red
sequence and blue cloud.


\begin{figure*}
  \begin{center}
    \includegraphics[height=12cm, angle=90]{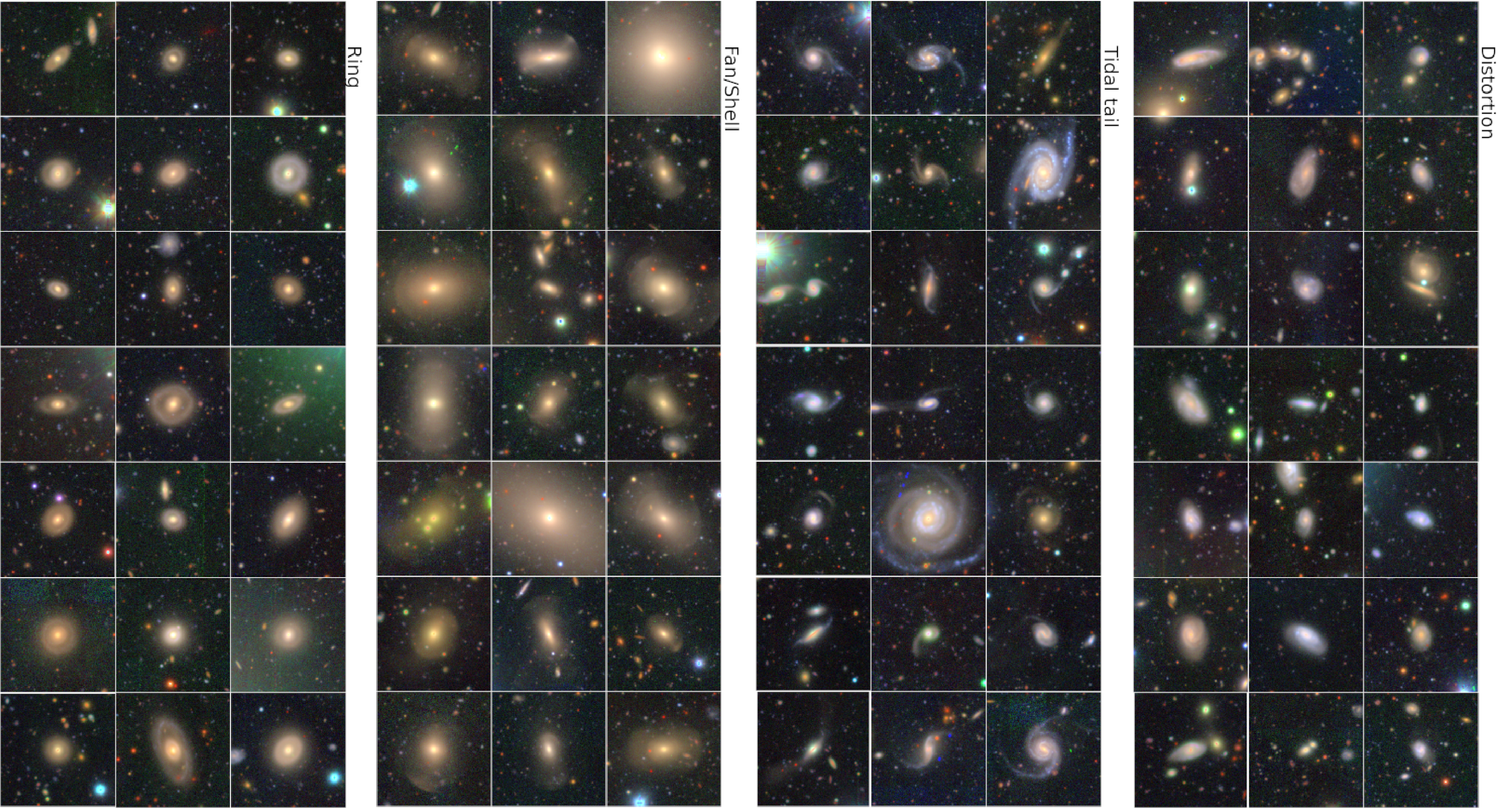}
  \end{center}
  \caption{
    Gallery of interacting galaxies.  Shown are those with highest probabilities for distorted shapes,
    tidal tails, fan/shell, and ring from top to bottom. Each panel is 1 arcmin on a side.
  }
  \label{fig:gallery2}
\end{figure*}

Fig.~\ref{fig:gallery2} shows galaxies with four distinct interaction features.  These
galaxies are first selected as $P(int.)>0.79$ and then 21 objects with the highest
probabilities for each category are shown in these figures.  It is impressive that
the classifications are made very well; all these individual features are nicely
captured by the participants.  Galaxies with distorted shapes tend to have a nearby
companion, which is likely interacting with the target galaxies.
We  find that some of the tidal tails seem to be distorted spiral arms.
It is in some cases difficult to distinguish tidal tails from distorted arms, but
this tendency of misclassifying distorted arms as tidal tails is a possible classification
bias in GALAXY CRUISE.  As for fans/shells, coherent caustic features are observed in all cases.
Finally, ring galaxies are also all nice ring galaxies.  Interestingly, they are all face-on
ring galaxies; there should be edge-on (or polar) ring galaxies, but they are not identified.
They might be confused with linear tails.  We will further discuss
ring galaxies in Section~\ref{sec:ring_galaxies}.

Let us also comment on galaxies with lower $P(int.)$. Visual inspections of a subset of $P(int.)\sim0.5$
cases suggest that they are indeed a mixture of objects; some show a weak hint of distorted shapes,
faint tails, etc, while others do not seem to exhibit clear interaction features. It seems wise to set
the threshold to securely identify interacting galaxies well above 0.5 as we do in this paper.
Another interesting class of objects is interacting galaxies with intermediate probabilities for
invididual interaction features.  We recall that we allow the participants to choose multiple interaction
features because some of the features can be difficult to distinguish and some galaxies actually
exhibit multiple features.  Visual inspections of interacting objects with the probability of
each feature between 0.15 and 0.35 indeed show complex morphologies with multiple features.
Tidal tails and shells are most frequently observed, but distorted shapes and ring-like features are
also observed.

\begin{figure*}
  \begin{center}
    \includegraphics[width=16cm]{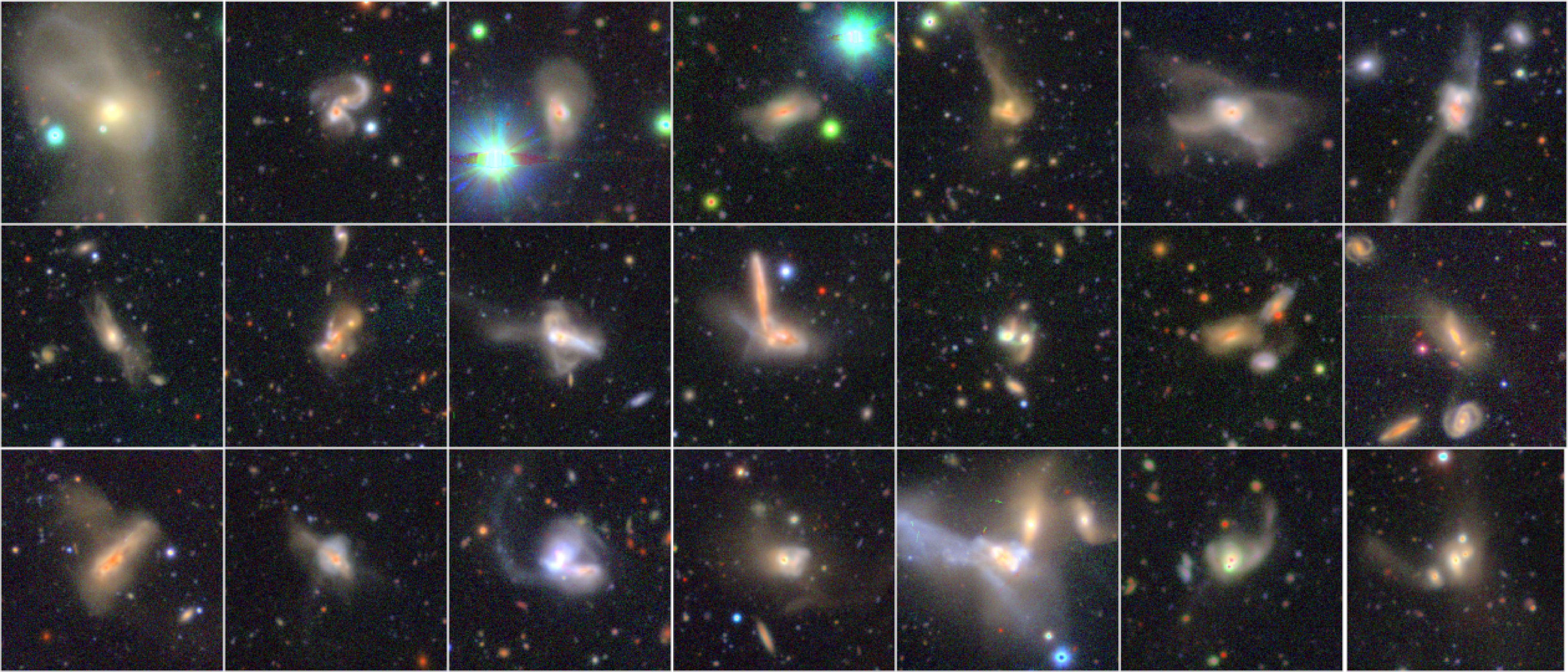}
  \end{center}
  \caption{
    Galaxies with $P(int.)>0.79$ and $P(not\_sure)>0.2$. Each panel is 1 arcmin on a side.
  }
  \label{fig:gallery3}
\end{figure*}

Finally, we show in Fig.~\ref{fig:gallery3} a collection of violent mergers.
This is constructed from a joint slection of $P(not\_sure)>0.2$ from the first
question and $P(int.)>0.79$ from the second question. As we discussed in Section~\ref{sec:combined_classifications},
the 'not sure' option was not used as we intended, but a certain fraction of
the participants followed our intention and sorted objects with significantly
disturbed morphology into the 'not sure' category. The mergers in the figure are
all in a violent merger phase and the original morphologies of the galaxies are
indeed difficult to classify.
In the analyses we present in Section~\ref{sec:properties_of_galaxies}, we primarily
focus on interacting galaxies defined as $P(int.)>0.79$, but we also include the violent merger
subsample defined here.
We note that the violent merger subsample comprises  8\% of the interacting galaxy sample.

\subsection{Statistics of Interacting Galaxies}

\begin{figure}
  \begin{center}
    \includegraphics[width=8cm]{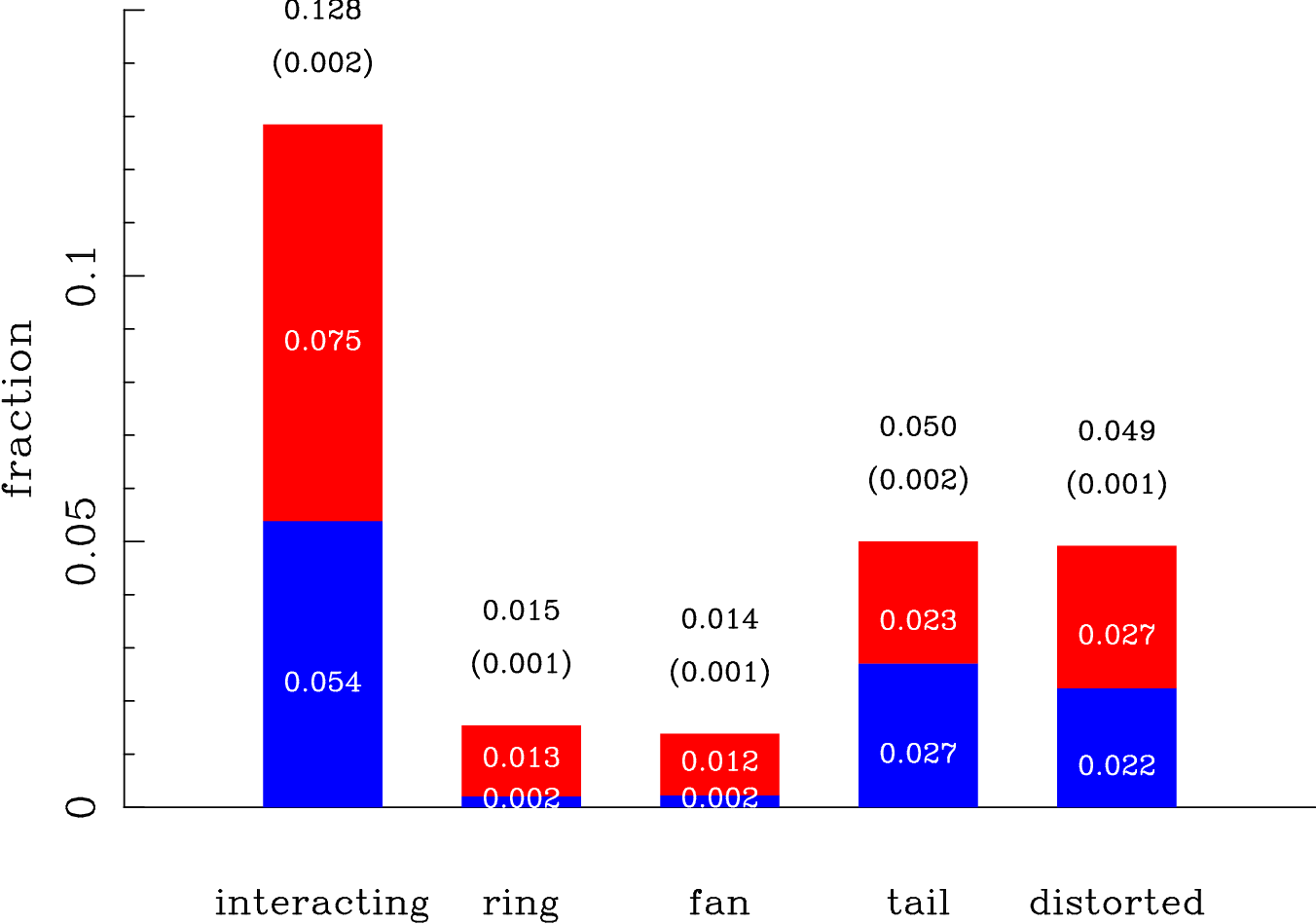}
  \end{center}
  \caption{
    The leftmost bar shows the fraction of interacting galaxies among the entire sample.
    The number at the top of the bar and the one in the brackets are the fraction and its uncertainty.
    The bar is split into the red and blue parts, which represent the elliptical and spiral fractions,
    respectively.
    The remaining bars are for the ring, fan, tail, and distortion features, and they are
    normalized such that the sum of their abundances equals to the abundance of interacting galaxies
    as shown in the leftmost bar.
    The meaning of the numbers and the red/blue parts are the same as the leftmost one.
    For reference, the fractions of elliptical and spiral galaxies that are not interacting are
    0.55 and 0.32, respectively (these numbers are normalized in the same way and hence the total
    fraction of non-interacting galaxies is 0.87).
  }
  \label{fig:stat1}
\end{figure}

Before we discuss properties of the target galaxies in detail in the next Section,
we briefly summarize statistics of interacting galaxies in Fig.~\ref{fig:stat1}.
Overall, we find that about 13\% of galaxies in our sample are interacting galaxies.
The sample is not volume-limited and the number should not be over-interpreted, but the numbers
here do not significantly change if we make a volume-limited sample with $z<0.1$ and $M_r<-20.5$.
Also, the number is somewhat sensitive to the threshold $P(int.)$ value adopted
(e.g., if we reduce it to 0.5, 36\% of galaxies are interacting).

%

It is interesting to note that the abundance spiral galaxies relative to elliptical galaxies
among the interacting galaxies is about 0.72 as shown in the leftmost bar, while this ratio
is about 0.60 for non-interacting galaxies. The same trend holds for a volume-limited sample
of $M_r<-20.5$ at $z<0.1$.
\citet{darg10a} found that the spiral to elliptical fraction among interacting galaxies is about
twice higher than the global population.  While direct comparisons cannot be made as the parent
samples are constructed in different ways, we observe a similar (but weaker) trend.

Turning to individual features, we find
that tidal tails and distortions are the most common interaction features and
they comprise about 3/4 of all interacting galaxies.
Spiral galaxies are more abundant than ellipticals for tidal tails, and
it is likely due to the misclassifications of distorted spiral arms as tails.
The remaining 1/4 of the interacting galaxies is split about equally into fan and ring
features.   Elliptical galaxies dominate over spiral galaxies here.  We suspect that
this is at least partly due to the general difficulty (not only for the participants but for
professional astronomers as well) of identifying fan and ring features in the presence
of spiral arms.  These features are much easier to identify around elliptical galaxies.
We are going to look at statistical properties of galaxies in GALAXY CRUISE in the next
section and we keep these biases in mind.

\section{Properties of Galaxies in the Local Universe Revisited}
\label{sec:properties_of_galaxies}

Based on the catalog constructed in the previous Section, we now move on to
discuss statistical properties of the local galaxies.  We first make a comparison with GZ2
\citep{willett13} in Section \ref{sec:comparison_with_galaxyzoo2}.
There is a more recent catalog
based on deeper DECals data \citep{walmsley22}, but here we focus on GZ2 as it is the most
widely used visual morphology catalog\footnote[2]{
Our results below remain essentially the same if we compare against DECals because HSC-SSP is deeper than DECals by more than $2$ mags.
}.
We then turn our attention to environmental dependence of galaxy morphology in Section \ref{sec:dependence_on_environment}.
In the following Section \ref{sec:agn_fraction}, we discuss AGNs and
revisit the classic picture of AGN activities being triggered during galaxy interactions.
Another activity that might be enhanced by mergers is
star formation and we look into it in Section \ref{sec:star_formation_rate}.
An interesting class of objects in GALAXY CRUISE
is ring galaxies; the participants' identification of ring galaxies is accurate as shown
in Fig.~\ref{fig:gallery2} and we discuss their statistical properties in Section \ref{sec:ring_galaxies}.
Finally, based on a number of simple assumptions, we infer the merger rate from GALAXY CRUISE and
discuss the mass growth rate of galaxies in Section~\ref{sec:merger_rate}.

\subsection{Comparison with GalaxyZoo2}
\label{sec:comparison_with_galaxyzoo2}

We cross-match our catalog with GZ2 'Main' catalog \citep{willett13} with positions.
8,354 objects\footnote[3]{
More than half of our targets do not match with GZ2, although both samples are based on the spectroscopic
redshifts from SDSS.  This is primarily due to GZ2's cut on the half-light Petrosian magnitude being brighter than 17.0 in the $r$-band
\citep{willett13}.
Our sample includes fainter galaxies.
}
are matched within 1.5 arcsec.
We first compare classifications from GZ2 with those from GALAXY CRUISE on an object-by-object basis.

The GZ2's classification scheme is different from ours, and
there is no elliptical vs. spiral question in GZ2.  However, the first question in GZ2,
{\it ``Is the galaxy smooth and round with no sign of a disk?''}, is sufficiently close and we use this question
to compute $P(spiral)$ from GZ2.  To be specific, a 'No' to the question indicates that a galaxy
is a spiral galaxy.   There is a more specific question about
the spiral feature later in the sequence of questions, {\it ``Is there any sign of a spiral arm pattern?''}.
Not all volunteers are asked this question.  We could compute a conditional probability for
spiral arm \citep{casteels13}, but we prefer to be simple here and use only the first question.

For interaction features, we use {\it ``Is there anything odd?''} and the subsequent question 
about observable features to estimate $P(int.)$.  As some of the features are not a result
of interaction (e.g., dust lane), we compute $P(int.)$ as a product of $P(there\ is\ something\ odd)$
and $P(ring) + P(disturbed) + P(irregular) + P(merger)$.

We use GZ2's classifications with weights to individual users (weighted fraction).  There is
a fraction corrected for the classification bias as a function of redshift (debiased fraction).
However, as noted in \citet{walmsley22}, these debiased fractions are useful only in a statistical
sense and are not appropriate for object-by-object comparisons.  For this reason, we focus
on the weighted fraction here.

\begin{figure*}
  \begin{center}
    \includegraphics[width=7.8cm]{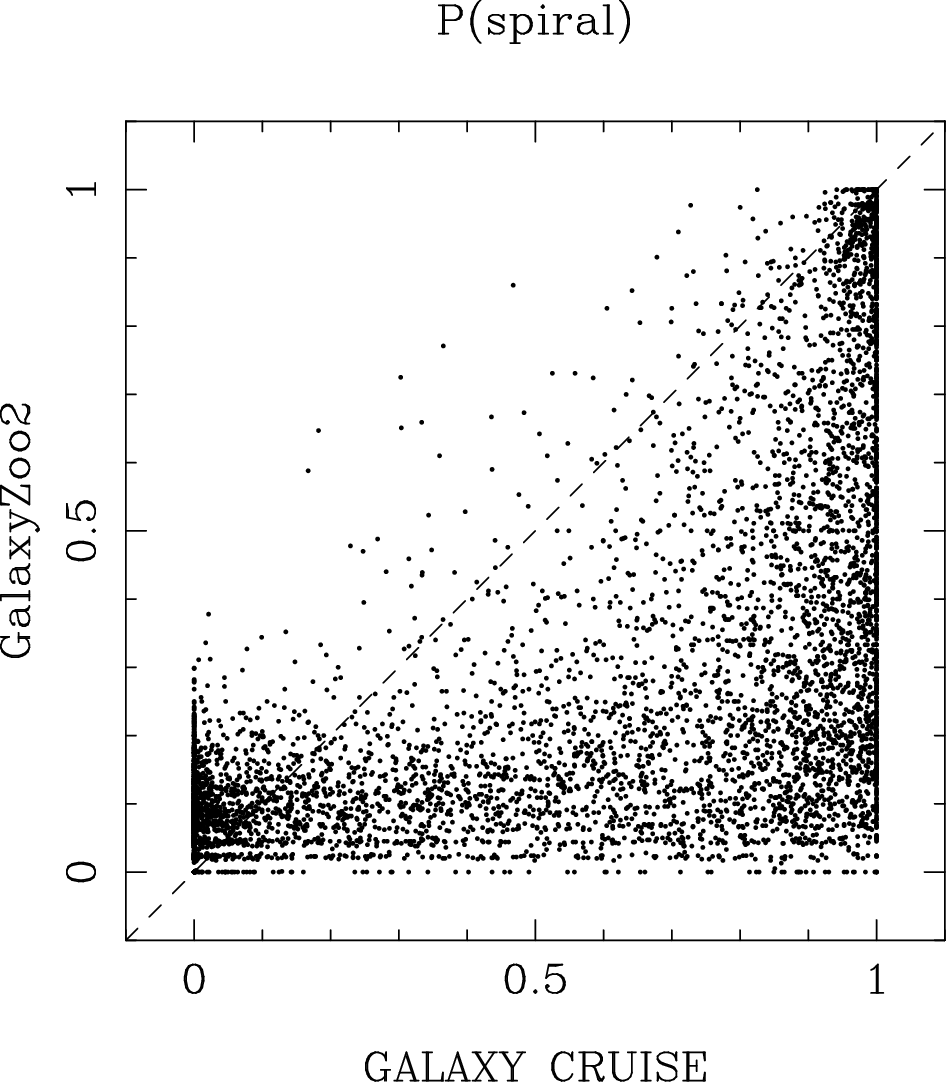}\hspace{1cm}
    \includegraphics[width=7.8cm]{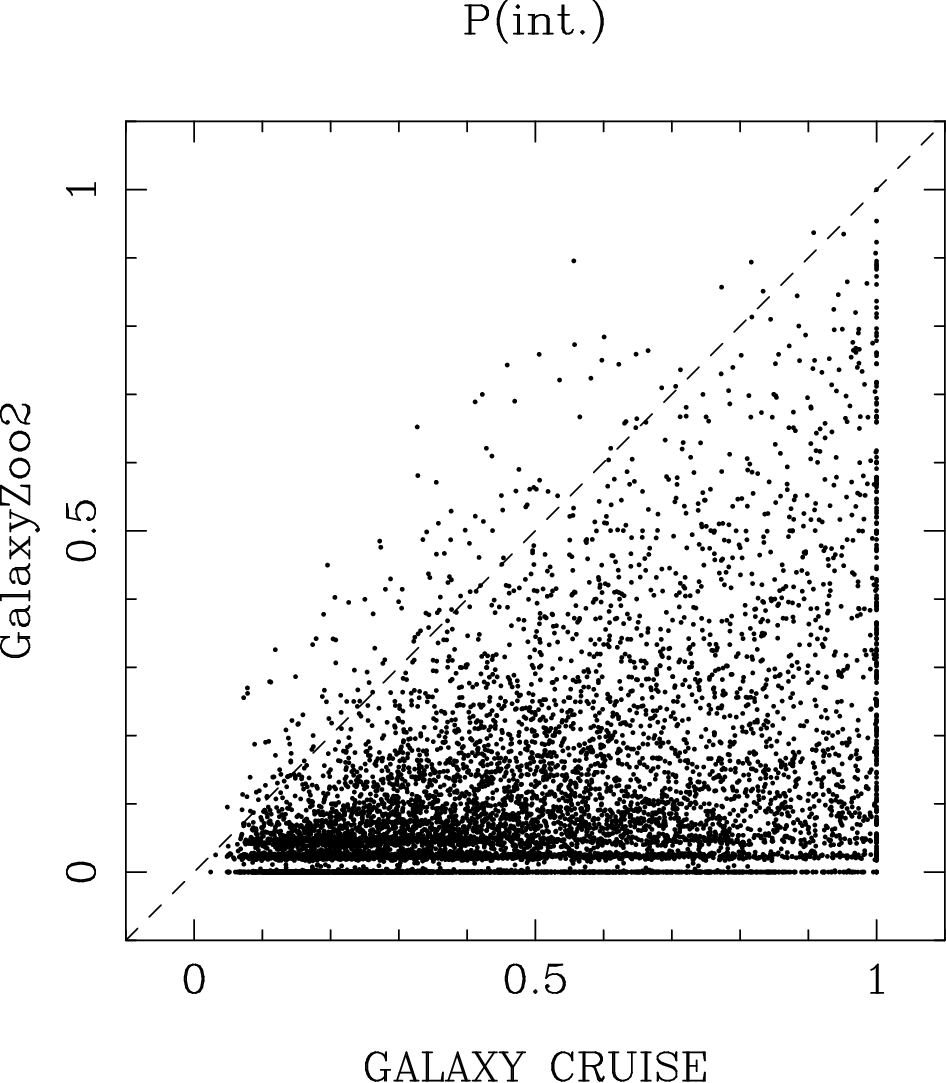}
  \end{center}
  \caption{
    Comparisons between GZ2 and GALAXY CRUISE.  The left figure shows $P(spiral)$ from GZ2 plotted against
    that from GALAXY CRUISE.  The dots are the weighted fraction from GZ2.  The dashed line shows the one to one correspondence.
    The right figure is for $P(int.)$ and the meanings of the symbols are the same as in the left figure.
  }
  \label{fig:gz2_comp}
\end{figure*}

\begin{figure*}
  \begin{center}
    \includegraphics[width=16cm]{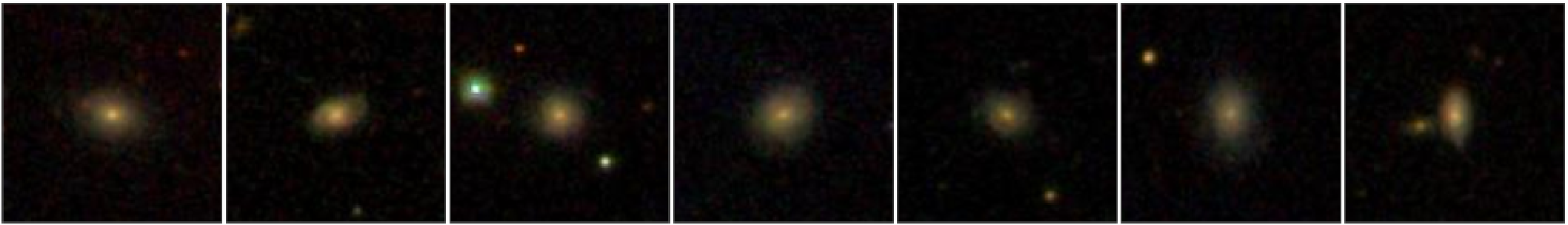}\\
    \includegraphics[width=16cm]{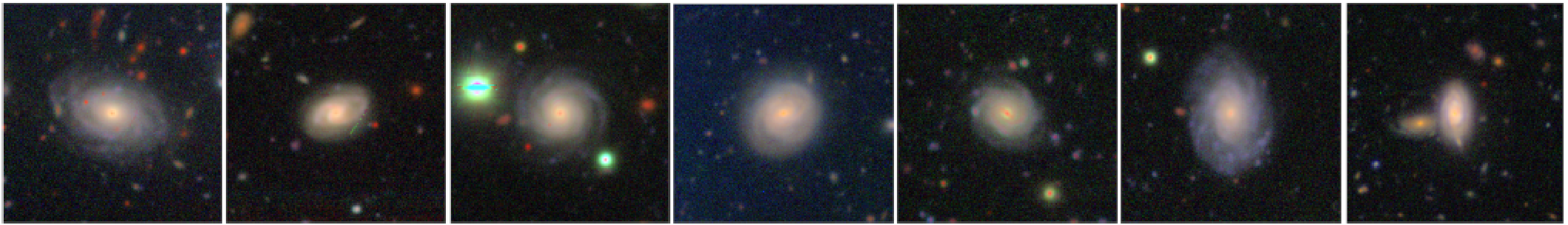}
  \end{center}
  \caption{
    Randomly drawn objects with $P_{GC}(spiral)\sim1$ and $P_{GZ2}(spiral)\sim0$.  The top row is
    the SDSS images used for GZ2 and the bottom row is GALAXY CRUISE for the same objects.  Each
    image is 48 arcsec on a side.
  }
  \label{fig:gz2_comp2}
\end{figure*}

The left panel in Fig.~\ref{fig:gz2_comp} compares $P(spiral)$.  There is a clump of objects at
$P\sim0$ (bottom-left corner) and 1 (top-right corner).  These are objects with obvious morphology and the classifications agree well.
However, there is a significant population of galaxies in the bottom-right part of the figure;
GALAXY CRUISE classifies them as spiral galaxies, while GZ2 classifies them as elliptical.
We randomly draw objects from the bottom-right corner and show their postage stamps in Fig.~\ref{fig:gz2_comp2}.
All of these galaxies are clearly spiral galaxies.  The reason for the discrepant classifications
is simply the image quality; the SDSS images do not have sufficient depth and resolution to identify 
the spiral feature.

\begin{figure*}
  \begin{center}
    \includegraphics[width=16cm]{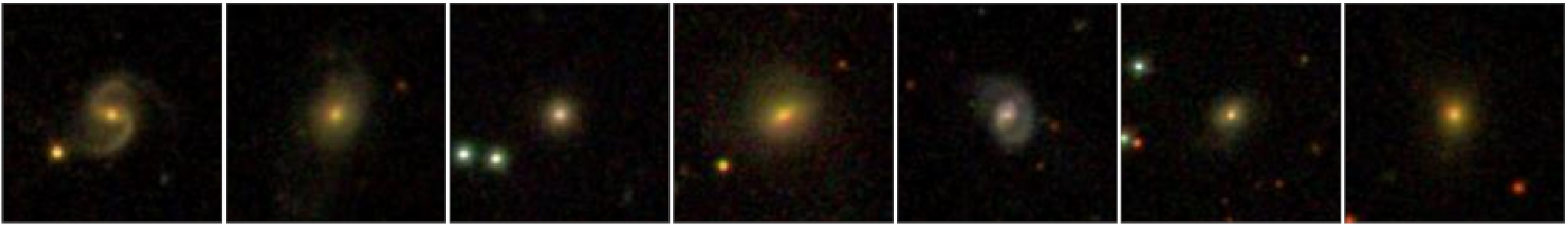}\\
    \includegraphics[width=16cm]{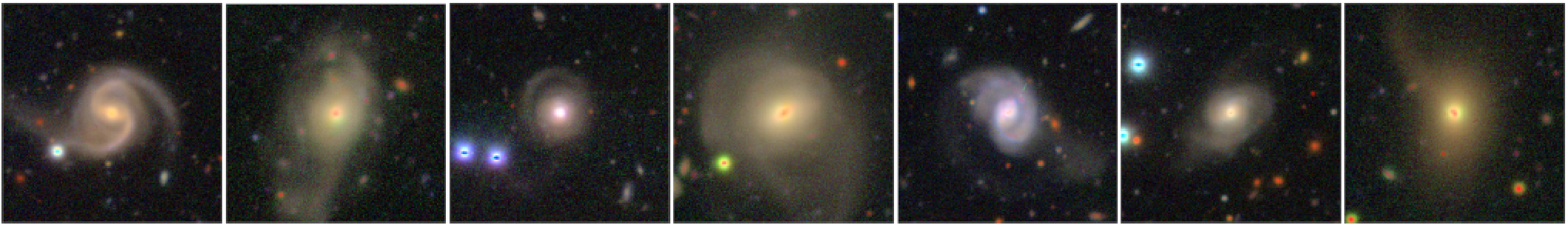}
  \end{center}
  \caption{
    As in Fig.~\ref{fig:gz2_comp2} but for interacting galaxies.
  }
  \label{fig:gz2_comp3}
\end{figure*}

The right panel in
Fig.~\ref{fig:gz2_comp} shows that there is also a significant scatter in the identifications of
interacting galaxies.  The top-left corner is empty and there is a large population again in
the bottom-right corner.  Fig.~\ref{fig:gz2_comp3} shows postage stamps of randomly drawn
objects from the bottom-right corner, and all these galaxies exhibit a clear feature to indicate
interactions.  The image depth is again the main reason for the discrepancy.  These figures
demonstrate that the classifications from GALAXY CRUISE are more accurate than those from GZ2.

\begin{figure*}
  \begin{center}
    \includegraphics[width=7.8cm]{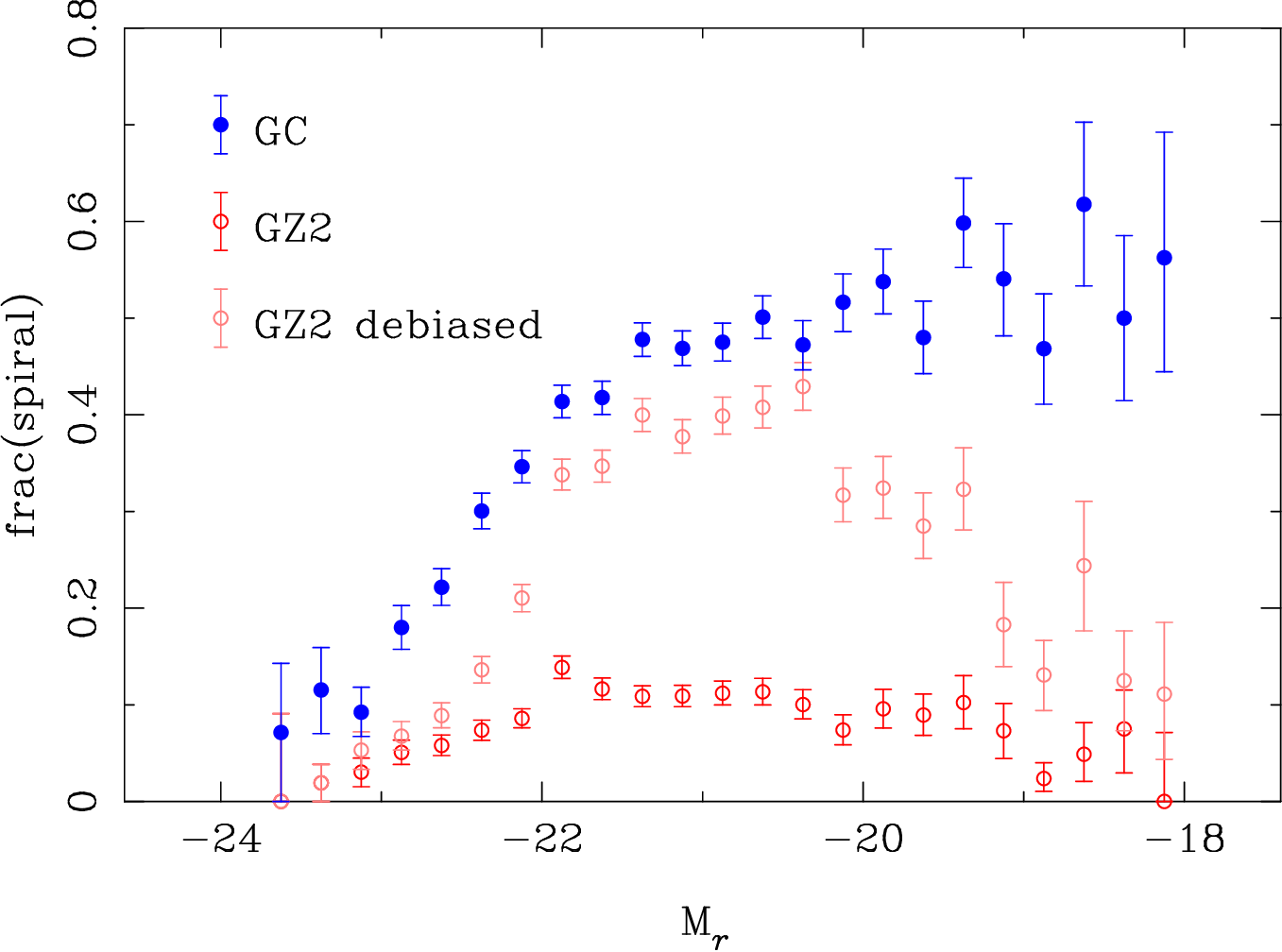}\hspace{1cm}
    \includegraphics[width=7.8cm]{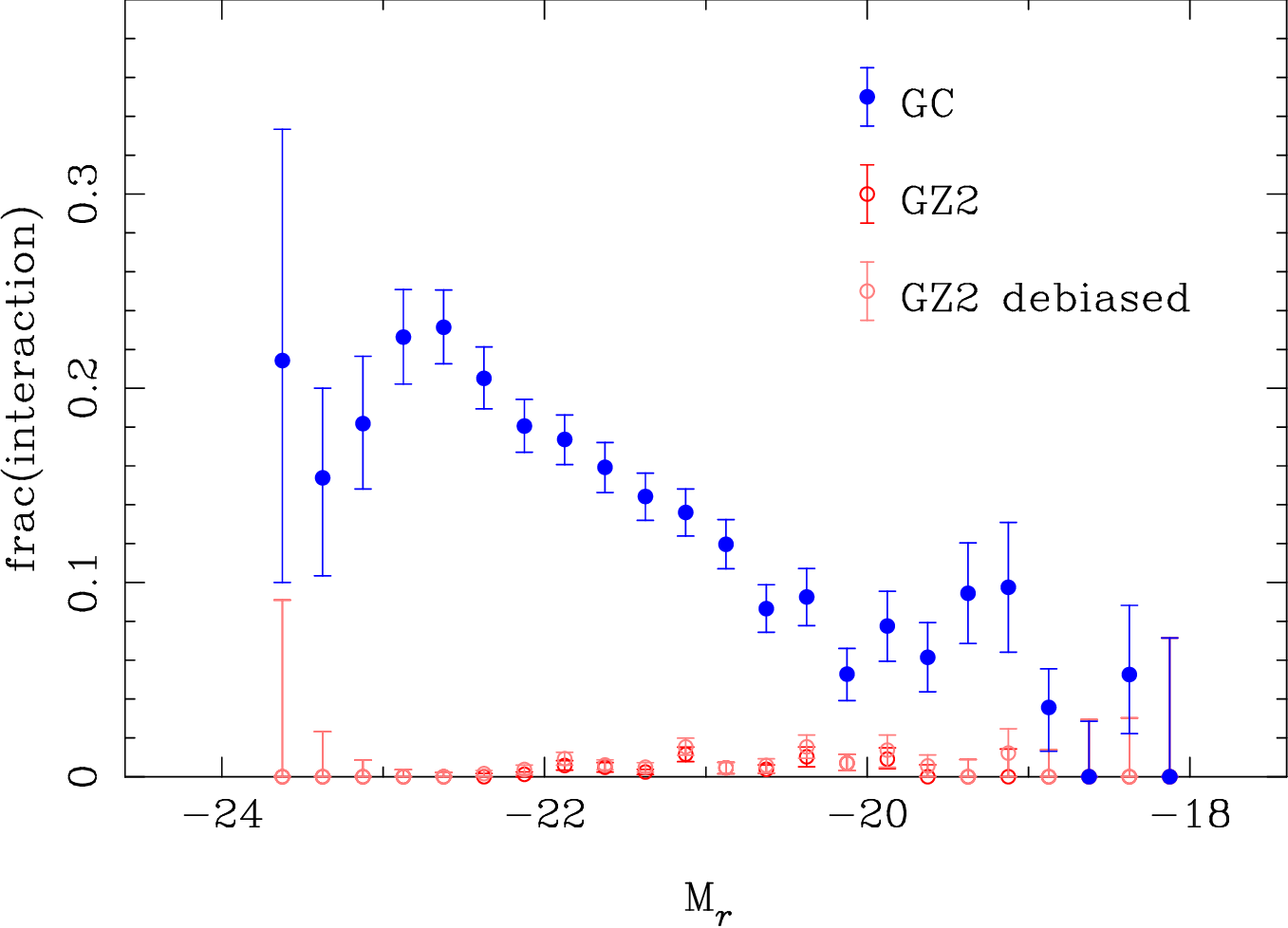}
  \end{center}
  \caption{
    {\bf Left:} Fraction of spiral galaxies plotted against $r$-band absolute magnitude.
    The blue filled circles are GALAXY CRUISE (GC), and the open circles are GZ2.
    For GZ2, we show the spiral fraction using the weighed fraction (red) and debiased fraction (pink).
    The error bars show the Poisson uncertainty.
    {\bf Right:} Fraction of interacting galaxies as a function of the $r$-band absolute magnitude.
    The meaning of the symbols are the same as in the left panel.
  }
  \label{fig:gz2_comp_absmag}
\end{figure*}

Finally, we look at the fraction of spiral galaxies and interacting galaxies as a function of
absolute magnitude in Fig.~\ref{fig:gz2_comp_absmag}. We recall that we define spiral galaxies
as galaxies with $P(spiral)>0.79$.  We adopt the same definition also for GZ2 to be fair.
The left panel shows the well-known trend that the intrinsically bright ($\sim$ massive) galaxies tend to be
early-type galaxies, while faint galaxies are more populated by late-type galaxies.  The overall normalization of the spiral
fraction is different between GALAXY CRUISE and GZ2 in the sense that the spiral galaxies are
more abundant in GALAXY CRUISE.  This trend holds even when we use the debiased fraction from GZ2
(we use the debiased fraction here because this is a statistical comparison).
We note that, if we define spiral galaxies as those with $P(spiral)>0.5$, GALAXY CRUISE and GZ2 debiased
fractions are in good agreement, except for the magnitude range at $M_r\gtrsim-21$, where GZ2 falls
below GALAXY CRUISE.
\citet{baldry04} examined the relative abundance of red and blue galaxies in the local Universe
and showed that the fraction of red galaxies rises slowly from $M_r=-18$ to $-22$, and then sharply
rises to almost unity at the brightest magnitudes.  While we plot the spiral fraction
($\sim$ fraction of blue galaxies) in Fig.~\ref{fig:gz2_comp_absmag}, the trend in the figure is
fully consistent with the finding by \citet{baldry04}.
Colors are a different property from morphology as noted earlier,
but the agreement here is reassuring because this strongly suggests that the participants'
classifications are accurate.

The right panel shows the fraction of interacting galaxies.  We recall again that the interacting galaxies
are conservatively defined as those with $P(int.)>0.79$ and the same definition is used for GZ2.
The same trend as in the left panel can be seen; the interaction fraction is higher in GALAXY CRUISE.
Interestingly, the weighted fraction and debiased fraction both yield similarly low fractions.
The fractions remain low even if we define interacting galaxies as $P(int.)>0.5$.

GALAXY CRUISE's interaction fraction seems to show a declining trend with magnitudes; fainter galaxies
are less likely to exhibit interaction features.  This luminosity dependence is consistent with previous work
(e.g., \cite{patton08}).  It is also consistent with \citet{rodriguez15} who examined the major merger rate in the Illustris Simulation
\citep{genel14,vogelsberger14} and showed that the rate is a strong function of stellar mass in the sense
that more massive galaxies experience more mergers.
The mass dependence of merger rate indicates that more massive galaxies have a larger fraction of accreted (ex-situ) stars \citep{rodriguez16}.
Our result here points to the same picture.


As described in Section \ref{sec:target_galaxies}, we mistakenly used HSC photometry to select targets.
A preliminary analysis of the data from the 2nd season shows that we tend to miss spiral galaxies and
interacting galaxies from the sample because they have significant structure.  The spiral and interaction
fractions discussed here should thus be interpreted as a lower limit.  The ongoing 2nd season of
GALAXY CRUISE will get rid of this problem and give us better estimates of these fractions.

\subsection{Dependence on Environment}
\label{sec:dependence_on_environment}

Another interesting question about our targets is their environmental dependence.
We characterize their local environment and discuss how the spiral fraction and
interaction fraction change with environment.  The former is the well-known morphology-density
relation \citep{dressler80} and we first check if the participants' classification
can reproduce it.

To compute the local environment, we use spectroscopic redshifts from SDSS and construct
a volume-limited sample at $z<0.1$ with $M_r<-20.5$.  We then estimate a distance to
the 5th nearest neighbor within a radial velocity slice of $\pm 1000\rm\ km\ s^{-1}$ from
each of our target, and translate it into local density as $5/(\pi r_5^2)$, where
$r_5$ is the projected physical distance to the 5th-nearest neighbor.
We apply the same redshift and magnitude cuts to our targets.

\begin{figure*}
  \begin{center}
    \includegraphics[width=7.8cm]{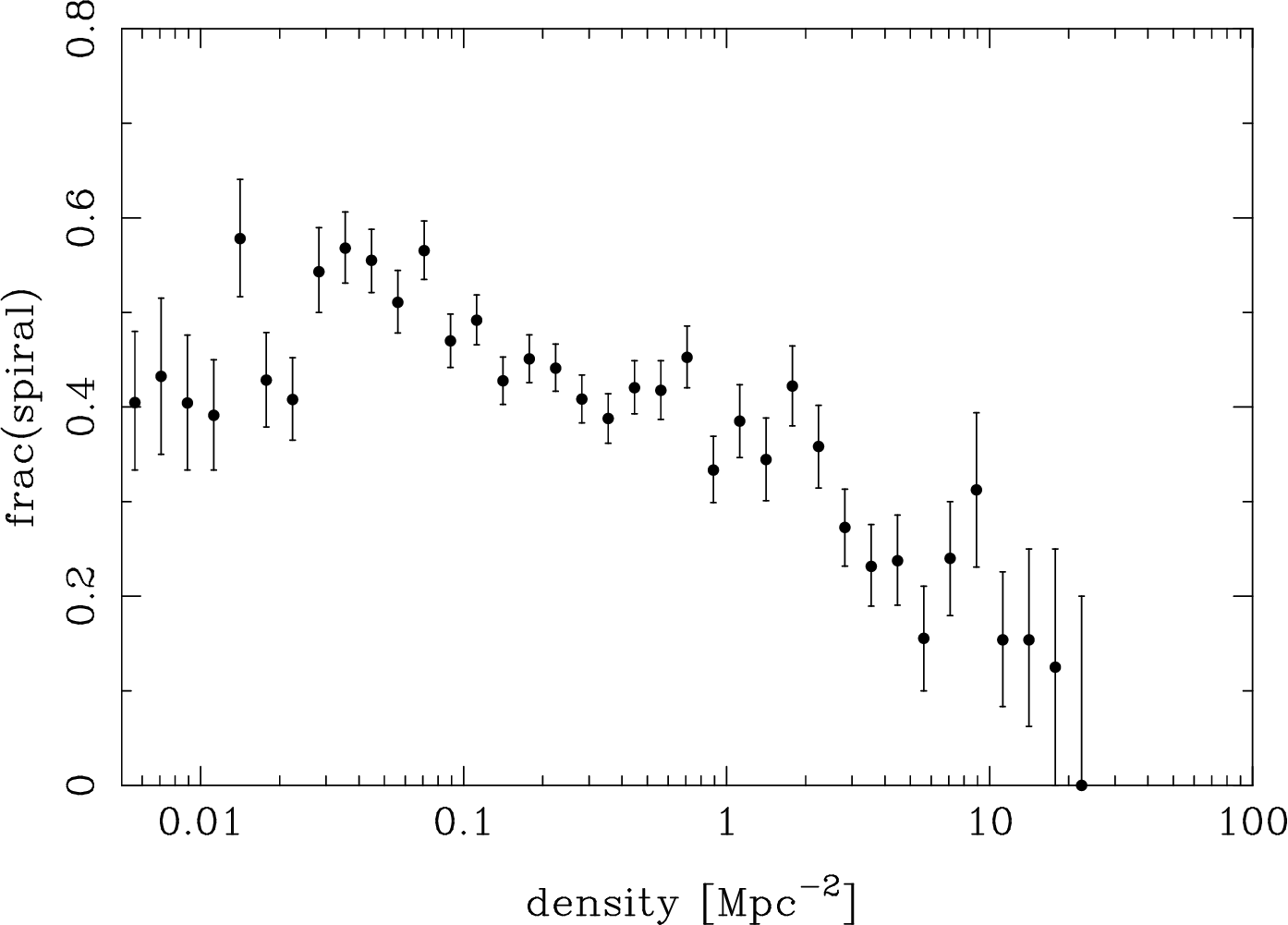}\hspace{1cm}
    \includegraphics[width=7.8cm]{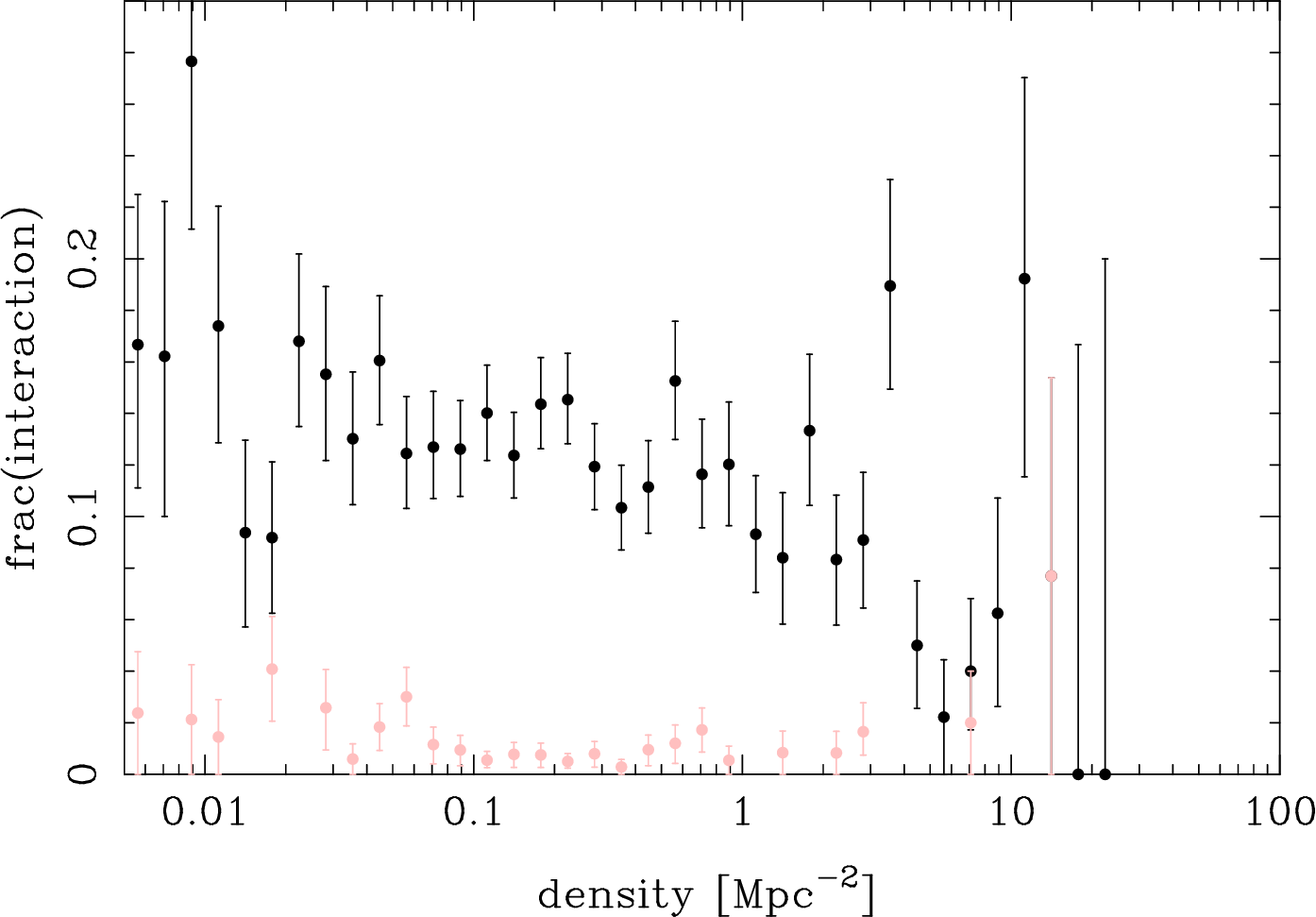}
  \end{center}
  \caption{
    {\bf Left:} Morphology-density relation.  The fraction of spiral galaxies is plotted against local density.
    Only bins with more than 5 objects are plotted.
    {\bf Right:} Fraction of interacting galaxies as a function of local density.
    The pink symbols are for the violent merger subsample.
  }
  \label{fig:morphology_density}
\end{figure*}

The left panel of Fig.
\ref{fig:morphology_density} shows the fraction of spiral galaxies as a function of local density.
We observe the spiral fraction decreasing with increasing local density.
In the densest region, the spiral fraction is below 20\%.
We observe a hint of a break in the relation at $density\sim2\ \rm Mpc^{-2}$, above which the fraction
decreases more rapidly.  The break density corresponds to the virial radius of groups and clusters \citep{tanaka04}.
The participants' classifications nicely reproduce the morphology-density relation.

The right panel of Fig. \ref{fig:morphology_density} shows the fraction of interacting galaxies
as a function of local density.  We observe a decreasing trend with increasing local density
as in the spiral fraction, although the scatter is larger. The trend remains the same
with different bin sizes.
The classical idea about interactions is that group environments is an efficient place for
interactions because relative velocities between galaxies is not very high, while
more massive clusters are not very efficient due to their high velocities.
This idea is supported by numerical simulations \citep{jian12}.
The local density alone cannot distinguish groups from clusters \citep{tanaka04} and
we have a mix of groups and clusters in Fig.~\ref{fig:morphology_density}.
Our result here implies that interactions occur, on average, less frequently in high density regions.

This possible decreasing trend is not consistent with some of the previous studies (e.g.,  \cite{lin10,deravel11}),
which reported an increasing fraction of interacting galaxies with increasing local density.
One might suspect that the difference is potentially due to different environments probed.
Some of the previous work is based on a relatively small data set such as zCOSMOS \citep{lilly07}
and DEEP2 \citep{newman13},
and thus their interacting galaxies are mostly located in groups rather than in clusters.
On the other hand, we use a larger data set and we have a larger fraction of cluster galaxies.
As a quick check, we perform the same analysis as in Fig.~\ref{fig:morphology_density} 
with density estimated with the 20th nearest neighbor so that we are more sensitive to
large-scale environment.  We find that the same trend holds; the interaction fraction
decreases in high density environments.  This suggests that the dominance of groups vs.
clusters is not the main driver of the observed trend.
We also examine the distribution of the violent mergers (Section~\ref{sec:gallery_of_selected_objects})
but they do not seem to prefer any specific environment.

The difference is at least in part due to different methods employed to identify interacting
galaxies.  Many of the previous studies are based on galaxy pairs, which probe the pre-merger
phase, while our work is based on visual signatures of interactions, which are also sensitive
to later phases of interactions.
Also, it is possible that the timescale over which tidal features are observable depends on
environment.  For instance, many galaxies are traveling fast in a small volume in clusters,
tidal features may be short-lived compared to those around isolated galaxies.
We speculate the combination of these reasons may be able to explain the observed difference.
The error bars in Fig.~\ref{fig:morphology_density} are still large and an increased sample of
interacting galaxies from, e.g., machine-learning based on the GALAXY CRUISE classifications
will allow us to further explore the origin of the reduced frequency of interactions in dense regions.

\subsection{AGN fraction}
\label{sec:agn_fraction}

Next, we look into the long-standing question of whether galaxy-galaxy interactions trigger AGN activities \citep{keel85}.
AGNs can be identified by a variety of methods, e.g., X-rays, IR colors, variability, etc.  As all of our
targets have spectra from SDSS and GAMA, we use emission line intensity ratios suggested by \citet{baldwin81}.
In order to separate AGNs from star forming galaxies, we use the threshold proposed by \citet{kauffmann03}.
This diagnostics involve 4 emission lines: H$\beta$, [OIII] ($\lambda5007$), [NII] ($\lambda6583$), and H$\alpha$.
We require that all of these 4 lines are measured at $>3\sigma$.
Among 20,686 galaxies GALAXY CRUISE targeted,
7,994 galaxies passed the condition.  The other galaxies are considered normal (i.e., non-AGN host) galaxies,
although some of them may harbor obscured or very weak AGNs.

\begin{figure}
  \begin{center}
    \includegraphics[width=8cm]{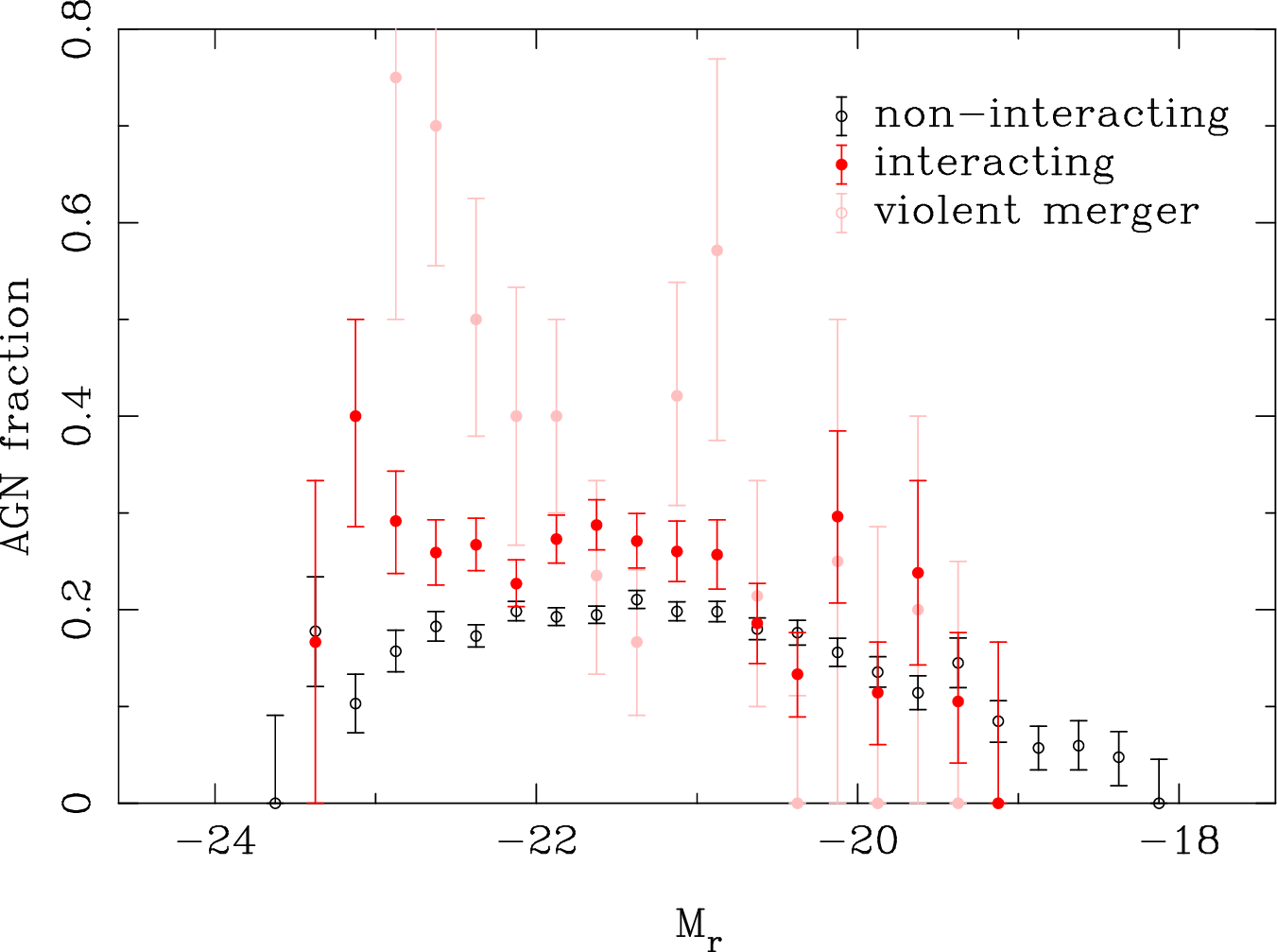}
  \end{center}
  \caption{
    AGN fraction plotted against $r$-band absolute magnitude.  The red and black circles for interacting and
    non-interacting galaxies, respectively.
    The pink points are the violent merger subsample.
    The error bars show the Poisson errors.  Only bins with
    more than 5 objects are plotted here.
  }
  \label{fig:mag_dep_agn}
\end{figure}

To get a rough picture of the AGN population in our sample, we first show in Fig.~\ref{fig:mag_dep_agn} the fraction of AGNs as a function
of magnitude.  The exact AGN fractions here should be taken with caution because they are not from a volume-limited sample.
Nonetheless, relative differences between the interacting and non-interacting galaxies are interesting;
AGN fraction is consistently higher for interacting galaxies than for non-interacting galaxies at relatively
bright magnitudes ($M_r\lesssim -21$),  
while the fraction is consistent at fainter magnitudes.  This trend is not sensitive to a particular choice
of the bin size.
The observed higher AGN fraction among interacting galaxies is in qualitative agreement with \citet{ellison13},
who observed that AGNs are more common in galaxy pairs with smaller angular separations.
We see a hint of the very high AGN fraction among bright violent mergers, which may also be in line with
\citet{ellison13} as the closest pairs are more likely to be strongly disturbed.
\citet{goulding18} addressed the AGN enhancement using the HSC-SSP imaging data and found that AGNs are
a factor of $2-8$ times more abundant.  
While the differences in the way mergers and AGNs are identified again make it difficult to make quantitative comparisons,
we are in qualitative agreement.

\begin{figure}
  \begin{center}
    \includegraphics[width=8cm]{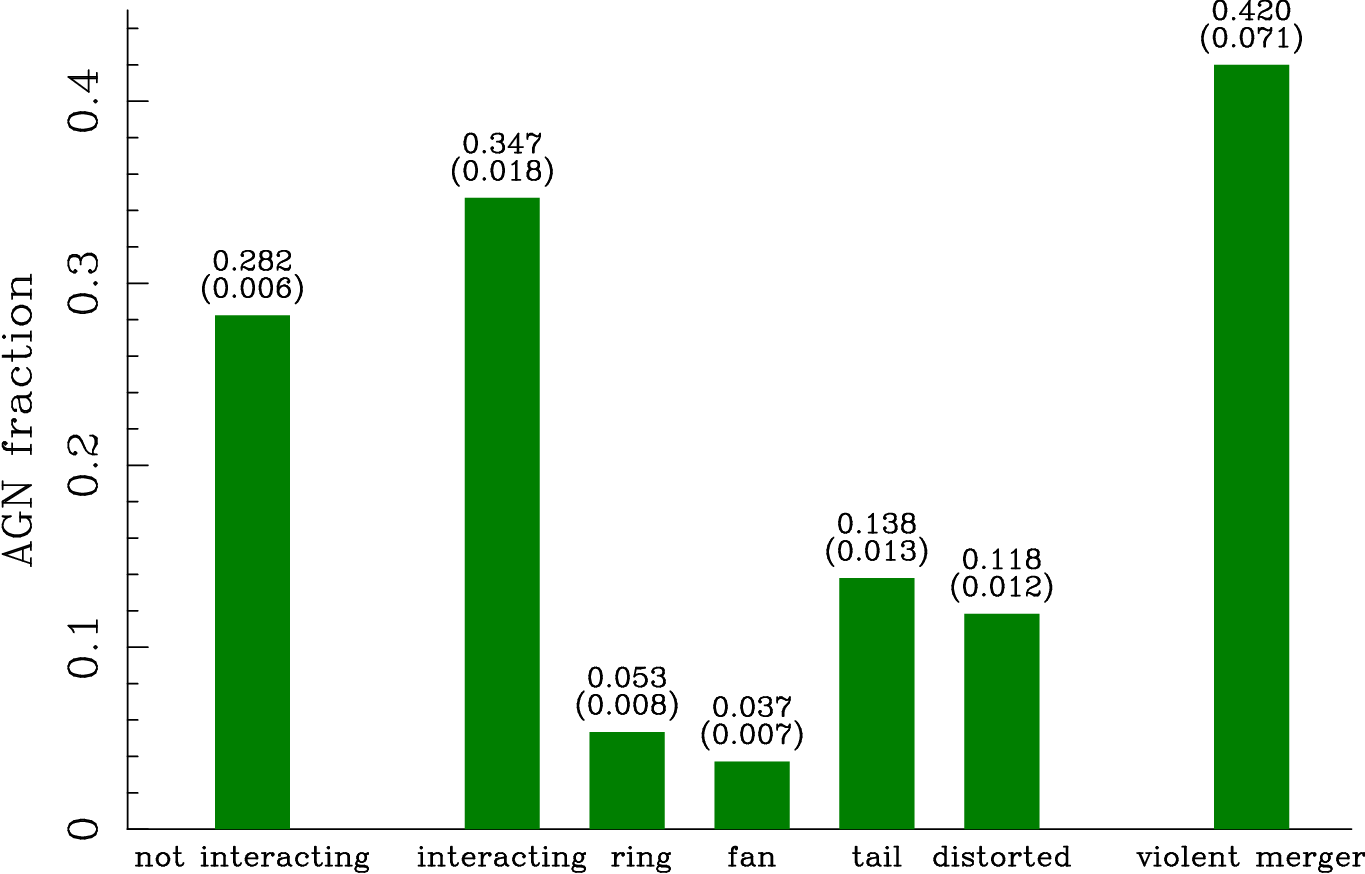}
  \end{center}
  \caption{
    Fraction of AGNs for interacting and non-interacting galaxies (left two bars).
    The overall AGN fractions are higher than those shown in Fig.~\ref{fig:mag_dep_agn}, but
    here we use a volume-limited sample at $z<0.1$.
    Interacting galaxies are split into different features as shown in the following four bars to the right.
    These feature fractions are normalized such that the sum of each feature equals the AGN fraction
    (i.e., the sum is 0.347).
    The rightmost bar is the violent merger subsample.
    The numbers on the top of each bar are the fraction and its statistical
    uncertainty, respectively.
  }
  \label{fig:stat2}
\end{figure}

We then make an attempt to see whether any of the specific interaction features is more important for triggering AGN activities.
Fig.~\ref{fig:stat2} shows AGN fractions for each interaction feature for a volume-limited sample
with $z<0.1$ and $M_r<-20.5$.  We find that the AGN fraction is clearly higher among interacting galaxies;
35\% of interacting galaxies exhibit a sign of AGN, while the fraction is 28\% for non-interacting galaxies.
This suggests that interactions can indeed trigger AGN activities at a certain rate.  If we turn our attention
to individual interaction features, we find that the relative fractions of fan, tail, and distortion features to
the total AGN fraction are similar to the overall feature fractions shown in Fig.~\ref{fig:stat1};
tail and distortions are about the same fraction and their sum comprises about 3/4 of the interacting galaxies,
and the ring and fan features are about one-third of each of these features.
A possible interpretation is that the orbital parameters of infalling galaxies are not important for
triggering AGN activities.  It may simply be the tidal disturbance that induces AGN activities.
We note that our interaction galaxies include many phases of mergers and interactions; some are in pre-merger phases,
while others are post-mergers.
There may be phase dependence of AGN activities; 
in fact, the violent mergers (Section~\ref{sec:gallery_of_selected_objects}) seem to show a higher AGN fraction of $0.42\pm0.07$,
indicating that the strong tidal field may be the key to trigger AGNs.
The uncertainty is large at this point, and more detailed investigations of merger phases may shed further light on
the physical process to activate AGNs.


Our analysis makes it clear that galaxy-galaxy interactions can trigger AGN activities, but galaxies without
any clear sign of interactions also exhibit AGN activities and their fraction is as high as 28\%.
This suggests that interactions are not the only mechanisms to trigger AGNs, and secular
processes are one of the primary ways to activate AGNs.
\citet{ellison19} reported that IR-selected AGNs show a sign of disturbance more frequently than
optically-selected AGNs.  It would be interesting to look at AGNs selected in various ways and investigate
this issue further. We leave it for future work.


\subsection{SFR enhancement during interaction}
\label{sec:star_formation_rate}

Another activity that may be enhanced during interactions is star formation
(e.g., \cite{nikolic04,ellison08,ellison13}) and
we focus on star formation activities of interacting vs. non-interacting 
galaxies in this subsection.  There are multiple ways to infer SFRs of galaxies, but here we
adopt the one based on the H$\alpha$ luminosity.  We translate the H$\alpha$ 
luminosity into SFR using the relationship from \citet{murphy11}.   We correct
for  the dust attenuation when H$\beta$ is measured at $>3\sigma$ using
the \citet{cardelli89} dust attenuation curve.  No correction is made for objects
with weaker H$\beta$.  We apply no correction for the fiber aperture loss as we are 
interested in relative differences between interacting and non-interacting
galaxies.

SFR is known to correlate significantly with stellar mass \citep{brinchmann04}.
As the interaction fraction is a function of absolute magnitude (and stellar mass as well)
as shown in Fig.~\ref{fig:gz2_comp_absmag}, a direct comparison of SFRs is not easy to
interpret.  We utilize the fact that the SFR-$M_*$ correlation is roughly linear  \citep{speagle14}
and we compare the distributions of specific SFR (sSFR) defined as SFR$/M_*$.
The relation is not completely linear, but we have confirmed the small non-linearity does
not alter our conclusion here.   We use stellar mass estimates from \citet{ahn14}
('GranadaWideDust') for SDSS galaxies.  We exclude the GAMA galaxies to avoid
potential systematic effects arising from different stellar mass estimates, but the fraction of GAMA galaxies
is only $\sim10$\% of the entire sample (Section~\ref{sec:target_galaxies}) and
whether we include them or not does not change our results.

\begin{figure}
  \begin{center}
    \includegraphics[width=8cm]{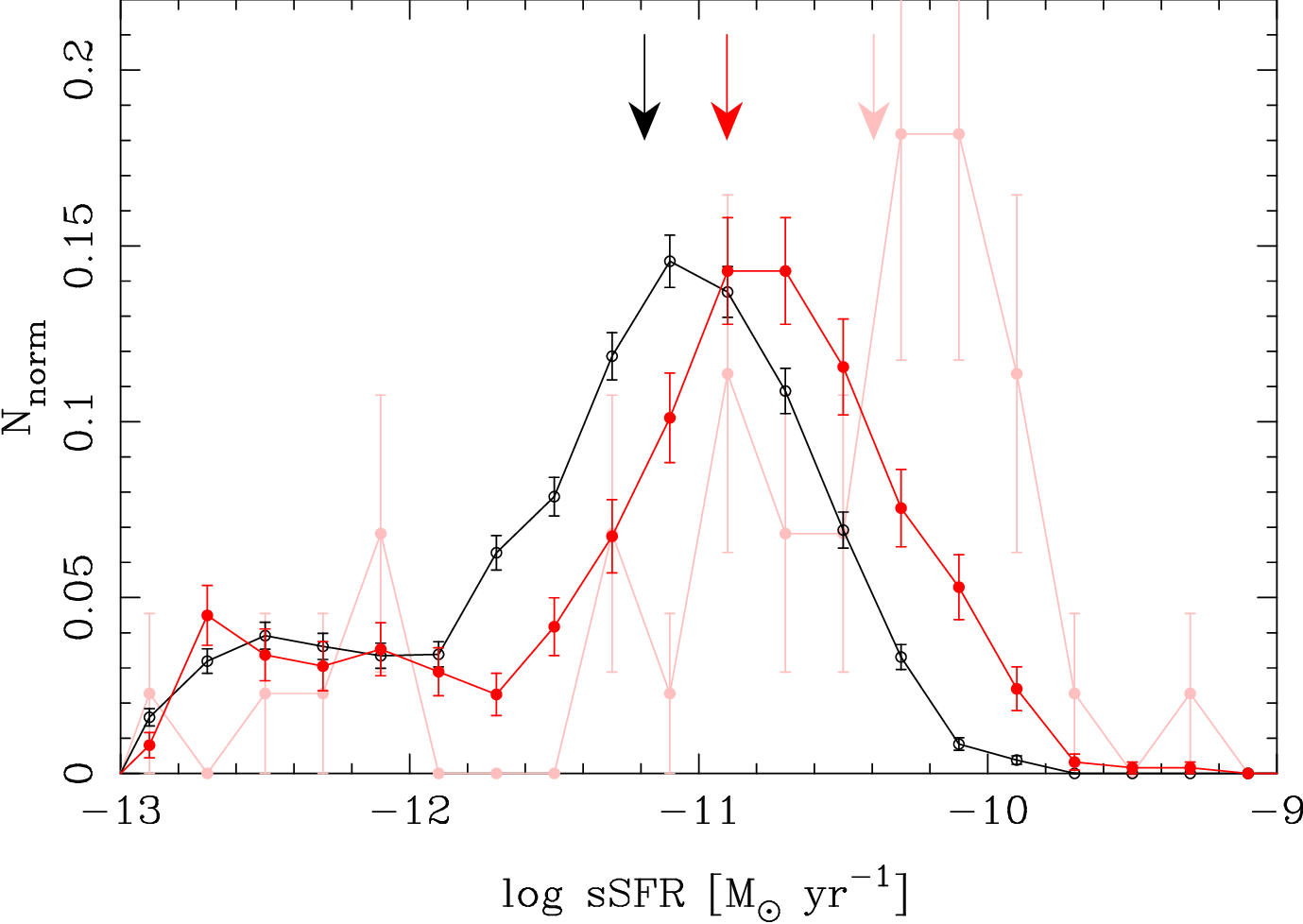}
  \end{center}
  \caption{
    Specific SFR distributions of interacting (red) and non-interacting (black) spiral galaxies.
    The violent merger subsample is shown in pink.
    The error bars show the Poisson error.  The arrows indicate the median of the distribution
    (-11.19, -10.88, and -10.39 for non-interacting, non-interacting, and violent mergers, respectively).
  }
  \label{fig:sfr_ms}
\end{figure}

Fig.~\ref{fig:sfr_ms} shows the sSFR distributions of the two populations.
For the non-interacting galaxies, we construct a mass-matched sample; we randomly
draw a subset of non-interacting galaxies so that
their stellar mass distribution is consistent with that of interacting galaxies.
The plot is for the entire sample, but the plot remains similar if we use
a volume-limited sample with $z<0.1$ and $M_*>10^{10.5}\ M_\odot$.
Note that AGNs identified in the previous subsection are excluded from
the figure so that we do not misinterpret AGN emission as star formation activity.
Note as well that we focus on spiral galaxies with $P(spiral)>0.79$ because
we are interested in the enhancement of star formation activities.  We confirm that
elliptical galaxies in our sample do not exhibit a significant sSFR enhancement.

The sSFR distributions are clearly different; interacting galaxies show an sSFR 
enhancement by about a factor of 2.  The overall shapes of the distributions are
similar, but there is a clear offset between them (the median offset is 0.31~dex).
This enhancement is consistent with the finding by \citet{darg10b}, \citet{ellison13},
\citet{bickley22} among others, although their interacting galaxies are selected in different ways.
The enhancement is also consistent with recent hydrodynamical simulations \citet{hani20}.
As we have classifications of various tidal features, we can address which feature
is more important.  It seems the offset is primarily driven by galaxies with
tidal tails.  Galaxies with distorted shapes also contribute to the overall trend.
Statistics is too poor for galaxies with ring and fan features to draw any conclusions about them.
As noted earlier, machine-learning techniques based on the current sample will
be able to make a larger sample, which will allow us to more securely characterize
the relative contribution of each feature.
However, it is fairly robust already from the current sample that sSFR is boosted
by a factor of $\sim2$ by interactions.
Interestingly, the violent mergers (Section~\ref{sec:gallery_of_selected_objects}) show
an even larger boost of a factor of $\sim6$ (0.8 dex), indicating that both star formation and AGN
activities (see the previous Section) are most significantly enhanced during the violent
merger phase.

\subsection{Ring Galaxies}
\label{sec:ring_galaxies}

\citet{few86} introduced two classes of ring galaxies: O-type, which is a smooth and regular ring with galaxy
nucleus at its center, and P-type, which is often knotted, asymmetric ring and the nucleus is not necessarily at
the center.  It is interesting that the ring galaxies shown in Fig~\ref{fig:gallery2} are mostly
O-type rings.  Some of them might be pseudo-rings \citep{shimakawa22} and there is certainly a bias towards
face-on rings in our classifications because there is no, e.g., polar-ring galaxy.
This is at least in part due to limited resolution of our imaging data on the physical scale (our sample extends
out to $z=0.2$).  \citet{buta17} discussed biases in ring galaxies selected from GZ2 and made a similar finding;
most ring galaxies are
those with bright, large outer rings with nearly face-on configuration.
While our resolution is better than SDSS on which GZ2 is based, our sample likely suffers from the same bias.
However, the subdominance of P-type rings is still interesting;
if there were P-type rings of similar angular sizes, they would have been identified.
Of course, we cannot deny the possibility that the subdominance of P-type rings is due
to classification bias that it is easier for the participants to identify smooth and symmetric
O-type rings. This caveat should be kept in mind.

\citet{elagali18} examined ring galaxies from the EAGLE simulation \citep{schaye15}.  They found that
about 80\% of the ring galaxies originate from interactions.  Interestingly, their ring galaxies seem
to be mostly P-type rings.  This may be expected as collisional rings may show star formation activities
induced by the collision.  The dominance of O-type rings in our sample is in contrast to their result.
We visually inspect galaxies with $P(int.)>0.79$ and $P(ring)>0.5$ and find that P-type rings comprise only about 10\%.
We further find that more than 80\% of the ring galaxies are classified as red galaxies on the basis of
the color-magnitude diagram, and the remaining galaxies tend to be located in the green valley.
This indicates that most ring galaxies are not actively forming stars.  O-type rings can be formed by
resonances due to the central bar, and the observed red color may be consistent with the secular formation
mechanism.  However, many of the ring galaxies in our sample do not seem to exhibit a clear bar structure.
\citet{elagali18} found that about 20\% of the ring galaxies hold the ring structure for as long as 2 Gyr
or more.  As our imaging data are fairly deep, we may be preferentially detecting these long-lived rings,
in which star formation activities have already ceased.

In any case, more detailed investigations are clearly needed to go beyond these speculations.
For instance, close comparisons with recent hydro-dynamical simulation will be useful.  Also, machine-learning techniques
make it possible to identify galaxies in simulations that look similar to a given observed galaxy.
By identifying counterparts of the ring galaxies in simulations and go back in time, we will be able to
get a better handle on the origin of these galaxies.

\subsection{Merger Rate}
\label{sec:merger_rate}

Finally, we discuss the merger rate.  It is not trivial to estimate the merger rate from
our data due to a number of observational complexities; our interacting galaxies include both
pre- and post-mergers, some of them might be undergoing fly-by interaction and will not merge, etc.
Here, we make simplified assumptions and attempt to make a rough estimate of the merger rate.

We infer merge rate as

\begin{equation}
  dN_{merger}=\frac{f_{int.}}{T} dt,
  \label{eq1}
\end{equation}

\noindent
where $f_{int.}$ is the observed interaction fraction and $T$ is the typical timescale over which
the interaction features can be observed with our data.  
Some of the merger features we have identified may be due to multiple merger events, but we assume
that they are due to a single event for simplicity.

It is difficult to make a robust estimate of
$T$ as the survival time of interaction features depend, for instance, on orbital parameters of
infalling galaxies.  A radial tail is likely short-lived because it can be destroyed as
it passes through the strong tidal field around the potential center, while tangential tails
(arcs) may live longer.  There may also be environmental dependence as we discussed earlier.
We assume that interaction features are observable for of order the dynamical timescale, $T\sim1$ Gyr.
Equation \ref{eq1} is then

\begin{equation}
  d N_{merger} = f_{int.}\left(\frac{1Gyr}{T}\right) \frac{dt}{Gyr}.
  \label{eq2}
\end{equation}

\begin{figure}
  \begin{center}
    \includegraphics[width=8cm]{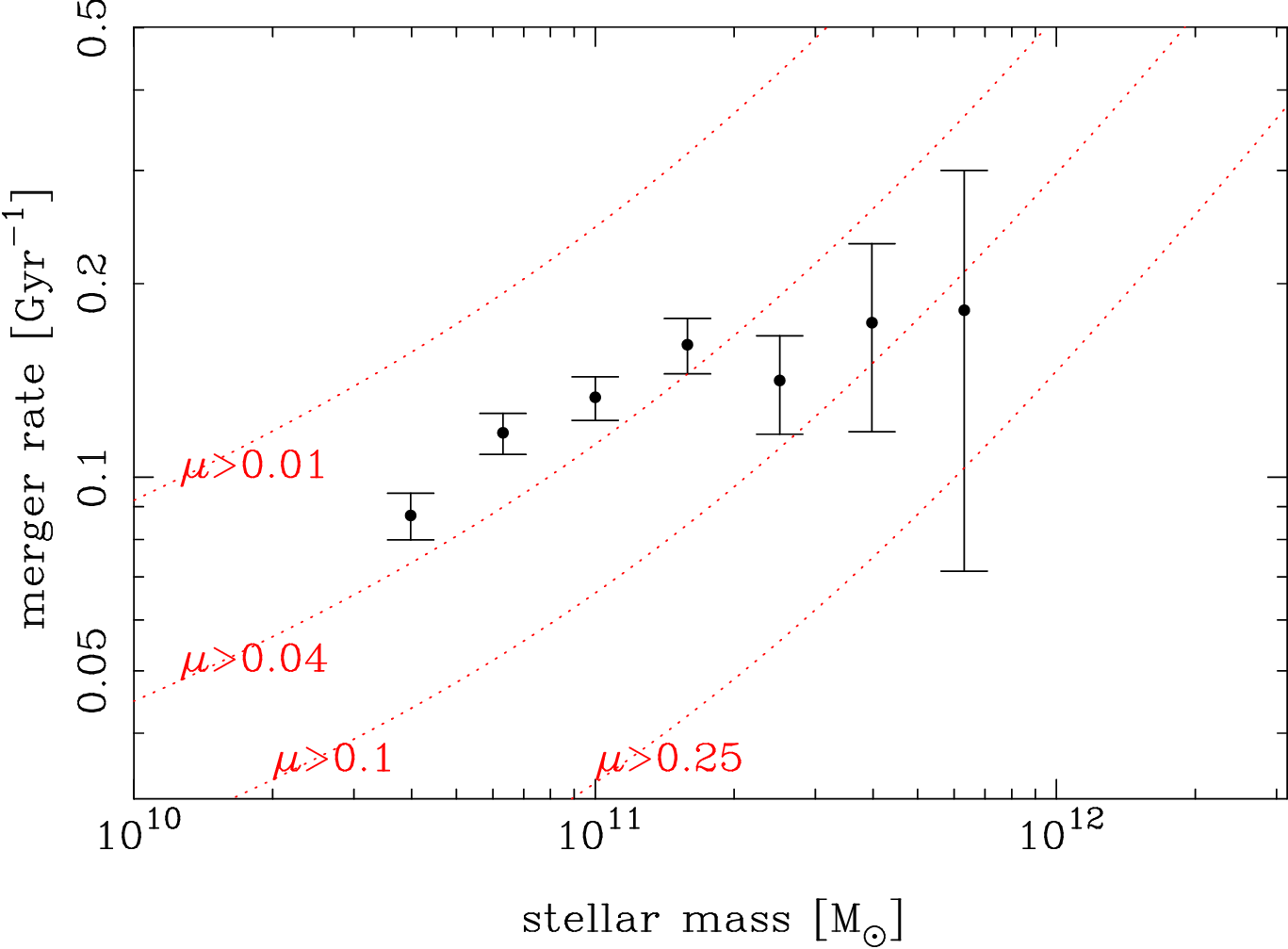}
  \end{center}
  \caption{
    Merger rate plotted against stellar mass.  The points show the observed merger rate inferred from
    Equation \ref{eq2}. The associated error bars include only the Poisson error.  The dotted curves are from
    \citet{rodriguez15} based on the Illustris simulation.  The different curves mean different
    range of mass ratios ($\mu$) between the merging galaxies considered.  For instance,
    the curve for $\mu>0.1$ is for all mergers with mass ratio of 0.1 or larger.
  }
  \label{fig:merger_rate}
\end{figure}

The inferred merger rate is shown in Fig.~\ref{fig:merger_rate} for a volume-limited sample with
$z<0.1$ and $M_*>10^{10.5}\rm M_\odot$.  For comparison, we show
predictions from the Illustris simulation \citep{rodriguez15}.  The merger rate is a strong
function of mass ratio between merging galaxies in the sense that minor mergers are more
common than major mergers.  Here, we find that our merger rate corresponds approximately to $\mu>0.04$
(i.e., mass ratios larger than $\sim1/25$).  This large mass ratio may not be too surprising because our imaging is very deep and
is able to detect faint features due to minor mergers.
Note that the overall normalization of our merger rate is sensitive to
the assumptions employed and it should be regarded only as a rough estimate.

Based on the merger rate, we can infer the mass growth rate as
\begin{eqnarray}
  \frac{d M_1}{dt} & = & <M_2> \frac{dN_{merger}}{dt}\nonumber\\
  \frac{d M_1}{M_1} & = & f_{int.}\left(\frac{<M_2>}{M_1}\right)\left(\frac{1Gyr}{T}\right) \frac{dt}{Gyr},
\end{eqnarray}

\noindent
where $M_1$ is the mass of the central galaxy, and $<M_2>$ is the typical mass of an accreting galaxy.
For $f_{int.}=0.13$ (which is the interaction fraction at $M_*=10^{11}\rm\ M_\odot$) and $<M_2>/M_1=0.04$,
the (fractional) growth rate inferred here is $dM_1/M_1\sim0.005\rm\ Gyr^{-1}$, which is more than
an order of magnitude smaller than that estimated by \citet{tal09}.  One of the main differences is
in $f_{int.}$; \citet{tal09} find that 70\% of galaxies they observed show a sign of interaction,
while we find 13\%.  As noted earlier, the overall normalization has a significant uncertainty
both in our work and also in \citet{tal09} and we do not peruse the difference further at this point.

Assuming the redshift evolution of the merger rate from \citet{rodriguez15} and consider all mergers
with mass ratio larger than $0.04$, we expect the mass growth of a factor of $\sim1.4$ since $z=1$ for
galaxies with $M_*=10^{11}\rm M_\odot$ at $z=0$.  This is in broad agreement with \citet{davidzon17} who
evaluated the evolution of the stellar mass function of galaxies and found the observed growth of the characteristic
stellar mass since $z\sim1$ is $\sim0.2$ dex.

Despite the simplified assumptions employed, the inferred mass growth rate roughly reproduces
the observed mass growth.  Our approach here is complementary to many of the literature results
based on galaxy pairs (e.g., \cite{bundy09,lopez-sanjuan13}) due to different systematic
effects.  Our merger rates are, however, sensitive to the assumptions employed as well as
the adopted threshold to define interacting galaxies, $P(int.)$.
As we discussed in Section~\ref{sec:color_magnitude_diagram}, we do now know the best $P(int.)$ to cut at.
The adopted threshold is rather conservative and more optimal threshold may change the overall normalization.

\section{Summary}
\label{sec:discussion_and_summary}

We have presented our community science project GALAXY CRUISE and its first science results.
We focus on morphologies of nearby galaxies with spectroscopic redshifts from SDSS and GAMA
with an emphasis on interaction features.  The combination of
the depth, seeing, and wide area coverage of the HSC-SSP data surpasses the previous community science projects
on similar subjects; we are able to identify spiral features and interaction features that
were missed in previous projects.  After careful screening and statistical corrections to
the classifications, the catalog is ready for various scientific explorations.

We have first compared with previous results from GZ2.  We confirm that we make more accurate
morphological classifications of galaxies than the previous work thanks to the better imaging data.
We successfully reproduce the morphology-density relation.  We find that 
tidal tails and distortion are the most frequently observed interaction features and they comprise about
3/4 of all the interaction features.  The remainder is equally shared between the ring and fan features.
The interacting galaxies seem to decrease in high density regions, which is not necessarily consistent with
previous findings.  We suspect that the different methods to used identify interacting galaxies
can at least partly explain the difference.

As all of our targets have spectra, we apply the BPT diagnostics to identify AGNs.
We demonstrate that interactions can trigger AGN activities; the fraction of AGN host
galaxies among interacting galaxies is clearly higher than that of non-interacting galaxies.
However, about a quarter of non-interacting galaxies harbor AGNs, which indicate that interactions
are not the only mechanism to trigger AGNs.
We also demonstrate that interactions enhance star formation activities of spiral galaxies
by about a factor of 2.
Both AGN and star formation activities are even more enhanced among violent mergers,
indicating that the strong tidal field is driving this enhancement.

Finally, we have made an attempt to infer the merger rate.  We find that our merger
rate is similar to that inferred from the Illustris simulation for mass ratios greater
than 1:25.  The inferred mass growth rate due to mergers since $z=1$ is in good agreement
with the observed evolution of the stellar mass function since $z=1$.  As many of the previous
studies inferred the merger rate from pair statistics, our analysis in this paper is
highly complementary because interacting galaxies are identified in a different way.

This work is based on classifications collected during the 1st season of GALAXY CRUISE.
We make the merged catalog used in the paper publicly available at the GALAXY CRUISE
website, so that the world-wide
community benefits from the large number of high-quality classifications.  As emphasized
throughout the paper, the HSC-SSP images are much deeper than those adopted in previous
projects on similar subjects.  Our catalog here can be used as a training data set to
develop machine-learning algorithms and apply them to galaxies that are not targeted
in this paper (e.g., those without spectroscopic redshifts and those covered in releases
after PDR2), which will significantly increase the sample of interacting galaxies.
This is just one example to use the catalog, and it can be used for a wide variety of purposes.

Currently, the 2nd season of GALAXY CRUISE is under way.
Given the good classification accuracy achieved in the 1st season, we choose to include
fainter galaxies.  We also eliminate the photometry problem that was erroneously introduced.
The 2nd season data will thus extend the magnitude range and help us unveil the nature of fainter galaxies
in the local Universe.  We will report on updated analyses using the 2nd season data in our future
paper, and the classifications will be released to the community in due course as well.




\begin{ack}

  This paper is based on morphological classifications of galaxies by the GALAXY CRUISE volunteers, without whom
  the work would not have been possible.  We deeply appreciate their efforts and contributions.
  We thank Miraikan, the National Museum of Emerging Science and Innovation, for their support and assistance
  during the public experiments.  The experiments were essential
  for us to define the classification scheme and and make a tutorial course.
  This work is supported by JSPS KAKENHI Grant Numbers JP 22H01270 and JSPS 22K14078.
  We thank the anonymous referee for constructive comments, which helped improve the paper.
  
  The Hyper Suprime-Cam (HSC) collaboration includes the astronomical communities of Japan and Taiwan,
  and Princeton University.  The HSC instrumentation and software were developed by the National
  Astronomical Observatory of Japan (NAOJ), the Kavli Institute for the Physics and Mathematics of
  the Universe (Kavli IPMU), the University of Tokyo, the High Energy Accelerator Research Organization (KEK),
  the Academia Sinica Institute for Astronomy and Astrophysics in Taiwan (ASIAA), and Princeton University.
  Funding was contributed by the FIRST program from Japanese Cabinet Office, the Ministry of Education,
  Culture, Sports, Science and Technology (MEXT), the Japan Society for the Promotion of Science (JSPS),
  Japan Science and Technology Agency  (JST),  the Toray Science  Foundation, NAOJ, Kavli IPMU, KEK, ASIAA,
  and Princeton University.
  
  This paper is based on data collected at the Subaru Telescope and retrieved from the HSC data archive system, which is operated by Subaru Telescope and Astronomy Data Center at NAOJ.  Data analysis was in part carried out with the cooperation of Center for Computational Astrophysics at NAOJ.  We are honored and grateful for the opportunity of observing the Universe from Maunakea, which has the cultural, historical and natural significance in Hawaii.
  
  This paper makes use of software developed for Vera C. Rubin Observatory. We thank the Rubin Observatory for
  making their code available as free software at http://pipelines.lsst.io/.
  
  The Pan-STARRS1 Surveys (PS1) and the PS1 public science archive have been made possible through contributions by the Institute for Astronomy, the University of Hawaii, the Pan-STARRS Project Office, the Max-Planck Society and its participating institutes, the Max Planck Institute for Astronomy, Heidelberg and the Max Planck Institute for Extraterrestrial Physics, Garching, The Johns Hopkins University, Durham University, the University of Edinburgh, the Queen's University Belfast, the Harvard-Smithsonian Center for Astrophysics, the Las Cumbres Observatory Global Telescope Network Incorporated, the National Central University of Taiwan, the Space Telescope Science Institute, the National Aeronautics and Space Administration under Grant No. NNX08AR22G issued through the Planetary Science Division of the NASA Science Mission Directorate, the National Science Foundation Grant No. AST-1238877, the University of Maryland, Eotvos Lorand University (ELTE), the Los Alamos National Laboratory, and the Gordon and Betty Moore Foundation.

  Funding for the Sloan Digital Sky 
Survey IV has been provided by the 
Alfred P. Sloan Foundation, the U.S. 
Department of Energy Office of 
Science, and the Participating 
Institutions. 

SDSS-IV acknowledges support and 
resources from the Center for High 
Performance Computing  at the 
University of Utah. The SDSS 
website is www.sdss4.org.

SDSS-IV is managed by the 
Astrophysical Research Consortium 
for the Participating Institutions 
of the SDSS Collaboration including 
the Brazilian Participation Group, 
the Carnegie Institution for Science, 
Carnegie Mellon University, Center for 
Astrophysics | Harvard \& 
Smithsonian, the Chilean Participation 
Group, the French Participation Group, 
Instituto de Astrof\'isica de 
Canarias, The Johns Hopkins 
University, Kavli Institute for the 
Physics and Mathematics of the 
Universe (IPMU) / University of 
Tokyo, the Korean Participation Group, 
Lawrence Berkeley National Laboratory, 
Leibniz Institut f\"ur Astrophysik 
Potsdam (AIP),  Max-Planck-Institut 
f\"ur Astronomie (MPIA Heidelberg), 
Max-Planck-Institut f\"ur 
Astrophysik (MPA Garching), 
Max-Planck-Institut f\"ur 
Extraterrestrische Physik (MPE), 
National Astronomical Observatories of 
China, New Mexico State University, 
New York University, University of 
Notre Dame, Observat\'ario 
Nacional / MCTI, The Ohio State 
University, Pennsylvania State 
University, Shanghai 
Astronomical Observatory, United 
Kingdom Participation Group, 
Universidad Nacional Aut\'onoma 
de M\'exico, University of Arizona, 
University of Colorado Boulder, 
University of Oxford, University of 
Portsmouth, University of Utah, 
University of Virginia, University 
of Washington, University of 
Wisconsin, Vanderbilt University, 
and Yale University.

   GAMA is a joint European-Australasian project based around a spectroscopic campaign using the Anglo-Australian Telescope. The GAMA input catalogue is based on data taken from the Sloan Digital Sky Survey and the UKIRT Infrared Deep Sky Survey. Complementary imaging of the GAMA regions is being obtained by a number of independent survey programmes including GALEX MIS, VST KiDS, VISTA VIKING, WISE, Herschel-ATLAS, GMRT and ASKAP providing UV to radio coverage. GAMA is funded by the STFC (UK), the ARC (Australia), the AAO, and the participating institutions. The GAMA website is http://www.gama-survey.org/ .

\end{ack}

\appendix 
\section*{Using probabilities as weights}

\begin{figure}
  \begin{center}
    \includegraphics[width=8cm]{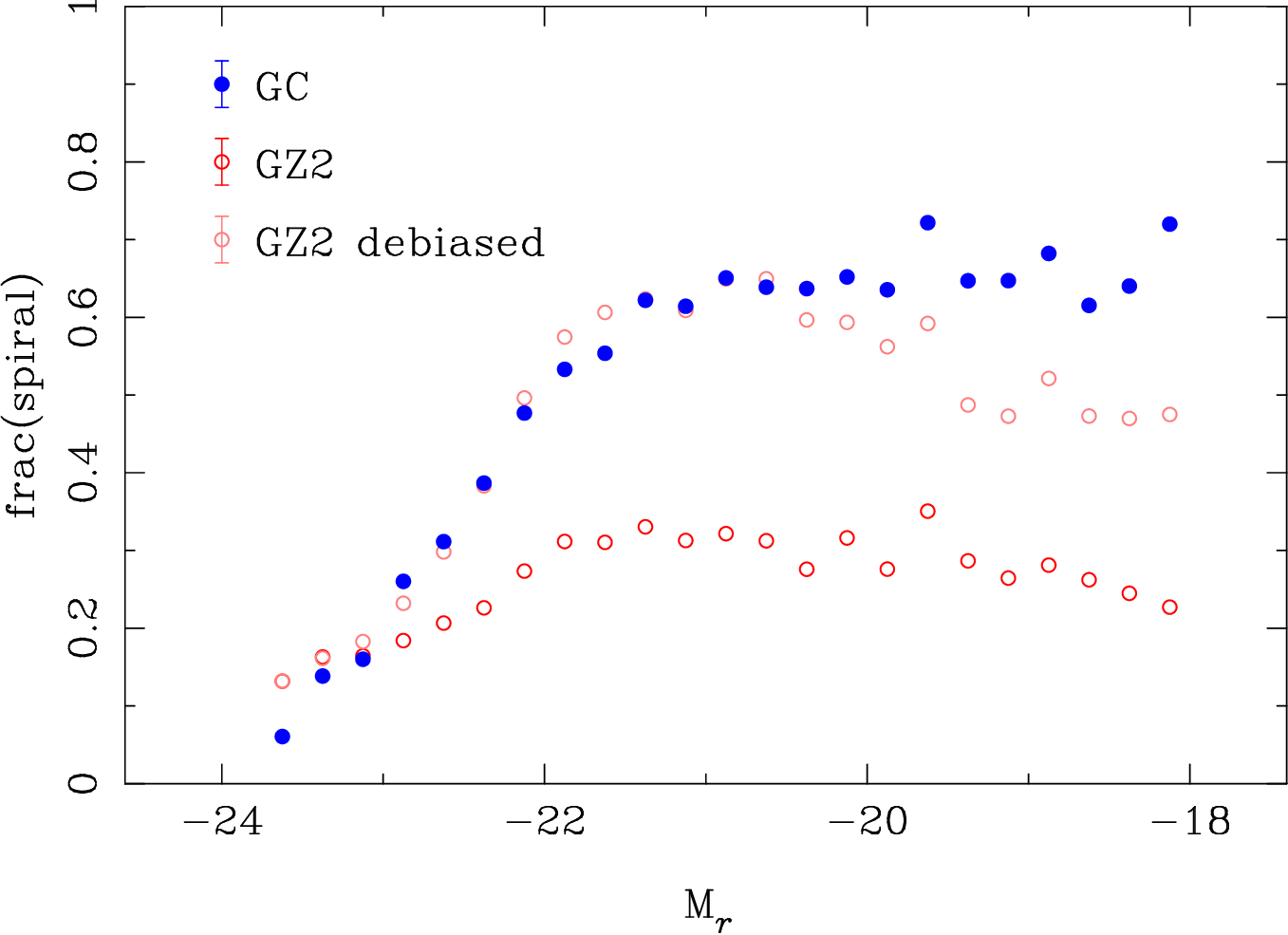}
  \end{center}
  \caption{
    Same as Fig.~\ref{fig:gz2_comp_absmag}, but here $P(spiral)$ is used as a weight instead of
    thresholding $P(spiral)$ to define spiral galaxies.
  }
  \label{fig:gz2_comp_absmag2}
\end{figure}

In the main body of the paper, we define galaxies with $P(spiral)>0.79$ as spiral galaxies.
Instead of thresholding the probability,
one can consider simply using the probability as weight.  For instance, a galaxy with
$P(spiral)=0.3$ can be counted as 0.3 spiral galaxy.  Which approach to adopt depends on
what one wishes to look at, but we here briefly discuss how the spiral fraction changes between
the two approaches.  Fig.~\ref{fig:gz2_comp_absmag2} shows the spiral fraction as a function of
$M_r$ using $P(spiral)$ as weight.  This figure can be compared with Fig.~\ref{fig:gz2_comp_absmag},
which is based on thresholding.  Overall, the spiral fraction from GALAXY CRUISE does not change
significantly.  This is not surprising because $P(spiral)$ is sharply peaked at 0 and 1 as shown
in Fig.~\ref{fig:color_magnitude2} and thresholding at $P(spiral)=0.79$ simply splits these
peaks into two.  On the other hand, $P(spiral)$ from GZ2 is more contiguous, and both weighted and
debiased fractions increase when $P(spiral)$ is used as weight.  The debiased fractions are in
better agreement with GALAXY CRUISE, but there is still the under-abundance of faint spirals.




\bibliographystyle{apj}
\bibliography{references}

\begin{thebibliography}{}
\expandafter\ifx\csname natexlab\endcsname\relax\def\natexlab#1{#1}\fi

\bibitem[{{Abraham} {et~al.}(1994){Abraham}, {Valdes}, {Yee}, \& {van den
  Bergh}}]{abraham94}
{Abraham}, R.~G., {Valdes}, F., {Yee}, H.~K.~C., \& {van den Bergh}, S. 1994,
  \apj, 432, 75

\bibitem[{{Abraham} {et~al.}(2003){Abraham}, {van den Bergh}, \&
  {Nair}}]{abraham03}
{Abraham}, R.~G., {van den Bergh}, S., \& {Nair}, P. 2003, \apj, 588, 218

\bibitem[{{Aguado} {et~al.}(2019){Aguado}, {Ahumada}, {Almeida}, {Anderson},
  {Andrews}, {Anguiano}, {Aquino Ort{\'\i}z}, {Arag{\'o}n-Salamanca},
  {Argudo-Fern{\'a}ndez}, {Aubert}, {Avila-Reese}, {Badenes}, {Barboza
  Rembold}, {Barger}, {Barrera-Ballesteros}, {Bates}, {Bautista}, {Beaton},
  {Beers}, {Belfiore}, {Bernardi}, {Bershady}, {Beutler}, {Bird}, {Bizyaev},
  {Blanc}, {Blanton}, {Blomqvist}, {Bolton}, {Boquien}, {Borissova}, {Bovy},
  {Brandt}, {Brinkmann}, {Brownstein}, {Bundy}, {Burgasser}, {Byler}, {Cano
  Diaz}, {Cappellari}, {Carrera}, {Cervantes Sodi}, {Chen}, {Cherinka}, {Choi},
  {Chung}, {Coffey}, {Comerford}, {Comparat}, {Covey}, {da Silva Ilha}, {da
  Costa}, {Dai}, {Damke}, {Darling}, {Davies}, {Dawson}, {de Sainte Agathe},
  {Deconto Machado}, {Del Moro}, {De Lee}, {Diamond-Stanic}, {Dom{\'\i}nguez
  S{\'a}nchez}, {Donor}, {Drory}, {du Mas des Bourboux}, {Duckworth}, {Dwelly},
  {Ebelke}, {Emsellem}, {Escoffier}, {Fern{\'a}ndez-Trincado}, {Feuillet},
  {Fischer}, {Fleming}, {Fraser-McKelvie}, {Freischlad}, {Frinchaboy}, {Fu},
  {Galbany}, {Garcia-Dias}, {Garc{\'\i}a-Hern{\'a}ndez}, {Garma Oehmichen},
  {Geimba Maia}, {Gil-Mar{\'\i}n}, {Grabowski}, {Gu}, {Guo}, {Ha},
  {Harrington}, {Hasselquist}, {Hayes}, {Hearty}, {Hernandez Toledo}, {Hicks},
  {Hogg}, {Holley-Bockelmann}, {Holtzman}, {Hsieh}, {Hunt}, {Hwang},
  {Ibarra-Medel}, {Jimenez Angel}, {Johnson}, {Jones}, {J{\"o}nsson},
  {Kinemuchi}, {Kollmeier}, {Krawczyk}, {Kreckel}, {Kruk}, {Lacerna}, {Lan},
  {Lane}, {Law}, {Lee}, {Li}, {Lian}, {Lin}, {Lin}, {Lintott}, {Long},
  {Longa-Pe{\~n}a}, {Mackereth}, {de la Macorra}, {Majewski}, {Malanushenko},
  {Manchado}, {Maraston}, {Mariappan}, {Marinelli}, {Marques-Chaves},
  {Masseron}, {Masters}, {McDermid}, {Medina Pe{\~n}a}, {Meneses-Goytia},
  {Merloni}, {Merrifield}, {Meszaros}, {Minniti}, {Minsley}, {Muna}, {Myers},
  {Nair}, {Correa do Nascimento}, {Newman}, {Nitschelm}, {Olmstead}, {Oravetz},
  {Oravetz}, {Ortega Minakata}, {Pace}, {Padilla}, {Palicio}, {Pan}, {Pan},
  {Parikh}, {Parker}, {Peirani}, {Penny}, {Percival}, {Perez-Fournon},
  {Peterken}, {Pinsonneault}, {Prakash}, {Raddick}, {Raichoor}, {Riffel},
  {Riffel}, {Rix}, {Robin}, {Roman-Lopes}, {Rose}, {Ross}, {Rossi}, {Rowlands},
  {Rubin}, {S{\'a}nchez}, {S{\'a}nchez-Gallego}, {Sayres}, {Schaefer},
  {Schiavon}, {Schimoia}, {Schlafly}, {Schlegel}, {Schneider}, {Schultheis},
  {Seo}, {Shamsi}, {Shao}, {Shen}, {Shetty}, {Simonian}, {Smethurst}, {Sobeck},
  {Souter}, {Spindler}, {Stark}, {Stassun}, {Steinmetz}, {Storchi-Bergmann},
  {Stringfellow}, {Su{\'a}rez}, {Sun}, {Taghizadeh-Popp}, {Talbot}, {Tayar},
  {Thakar}, {Thomas}, {Tissera}, {Tojeiro}, {Troup}, {Unda-Sanzana},
  {Valenzuela}, {Vargas-Maga{\~n}a}, {V{\'a}zquez-Mata}, {Wake}, {Weaver},
  {Weijmans}, {Westfall}, {Wild}, {Wilson}, {Woods}, {Yan}, {Yang}, {Zamora},
  {Zasowski}, {Zhang}, {Zheng}, {Zheng}, {Zhu}, {Zinn}, \& {Zou}}]{aguado19}
{Aguado}, D.~S., {Ahumada}, R., {Almeida}, A., {et~al.} 2019, \apjs, 240, 23

\bibitem[{{Ahn} {et~al.}(2014){Ahn}, {Alexandroff}, {Allende Prieto}, {Anders},
  {Anderson}, {Anderton}, {Andrews}, {Aubourg}, {Bailey}, {Bastien}, \&
  et~al.}]{ahn14}
{Ahn}, C.~P., {Alexandroff}, R., {Allende Prieto}, C., {et~al.} 2014, \apjs,
  211, 17

\bibitem[{{Aihara} {et~al.}(2018{\natexlab{a}}){Aihara}, {Armstrong},
  {Bickerton}, {Bosch}, {Coupon}, {Furusawa}, {Hayashi}, {Ikeda}, {Kamata},
  {Karoji}, {Kawanomoto}, {Koike}, {Komiyama}, {Lang}, {Lupton}, {Mineo},
  {Miyatake}, {Miyazaki}, {Morokuma}, {Obuchi}, {Oishi}, {Okura}, {Price},
  {Takata}, {Tanaka}, {Tanaka}, {Tanaka}, {Uchida}, {Uraguchi}, {Utsumi},
  {Wang}, {Yamada}, {Yamanoi}, {Yasuda}, {Arimoto}, {Chiba}, {Finet},
  {Fujimori}, {Fujimoto}, {Furusawa}, {Goto}, {Goulding}, {Gunn}, {Harikane},
  {Hattori}, {Hayashi}, {He{\l}miniak}, {Higuchi}, {Hikage}, {Ho}, {Hsieh},
  {Huang}, {Huang}, {Imanishi}, {Iwata}, {Jaelani}, {Jian}, {Kashikawa},
  {Katayama}, {Kojima}, {Konno}, {Koshida}, {Kusakabe}, {Leauthaud}, {Lee},
  {Lin}, {Lin}, {Mandelbaum}, {Matsuoka}, {Medezinski}, {Miyama}, {Momose},
  {More}, {More}, {Mukae}, {Murata}, {Murayama}, {Nagao}, {Nakata}, {Niida},
  {Niikura}, {Nishizawa}, {Oguri}, {Okabe}, {Ono}, {Onodera}, {Onoue}, {Ouchi},
  {Pyo}, {Shibuya}, {Shimasaku}, {Simet}, {Speagle}, {Spergel}, {Strauss},
  {Sugahara}, {Sugiyama}, {Suto}, {Suzuki}, {Tait}, {Takada}, {Terai}, {Toba},
  {Turner}, {Uchiyama}, {Umetsu}, {Urata}, {Usuda}, {Yeh}, \&
  {Yuma}}]{aihara18a}
{Aihara}, H., {Armstrong}, R., {Bickerton}, S., {et~al.} 2018{\natexlab{a}},
  \pasj, 70, S8

\bibitem[{{Aihara} {et~al.}(2018{\natexlab{b}}){Aihara}, {Arimoto},
  {Armstrong}, {Arnouts}, {Bahcall}, {Bickerton}, {Bosch}, {Bundy}, {Capak},
  {Chan}, {Chiba}, {Coupon}, {Egami}, {Enoki}, {Finet}, {Fujimori}, {Fujimoto},
  {Furusawa}, {Furusawa}, {Goto}, {Goulding}, {Greco}, {Greene}, {Gunn},
  {Hamana}, {Harikane}, {Hashimoto}, {Hattori}, {Hayashi}, {Hayashi},
  {He{\l}miniak}, {Higuchi}, {Hikage}, {Ho}, {Hsieh}, {Huang}, {Huang},
  {Ikeda}, {Imanishi}, {Inoue}, {Iwasawa}, {Iwata}, {Jaelani}, {Jian},
  {Kamata}, {Karoji}, {Kashikawa}, {Katayama}, {Kawanomoto}, {Kayo}, {Koda},
  {Koike}, {Kojima}, {Komiyama}, {Konno}, {Koshida}, {Koyama}, {Kusakabe},
  {Leauthaud}, {Lee}, {Lin}, {Lin}, {Lupton}, {Mandelbaum}, {Matsuoka},
  {Medezinski}, {Mineo}, {Miyama}, {Miyatake}, {Miyazaki}, {Momose}, {More},
  {More}, {Moritani}, {Moriya}, {Morokuma}, {Mukae}, {Murata}, {Murayama},
  {Nagao}, {Nakata}, {Niida}, {Niikura}, {Nishizawa}, {Obuchi}, {Oguri},
  {Oishi}, {Okabe}, {Okamoto}, {Okura}, {Ono}, {Onodera}, {Onoue}, {Osato},
  {Ouchi}, {Price}, {Pyo}, {Sako}, {Sawicki}, {Shibuya}, {Shimasaku},
  {Shimono}, {Shirasaki}, {Silverman}, {Simet}, {Speagle}, {Spergel},
  {Strauss}, {Sugahara}, {Sugiyama}, {Suto}, {Suyu}, {Suzuki}, {Tait},
  {Takada}, {Takata}, {Tamura}, {Tanaka}, {Tanaka}, {Tanaka}, {Tanaka},
  {Terai}, {Terashima}, {Toba}, {Tominaga}, {Toshikawa}, {Turner}, {Uchida},
  {Uchiyama}, {Umetsu}, {Uraguchi}, {Urata}, {Usuda}, {Utsumi}, {Wang}, {Wang},
  {Wong}, {Yabe}, {Yamada}, {Yamanoi}, {Yasuda}, {Yeh}, {Yonehara}, \&
  {Yuma}}]{aihara18b}
{Aihara}, H., {Arimoto}, N., {Armstrong}, R., {et~al.} 2018{\natexlab{b}},
  \pasj, 70, S4

\bibitem[{{Aihara} {et~al.}(2019){Aihara}, {AlSayyad}, {Ando}, {Armstrong},
  {Bosch}, {Egami}, {Furusawa}, {Furusawa}, {Goulding}, {Harikane}, {Hikage},
  {Ho}, {Hsieh}, {Huang}, {Ikeda}, {Imanishi}, {Ito}, {Iwata}, {Jaelani},
  {Kakuma}, {Kawana}, {Kikuta}, {Kobayashi}, {Koike}, {Komiyama}, {Li},
  {Liang}, {Lin}, {Luo}, {Lupton}, {Lust}, {MacArthur}, {Matsuoka}, {Mineo},
  {Miyatake}, {Miyazaki}, {More}, {Murata}, {Namiki}, {Nishizawa}, {Oguri},
  {Okabe}, {Okamoto}, {Okura}, {Ono}, {Onodera}, {Onoue}, {Osato}, {Ouchi},
  {Shibuya}, {Strauss}, {Sugiyama}, {Suto}, {Takada}, {Takagi}, {Takata},
  {Takita}, {Tanaka}, {Terai}, {Toba}, {Uchiyama}, {Utsumi}, {Wang}, {Wang}, \&
  {Yamada}}]{aihara19}
{Aihara}, H., {AlSayyad}, Y., {Ando}, M., {et~al.} 2019, \pasj, 71, 114

\bibitem[{{Baldry} {et~al.}(2004){Baldry}, {Glazebrook}, {Brinkmann},
  {Ivezi{\'c}}, {Lupton}, {Nichol}, \& {Szalay}}]{baldry04}
{Baldry}, I.~K., {Glazebrook}, K., {Brinkmann}, J., {et~al.} 2004, \apj, 600,
  681

\bibitem[{{Baldwin} {et~al.}(1981){Baldwin}, {Phillips}, \&
  {Terlevich}}]{baldwin81}
{Baldwin}, J.~A., {Phillips}, M.~M., \& {Terlevich}, R. 1981, \pasp, 93, 5

\bibitem[{{Bickley} {et~al.}(2022){Bickley}, {Ellison}, {Patton}, {Bottrell},
  {Gwyn}, \& {Hudson}}]{bickley22}
{Bickley}, R.~W., {Ellison}, S.~L., {Patton}, D.~R., {et~al.} 2022, \mnras,
  514, 3294

\bibitem[{{Blanton} \& {Roweis}(2007)}]{blanton07}
{Blanton}, M.~R., \& {Roweis}, S. 2007, \aj, 133, 734

\bibitem[{{Bosch} {et~al.}(2018){Bosch}, {Armstrong}, {Bickerton}, {Furusawa},
  {Ikeda}, {Koike}, {Lupton}, {Mineo}, {Price}, {Takata}, {Tanaka}, {Yasuda},
  {AlSayyad}, {Becker}, {Coulton}, {Coupon}, {Garmilla}, {Huang}, {Krughoff},
  {Lang}, {Leauthaud}, {Lim}, {Lust}, {MacArthur}, {Mandelbaum}, {Miyatake},
  {Miyazaki}, {Murata}, {More}, {Okura}, {Owen}, {Swinbank}, {Strauss},
  {Yamada}, \& {Yamanoi}}]{bosch18}
{Bosch}, J., {Armstrong}, R., {Bickerton}, S., {et~al.} 2018, \pasj, 70, S5

\bibitem[{{Bottrell} {et~al.}(2019){Bottrell}, {Simard}, {Mendel}, \&
  {Ellison}}]{bottrell19}
{Bottrell}, C., {Simard}, L., {Mendel}, J.~T., \& {Ellison}, S.~L. 2019,
  \mnras, 486, 390

\bibitem[{{Brinchmann} {et~al.}(2004){Brinchmann}, {Charlot}, {White},
  {Tremonti}, {Kauffmann}, {Heckman}, \& {Brinkmann}}]{brinchmann04}
{Brinchmann}, J., {Charlot}, S., {White}, S.~D.~M., {et~al.} 2004, \mnras, 351,
  1151

\bibitem[{{Bundy} {et~al.}(2009){Bundy}, {Fukugita}, {Ellis}, {Targett},
  {Belli}, \& {Kodama}}]{bundy09}
{Bundy}, K., {Fukugita}, M., {Ellis}, R.~S., {et~al.} 2009, \apj, 697, 1369

\bibitem[{{Buta}(2017)}]{buta17}
{Buta}, R.~J. 2017, \mnras, 471, 4027

\bibitem[{{Cardelli} {et~al.}(1989){Cardelli}, {Clayton}, \&
  {Mathis}}]{cardelli89}
{Cardelli}, J.~A., {Clayton}, G.~C., \& {Mathis}, J.~S. 1989, \apj, 345, 245

\bibitem[{{Casteels} {et~al.}(2013){Casteels}, {Bamford}, {Skibba}, {Masters},
  {Lintott}, {Keel}, {Schawinski}, {Nichol}, \& {Smith}}]{casteels13}
{Casteels}, K. R.~V., {Bamford}, S.~P., {Skibba}, R.~A., {et~al.} 2013, \mnras,
  429, 1051

\bibitem[{{Chambers} {et~al.}(2016){Chambers}, {Magnier}, {Metcalfe},
  {Flewelling}, {Huber}, {Waters}, {Denneau}, {Draper}, {Farrow}, {Finkbeiner},
  {Holmberg}, {Koppenhoefer}, {Price}, {Rest}, {Saglia}, {Schlafly}, {Smartt},
  {Sweeney}, {Wainscoat}, {Burgett}, {Chastel}, {Grav}, {Heasley}, {Hodapp},
  {Jedicke}, {Kaiser}, {Kudritzki}, {Luppino}, {Lupton}, {Monet}, {Morgan},
  {Onaka}, {Shiao}, {Stubbs}, {Tonry}, {White}, {Ba{\~n}ados}, {Bell},
  {Bender}, {Bernard}, {Boegner}, {Boffi}, {Botticella}, {Calamida},
  {Casertano}, {Chen}, {Chen}, {Cole}, {Deacon}, {Frenk}, {Fitzsimmons},
  {Gezari}, {Gibbs}, {Goessl}, {Goggia}, {Gourgue}, {Goldman}, {Grant},
  {Grebel}, {Hambly}, {Hasinger}, {Heavens}, {Heckman}, {Henderson}, {Henning},
  {Holman}, {Hopp}, {Ip}, {Isani}, {Jackson}, {Keyes}, {Koekemoer}, {Kotak},
  {Le}, {Liska}, {Long}, {Lucey}, {Liu}, {Martin}, {Masci}, {McLean}, {Mindel},
  {Misra}, {Morganson}, {Murphy}, {Obaika}, {Narayan}, {Nieto-Santisteban},
  {Norberg}, {Peacock}, {Pier}, {Postman}, {Primak}, {Rae}, {Rai}, {Riess},
  {Riffeser}, {Rix}, {R{\"o}ser}, {Russel}, {Rutz}, {Schilbach}, {Schultz},
  {Scolnic}, {Strolger}, {Szalay}, {Seitz}, {Small}, {Smith}, {Soderblom},
  {Taylor}, {Thomson}, {Taylor}, {Thakar}, {Thiel}, {Thilker}, {Unger},
  {Urata}, {Valenti}, {Wagner}, {Walder}, {Walter}, {Watters}, {Werner},
  {Wood-Vasey}, \& {Wyse}}]{chambers16}
{Chambers}, K.~C., {Magnier}, E.~A., {Metcalfe}, N., {et~al.} 2016, arXiv
  e-prints, arXiv:1612.05560

\bibitem[{{Couch} {et~al.}(1998){Couch}, {Barger}, {Smail}, {Ellis}, \&
  {Sharples}}]{couch98}
{Couch}, W.~J., {Barger}, A.~J., {Smail}, I., {Ellis}, R.~S., \& {Sharples},
  R.~M. 1998, \apj, 497, 188

\bibitem[{{Darg} {et~al.}(2010{\natexlab{a}}){Darg}, {Kaviraj}, {Lintott},
  {Schawinski}, {Sarzi}, {Bamford}, {Silk}, {Proctor}, {Andreescu}, {Murray},
  {Nichol}, {Raddick}, {Slosar}, {Szalay}, {Thomas}, \& {Vandenberg}}]{darg10a}
{Darg}, D.~W., {Kaviraj}, S., {Lintott}, C.~J., {et~al.} 2010{\natexlab{a}},
  \mnras, 401, 1043

\bibitem[{{Darg} {et~al.}(2010{\natexlab{b}}){Darg}, {Kaviraj}, {Lintott},
  {Schawinski}, {Sarzi}, {Bamford}, {Silk}, {Andreescu}, {Murray}, {Nichol},
  {Raddick}, {Slosar}, {Szalay}, {Thomas}, \& {Vandenberg}}]{darg10b}
---. 2010{\natexlab{b}}, \mnras, 401, 1552

\bibitem[{{Davidzon} {et~al.}(2017){Davidzon}, {Ilbert}, {Laigle}, {Coupon},
  {McCracken}, {Delvecchio}, {Masters}, {Capak}, {Hsieh}, {Le F{\`e}vre},
  {Tresse}, {Bethermin}, {Chang}, {Faisst}, {Le Floc'h}, {Steinhardt}, {Toft},
  {Aussel}, {Dubois}, {Hasinger}, {Salvato}, {Sanders}, {Scoville}, \&
  {Silverman}}]{davidzon17}
{Davidzon}, I., {Ilbert}, O., {Laigle}, C., {et~al.} 2017, \aap, 605, A70

\bibitem[{{de Ravel} {et~al.}(2011){de Ravel}, {Kampczyk}, {Le F{\`e}vre},
  {Lilly}, {Tasca}, {Tresse}, {Lopez-Sanjuan}, {Bolzonella}, {Kovac}, {Abbas},
  {Bardelli}, {Bongiorno}, {Caputi}, {Contini}, {Coppa}, {Cucciati}, {de la
  Torre}, {Dunlop}, {Franzetti}, {Garilli}, {Iovino}, {Kneib}, {Koekemoer},
  {Knobel}, {Lamareille}, {Le Borgne}, {Le Brun}, {Leauthaud}, {Maier},
  {Mainieri}, {Mignoli}, {Pello}, {Peng}, {Perez Montero}, {Ricciardelli},
  {Scodeggio}, {Silverman}, {Tanaka}, {Vergani}, {Zamorani}, {Zucca},
  {Bottini}, {Cappi}, {Carollo}, {Cassata}, {Cimatti}, {Fumana}, {Guzzo},
  {Maccagni}, {Marinoni}, {McCracken}, {Memeo}, {Meneux}, {Oesch}, {Porciani},
  {Pozzetti}, {Renzini}, {Scaramella}, \& {Scarlata}}]{deravel11}
{de Ravel}, L., {Kampczyk}, P., {Le F{\`e}vre}, O., {et~al.} 2011, arXiv
  e-prints, arXiv:1104.5470

\bibitem[{{Doi} {et~al.}(1993){Doi}, {Fukugita}, \& {Okamura}}]{doi93}
{Doi}, M., {Fukugita}, M., \& {Okamura}, S. 1993, \mnras, 264, 832

\bibitem[{{Dressler}(1980)}]{dressler80}
{Dressler}, A. 1980, \apj, 236, 351

\bibitem[{{Driver} {et~al.}(2009){Driver}, {Norberg}, {Baldry}, {Bamford},
  {Hopkins}, {Liske}, {Loveday}, {Peacock}, {Hill}, {Kelvin}, {Robotham},
  {Cross}, {Parkinson}, {Prescott}, {Conselice}, {Dunne}, {Brough}, {Jones},
  {Sharp}, {van Kampen}, {Oliver}, {Roseboom}, {Bland-Hawthorn}, {Croom},
  {Ellis}, {Cameron}, {Cole}, {Frenk}, {Couch}, {Graham}, {Proctor}, {De
  Propris}, {Doyle}, {Edmondson}, {Nichol}, {Thomas}, {Eales}, {Jarvis},
  {Kuijken}, {Lahav}, {Madore}, {Seibert}, {Meyer}, {Staveley-Smith},
  {Phillipps}, {Popescu}, {Sansom}, {Sutherland}, {Tuffs}, \&
  {Warren}}]{driver09}
{Driver}, S.~P., {Norberg}, P., {Baldry}, I.~K., {et~al.} 2009, Astronomy and
  Geophysics, 50, 5.12

\bibitem[{{Duc} {et~al.}(2015){Duc}, {Cuillandre}, {Karabal}, {Cappellari},
  {Alatalo}, {Blitz}, {Bournaud}, {Bureau}, {Crocker}, {Davies}, {Davis}, {de
  Zeeuw}, {Emsellem}, {Khochfar}, {Krajnovi{\'c}}, {Kuntschner}, {McDermid},
  {Michel-Dansac}, {Morganti}, {Naab}, {Oosterloo}, {Paudel}, {Sarzi}, {Scott},
  {Serra}, {Weijmans}, \& {Young}}]{duc15}
{Duc}, P.-A., {Cuillandre}, J.-C., {Karabal}, E., {et~al.} 2015, \mnras, 446,
  120

\bibitem[{{Elagali} {et~al.}(2018){Elagali}, {Lagos}, {Wong}, {Staveley-Smith},
  {Trayford}, {Schaller}, {Yuan}, \& {Abadi}}]{elagali18}
{Elagali}, A., {Lagos}, C. D.~P., {Wong}, O.~I., {et~al.} 2018, \mnras, 481,
  2951

\bibitem[{{Ellison} {et~al.}(2013){Ellison}, {Mendel}, {Patton}, \&
  {Scudder}}]{ellison13}
{Ellison}, S.~L., {Mendel}, J.~T., {Patton}, D.~R., \& {Scudder}, J.~M. 2013,
  \mnras, 435, 3627

\bibitem[{{Ellison} {et~al.}(2008){Ellison}, {Patton}, {Simard}, \&
  {McConnachie}}]{ellison08}
{Ellison}, S.~L., {Patton}, D.~R., {Simard}, L., \& {McConnachie}, A.~W. 2008,
  \aj, 135, 1877

\bibitem[{{Ellison} {et~al.}(2019){Ellison}, {Viswanathan}, {Patton},
  {Bottrell}, {McConnachie}, {Gwyn}, \& {Cuillandre}}]{ellison19}
{Ellison}, S.~L., {Viswanathan}, A., {Patton}, D.~R., {et~al.} 2019, \mnras,
  487, 2491

\bibitem[{{Eyre} \& {Binney}(2011)}]{eyre11}
{Eyre}, A., \& {Binney}, J. 2011, \mnras, 413, 1852

\bibitem[{{Few} \& {Madore}(1986)}]{few86}
{Few}, J. M.~A., \& {Madore}, B.~F. 1986, \mnras, 222, 673

\bibitem[{{Fukugita} {et~al.}(2007){Fukugita}, {Nakamura}, {Okamura}, {Yasuda},
  {Barentine}, {Brinkmann}, {Gunn}, {Harvanek}, {Ichikawa}, {Lupton},
  {Schneider}, {Strauss}, \& {York}}]{fukugita07}
{Fukugita}, M., {Nakamura}, O., {Okamura}, S., {et~al.} 2007, \aj, 134, 579

\bibitem[{{Genel} {et~al.}(2014){Genel}, {Vogelsberger}, {Springel}, {Sijacki},
  {Nelson}, {Snyder}, {Rodriguez-Gomez}, {Torrey}, \& {Hernquist}}]{genel14}
{Genel}, S., {Vogelsberger}, M., {Springel}, V., {et~al.} 2014, \mnras, 445,
  175

\bibitem[{{Goulding} {et~al.}(2018){Goulding}, {Greene}, {Bezanson}, {Greco},
  {Johnson}, {Leauthaud}, {Matsuoka}, {Medezinski}, \&
  {Price-Whelan}}]{goulding18}
{Goulding}, A.~D., {Greene}, J.~E., {Bezanson}, R., {et~al.} 2018, \pasj, 70,
  S37

\bibitem[{{Hani} {et~al.}(2020){Hani}, {Gosain}, {Ellison}, {Patton}, \&
  {Torrey}}]{hani20}
{Hani}, M.~H., {Gosain}, H., {Ellison}, S.~L., {Patton}, D.~R., \& {Torrey}, P.
  2020, \mnras, 493, 3716

\bibitem[{{Jian} {et~al.}(2012){Jian}, {Lin}, \& {Chiueh}}]{jian12}
{Jian}, H.-Y., {Lin}, L., \& {Chiueh}, T. 2012, \apj, 754, 26

\bibitem[{{Kauffmann} {et~al.}(2003){Kauffmann}, {Heckman}, {Tremonti},
  {Brinchmann}, {Charlot}, {White}, {Ridgway}, {Brinkmann}, {Fukugita}, {Hall},
  {Ivezi{\'c}}, {Richards}, \& {Schneider}}]{kauffmann03}
{Kauffmann}, G., {Heckman}, T.~M., {Tremonti}, C., {et~al.} 2003, \mnras, 346,
  1055

\bibitem[{{Keel} {et~al.}(1985){Keel}, {Kennicutt}, {Hummel}, \& {van der
  Hulst}}]{keel85}
{Keel}, W.~C., {Kennicutt}, R.~C., J., {Hummel}, E., \& {van der Hulst}, J.~M.
  1985, \aj, 90, 708

\bibitem[{{Lilly} {et~al.}(2007){Lilly}, {Le F{\`e}vre}, {Renzini}, {Zamorani},
  {Scodeggio}, {Contini}, {Carollo}, {Hasinger}, {Kneib}, {Iovino}, {Le Brun},
  {Maier}, {Mainieri}, {Mignoli}, {Silverman}, {Tasca}, {Bolzonella},
  {Bongiorno}, {Bottini}, {Capak}, {Caputi}, {Cimatti}, {Cucciati}, {Daddi},
  {Feldmann}, {Franzetti}, {Garilli}, {Guzzo}, {Ilbert}, {Kampczyk}, {Kovac},
  {Lamareille}, {Leauthaud}, {Le Borgne}, {McCracken}, {Marinoni}, {Pello},
  {Ricciardelli}, {Scarlata}, {Vergani}, {Sanders}, {Schinnerer}, {Scoville},
  {Taniguchi}, {Arnouts}, {Aussel}, {Bardelli}, {Brusa}, {Cappi}, {Ciliegi},
  {Finoguenov}, {Foucaud}, {Franceschini}, {Halliday}, {Impey}, {Knobel},
  {Koekemoer}, {Kurk}, {Maccagni}, {Maddox}, {Marano}, {Marconi}, {Meneux},
  {Mobasher}, {Moreau}, {Peacock}, {Porciani}, {Pozzetti}, {Scaramella},
  {Schiminovich}, {Shopbell}, {Smail}, {Thompson}, {Tresse}, {Vettolani},
  {Zanichelli}, \& {Zucca}}]{lilly07}
{Lilly}, S.~J., {Le F{\`e}vre}, O., {Renzini}, A., {et~al.} 2007, \apjs, 172,
  70

\bibitem[{{Lin} {et~al.}(2004){Lin}, {Koo}, {Willmer}, {Patton}, {Conselice},
  {Yan}, {Coil}, {Cooper}, {Davis}, {Faber}, {Gerke}, {Guhathakurta}, \&
  {Newman}}]{lin04}
{Lin}, L., {Koo}, D.~C., {Willmer}, C. N.~A., {et~al.} 2004, \apjl, 617, L9

\bibitem[{{Lin} {et~al.}(2010){Lin}, {Cooper}, {Jian}, {Koo}, {Patton}, {Yan},
  {Willmer}, {Coil}, {Chiueh}, {Croton}, {Gerke}, {Lotz}, {Guhathakurta}, \&
  {Newman}}]{lin10}
{Lin}, L., {Cooper}, M.~C., {Jian}, H.-Y., {et~al.} 2010, \apj, 718, 1158

\bibitem[{{Lintott} {et~al.}(2008){Lintott}, {Schawinski}, {Slosar}, {Land},
  {Bamford}, {Thomas}, {Raddick}, {Nichol}, {Szalay}, {Andreescu}, {Murray}, \&
  {Vandenberg}}]{lintott08}
{Lintott}, C.~J., {Schawinski}, K., {Slosar}, A., {et~al.} 2008, \mnras, 389,
  1179

\bibitem[{{Liske} {et~al.}(2015){Liske}, {Baldry}, {Driver}, {Tuffs},
  {Alpaslan}, {Andrae}, {Brough}, {Cluver}, {Grootes}, {Gunawardhana},
  {Kelvin}, {Loveday}, {Robotham}, {Taylor}, {Bamford}, {Bland-Hawthorn},
  {Brown}, {Drinkwater}, {Hopkins}, {Meyer}, {Norberg}, {Peacock}, {Agius},
  {Andrews}, {Bauer}, {Ching}, {Colless}, {Conselice}, {Croom}, {Davies}, {De
  Propris}, {Dunne}, {Eardley}, {Ellis}, {Foster}, {Frenk}, {H{\"a}u{\ss}ler},
  {Holwerda}, {Howlett}, {Ibarra}, {Jarvis}, {Jones}, {Kafle}, {Lacey},
  {Lange}, {Lara-L{\'o}pez}, {L{\'o}pez-S{\'a}nchez}, {Maddox}, {Madore},
  {McNaught-Roberts}, {Moffett}, {Nichol}, {Owers}, {Palamara}, {Penny},
  {Phillipps}, {Pimbblet}, {Popescu}, {Prescott}, {Proctor}, {Sadler},
  {Sansom}, {Seibert}, {Sharp}, {Sutherland}, {V{\'a}zquez-Mata}, {van Kampen},
  {Wilkins}, {Williams}, \& {Wright}}]{lieske15}
{Liske}, J., {Baldry}, I.~K., {Driver}, S.~P., {et~al.} 2015, \mnras, 452, 2087

\bibitem[{{L{\'o}pez-Sanjuan} {et~al.}(2013){L{\'o}pez-Sanjuan}, {Le
  F{\`e}vre}, {Tasca}, {Epinat}, {Amram}, {Contini}, {Garilli},
  {Kissler-Patig}, {Moultaka}, {Paioro}, {Perret}, {Queyrel}, {Tresse},
  {Vergani}, \& {Divoy}}]{lopez-sanjuan13}
{L{\'o}pez-Sanjuan}, C., {Le F{\`e}vre}, O., {Tasca}, L.~A.~M., {et~al.} 2013,
  \aap, 553, A78

\bibitem[{{Lotz} {et~al.}(2008){Lotz}, {Davis}, {Faber}, {Guhathakurta},
  {Gwyn}, {Huang}, {Koo}, {Le Floc'h}, {Lin}, {Newman}, {Noeske}, {Papovich},
  {Willmer}, {Coil}, {Conselice}, {Cooper}, {Hopkins}, {Metevier}, {Primack},
  {Rieke}, \& {Weiner}}]{lotz08}
{Lotz}, J.~M., {Davis}, M., {Faber}, S.~M., {et~al.} 2008, \apj, 672, 177

\bibitem[{{Lupton} {et~al.}(2004){Lupton}, {Blanton}, {Fekete}, {Hogg},
  {O'Mullane}, {Szalay}, \& {Wherry}}]{lupton04}
{Lupton}, R., {Blanton}, M.~R., {Fekete}, G., {et~al.} 2004, \pasp, 116, 133

\bibitem[{{Masters} {et~al.}(2010){Masters}, {Mosleh}, {Romer}, {Nichol},
  {Bamford}, {Schawinski}, {Lintott}, {Andreescu}, {Campbell}, {Crowcroft},
  {Doyle}, {Edmondson}, {Murray}, {Raddick}, {Slosar}, {Szalay}, \&
  {Vandenberg}}]{masters10}
{Masters}, K.~L., {Mosleh}, M., {Romer}, A.~K., {et~al.} 2010, \mnras, 405, 783

\bibitem[{{Morgan}(1958)}]{morgan1958}
{Morgan}, W.~W. 1958, \pasp, 70, 364

\bibitem[{{Murphy} {et~al.}(2011){Murphy}, {Condon}, {Schinnerer}, {Kennicutt},
  {Calzetti}, {Armus}, {Helou}, {Turner}, {Aniano}, {Beir{\~a}o}, {Bolatto},
  {Brandl}, {Croxall}, {Dale}, {Donovan Meyer}, {Draine}, {Engelbracht},
  {Hunt}, {Hao}, {Koda}, {Roussel}, {Skibba}, \& {Smith}}]{murphy11}
{Murphy}, E.~J., {Condon}, J.~J., {Schinnerer}, E., {et~al.} 2011, \apj, 737,
  67

\bibitem[{{Newman} {et~al.}(2013){Newman}, {Cooper}, {Davis}, {Faber}, {Coil},
  {Guhathakurta}, {Koo}, {Phillips}, {Conroy}, {Dutton}, {Finkbeiner}, {Gerke},
  {Rosario}, {Weiner}, {Willmer}, {Yan}, {Harker}, {Kassin}, {Konidaris},
  {Lai}, {Madgwick}, {Noeske}, {Wirth}, {Connolly}, {Kaiser}, {Kirby},
  {Lemaux}, {Lin}, {Lotz}, {Luppino}, {Marinoni}, {Matthews}, {Metevier}, \&
  {Schiavon}}]{newman13}
{Newman}, J.~A., {Cooper}, M.~C., {Davis}, M., {et~al.} 2013, \apjs, 208, 5

\bibitem[{{Nikolic} {et~al.}(2004){Nikolic}, {Cullen}, \&
  {Alexander}}]{nikolic04}
{Nikolic}, B., {Cullen}, H., \& {Alexander}, P. 2004, \mnras, 355, 874

\bibitem[{{Oke} \& {Gunn}(1983)}]{oke83}
{Oke}, J.~B., \& {Gunn}, J.~E. 1983, \apj, 266, 713

\bibitem[{{Pasachoff}(1999)}]{pasachoff99}
{Pasachoff}, J.~M. 1999, Journal of Astronomical History and Heritage, 2, 39

\bibitem[{{Patton} \& {Atfield}(2008)}]{patton08}
{Patton}, D.~R., \& {Atfield}, J.~E. 2008, \apj, 685, 235

\bibitem[{{Patton} {et~al.}(2000){Patton}, {Carlberg}, {Marzke}, {Pritchet},
  {da Costa}, \& {Pellegrini}}]{patton00}
{Patton}, D.~R., {Carlberg}, R.~G., {Marzke}, R.~O., {et~al.} 2000, \apj, 536,
  153

\bibitem[{{Patton} {et~al.}(2002){Patton}, {Pritchet}, {Carlberg}, {Marzke},
  {Yee}, {Hall}, {Lin}, {Morris}, {Sawicki}, {Shepherd}, \& {Wirth}}]{patton02}
{Patton}, D.~R., {Pritchet}, C.~J., {Carlberg}, R.~G., {et~al.} 2002, \apj,
  565, 208

\bibitem[{{Pop} {et~al.}(2018){Pop}, {Pillepich}, {Amorisco}, \&
  {Hernquist}}]{pop18}
{Pop}, A.-R., {Pillepich}, A., {Amorisco}, N.~C., \& {Hernquist}, L. 2018,
  \mnras, 480, 1715

\bibitem[{{Rodriguez-Gomez} {et~al.}(2015){Rodriguez-Gomez}, {Genel},
  {Vogelsberger}, {Sijacki}, {Pillepich}, {Sales}, {Torrey}, {Snyder},
  {Nelson}, {Springel}, {Ma}, \& {Hernquist}}]{rodriguez15}
{Rodriguez-Gomez}, V., {Genel}, S., {Vogelsberger}, M., {et~al.} 2015, \mnras,
  449, 49

\bibitem[{{Rodriguez-Gomez} {et~al.}(2016){Rodriguez-Gomez}, {Pillepich},
  {Sales}, {Genel}, {Vogelsberger}, {Zhu}, {Wellons}, {Nelson}, {Torrey},
  {Springel}, {Ma}, \& {Hernquist}}]{rodriguez16}
{Rodriguez-Gomez}, V., {Pillepich}, A., {Sales}, L.~V., {et~al.} 2016, \mnras,
  458, 2371

\bibitem[{{Schawinski} {et~al.}(2009){Schawinski}, {Lintott}, {Thomas},
  {Sarzi}, {Andreescu}, {Bamford}, {Kaviraj}, {Khochfar}, {Land}, {Murray},
  {Nichol}, {Raddick}, {Slosar}, {Szalay}, {VandenBerg}, \&
  {Yi}}]{schawinski09}
{Schawinski}, K., {Lintott}, C., {Thomas}, D., {et~al.} 2009, \mnras, 396, 818

\bibitem[{{Schaye} {et~al.}(2015){Schaye}, {Crain}, {Bower}, {Furlong},
  {Schaller}, {Theuns}, {Dalla Vecchia}, {Frenk}, {McCarthy}, {Helly},
  {Jenkins}, {Rosas-Guevara}, {White}, {Baes}, {Booth}, {Camps}, {Navarro},
  {Qu}, {Rahmati}, {Sawala}, {Thomas}, \& {Trayford}}]{schaye15}
{Schaye}, J., {Crain}, R.~A., {Bower}, R.~G., {et~al.} 2015, \mnras, 446, 521

\bibitem[{{Sersic}(1968)}]{sersic68}
{Sersic}, J.~L. 1968, {Atlas de Galaxias Australes}

\bibitem[{{Shimakawa} {et~al.}(2022){Shimakawa}, {Tanaka}, {Bottrell}, {Wu},
  {Chang}, {Toba}, \& {Ali}}]{shimakawa22}
{Shimakawa}, R., {Tanaka}, M., {Bottrell}, C., {et~al.} 2022, \pasj, 74, 612

\bibitem[{{Speagle} {et~al.}(2014){Speagle}, {Steinhardt}, {Capak}, \&
  {Silverman}}]{speagle14}
{Speagle}, J.~S., {Steinhardt}, C.~L., {Capak}, P.~L., \& {Silverman}, J.~D.
  2014, \apjs, 214, 15

\bibitem[{{Strauss} {et~al.}(2002){Strauss}, {Weinberg}, {Lupton}, {Narayanan},
  {Annis}, {Bernardi}, {Blanton}, {Burles}, {Connolly}, {Dalcanton}, {Doi},
  {Eisenstein}, {Frieman}, {Fukugita}, {Gunn}, {Ivezi{\'c}}, {Kent}, {Kim},
  {Knapp}, {Kron}, {Munn}, {Newberg}, {Nichol}, {Okamura}, {Quinn}, {Richmond},
  {Schlegel}, {Shimasaku}, {SubbaRao}, {Szalay}, {Vanden Berk}, {Vogeley},
  {Yanny}, {Yasuda}, {York}, \& {Zehavi}}]{strauss02}
{Strauss}, M.~A., {Weinberg}, D.~H., {Lupton}, R.~H., {et~al.} 2002, \aj, 124,
  1810

\bibitem[{{Tadaki} {et~al.}(2020){Tadaki}, {Iye}, {Fukumoto}, {Hayashi},
  {Rusu}, {Shimakawa}, \& {Tosaki}}]{tadaki20}
{Tadaki}, K.-i., {Iye}, M., {Fukumoto}, H., {et~al.} 2020, \mnras, 496, 4276

\bibitem[{{Tal} {et~al.}(2009){Tal}, {van Dokkum}, {Nelan}, \&
  {Bezanson}}]{tal09}
{Tal}, T., {van Dokkum}, P.~G., {Nelan}, J., \& {Bezanson}, R. 2009, \aj, 138,
  1417

\bibitem[{{Tanaka} {et~al.}(2004){Tanaka}, {Goto}, {Okamura}, {Shimasaku}, \&
  {Brinkmann}}]{tanaka04}
{Tanaka}, M., {Goto}, T., {Okamura}, S., {Shimasaku}, K., \& {Brinkmann}, J.
  2004, \aj, 128, 2677

\bibitem[{{The Dark Energy Survey Collaboration}(2005)}]{des05}
{The Dark Energy Survey Collaboration}. 2005, arXiv e-prints, astro

\bibitem[{{van den Bergh}(1976)}]{vandenbergh76}
{van den Bergh}, S. 1976, \apj, 206, 883

\bibitem[{{Vogelsberger} {et~al.}(2014){Vogelsberger}, {Genel}, {Springel},
  {Torrey}, {Sijacki}, {Xu}, {Snyder}, {Nelson}, \&
  {Hernquist}}]{vogelsberger14}
{Vogelsberger}, M., {Genel}, S., {Springel}, V., {et~al.} 2014, \mnras, 444,
  1518

\bibitem[{{Walmsley} {et~al.}(2022){Walmsley}, {Lintott}, {G{\'e}ron}, {Kruk},
  {Krawczyk}, {Willett}, {Bamford}, {Kelvin}, {Fortson}, {Gal}, {Keel},
  {Masters}, {Mehta}, {Simmons}, {Smethurst}, {Smith}, {Baeten}, \&
  {Macmillan}}]{walmsley22}
{Walmsley}, M., {Lintott}, C., {G{\'e}ron}, T., {et~al.} 2022, \mnras, 509,
  3966

\bibitem[{{Wilkinson} {et~al.}(2022){Wilkinson}, {Ellison}, {Bottrell},
  {Bickley}, {Gwyn}, {Cuillandre}, \& {Wild}}]{wilkinson22}
{Wilkinson}, S., {Ellison}, S.~L., {Bottrell}, C., {et~al.} 2022, \mnras, 516,
  4354

\bibitem[{{Willett} {et~al.}(2013){Willett}, {Lintott}, {Bamford}, {Masters},
  {Simmons}, {Casteels}, {Edmondson}, {Fortson}, {Kaviraj}, {Keel}, {Melvin},
  {Nichol}, {Raddick}, {Schawinski}, {Simpson}, {Skibba}, {Smith}, \&
  {Thomas}}]{willett13}
{Willett}, K.~W., {Lintott}, C.~J., {Bamford}, S.~P., {et~al.} 2013, \mnras,
  435, 2835

\bibitem[{{York} {et~al.}(2000){York}, {Adelman}, {Anderson}, {Anderson},
  {Annis}, {Bahcall}, {Bakken}, {Barkhouser}, {Bastian}, {Berman}, {Boroski},
  {Bracker}, {Briegel}, {Briggs}, {Brinkmann}, {Brunner}, {Burles}, {Carey},
  {Carr}, {Castander}, {Chen}, {Colestock}, {Connolly}, {Crocker}, {Csabai},
  {Czarapata}, {Davis}, {Doi}, {Dombeck}, {Eisenstein}, {Ellman}, {Elms},
  {Evans}, {Fan}, {Federwitz}, {Fiscelli}, {Friedman}, {Frieman}, {Fukugita},
  {Gillespie}, {Gunn}, {Gurbani}, {de Haas}, {Haldeman}, {Harris}, {Hayes},
  {Heckman}, {Hennessy}, {Hindsley}, {Holm}, {Holmgren}, {Huang}, {Hull},
  {Husby}, {Ichikawa}, {Ichikawa}, {Ivezi{\'c}}, {Kent}, {Kim}, {Kinney},
  {Klaene}, {Kleinman}, {Kleinman}, {Knapp}, {Korienek}, {Kron}, {Kunszt},
  {Lamb}, {Lee}, {Leger}, {Limmongkol}, {Lindenmeyer}, {Long}, {Loomis},
  {Loveday}, {Lucinio}, {Lupton}, {MacKinnon}, {Mannery}, {Mantsch}, {Margon},
  {McGehee}, {McKay}, {Meiksin}, {Merelli}, {Monet}, {Munn}, {Narayanan},
  {Nash}, {Neilsen}, {Neswold}, {Newberg}, {Nichol}, {Nicinski}, {Nonino},
  {Okada}, {Okamura}, {Ostriker}, {Owen}, {Pauls}, {Peoples}, {Peterson},
  {Petravick}, {Pier}, {Pope}, {Pordes}, {Prosapio}, {Rechenmacher}, {Quinn},
  {Richards}, {Richmond}, {Rivetta}, {Rockosi}, {Ruthmansdorfer}, {Sandford},
  {Schlegel}, {Schneider}, {Sekiguchi}, {Sergey}, {Shimasaku}, {Siegmund},
  {Smee}, {Smith}, {Snedden}, {Stone}, {Stoughton}, {Strauss}, {Stubbs},
  {SubbaRao}, {Szalay}, {Szapudi}, {Szokoly}, {Thakar}, {Tremonti}, {Tucker},
  {Uomoto}, {Vanden Berk}, {Vogeley}, {Waddell}, {Wang}, {Watanabe},
  {Weinberg}, {Yanny}, {Yasuda}, \& {SDSS Collaboration}}]{york00}
{York}, D.~G., {Adelman}, J., {Anderson}, John~E., J., {et~al.} 2000, \aj, 120,
  1579

\end{thebibliography}

\end{document}